\documentclass[prl,reprint,superscriptaddress]{revtex4-2}

\usepackage{amsmath}
\usepackage{mathtools}
\usepackage{bbm}
\usepackage{overpic}
\usepackage{graphicx}

\usepackage[T1]{fontenc}
\usepackage[utf8]{inputenc}

\usepackage[dvipsnames]{xcolor}
\definecolor{dark-blue}{rgb}{0,0.2,0.6}

\usepackage[
    colorlinks,
    pdfusetitle,
    urlcolor=dark-blue,
    citecolor=dark-blue,
    linkcolor=dark-blue
]{hyperref}

\usepackage{comment}
\usepackage{newtx}
\usepackage{bm}

\DeclareMathAlphabet{\mathcal}{OMS}{cmsy}{m}{n}
\DeclareSymbolFont{CMAlt}{OMX}{cmex}{m}{n}
\DeclareMathSymbol{\sumop}{\mathop}{CMAlt}{"50}
\DeclareMathSymbol{\intop}{\mathop}{CMAlt}{"52}

\usepackage{microtype}
\usepackage{dcolumn}
\usepackage{booktabs}
\usepackage{units}
\usepackage[normalem]{ulem}
\usepackage{braket}
\usepackage{tabularx}
\usepackage{array}

\allowdisplaybreaks

\begin{document}

\title{A magnetic monopole in a superfluid bubble}

\author{Marianna Sorba}
\email[E-mail address: ]{marianna.sorba@ino.cnr.it}
\affiliation{CNR-INO, Area Science Park, Basovizza, 34149 Trieste, Italy}
\author{Andrea Richaud}
\email[E-mail address: ]{andrea.richaud@upc.edu}
\affiliation{Departament de F\'isica, Universitat Polit\`ecnica de Catalunya, Campus Nord B4-B5, E-08034 Barcelona, Spain}

\begin{abstract}
Magnetic monopoles lie at the crossroads of gauge fields, topology, geometric phases, and charge quantization, yet they remain elusive as fundamental particles. Here we show that an emergent Dirac-monopole framework arises naturally from the dynamics of massive quantum vortices in a spherical superfluid shell. Their dynamics is formally equivalent to that of interacting charged particles constrained to a sphere in the field of a magnetic monopole. The monopole charge is fixed by the superfluid density and automatically satisfies Dirac’s quantization condition. The emergent monopole description predicts cyclotron-like vortex motion, in quantitative agreement with Gross--Pitaevskii simulations. We further show that topological frustration induced by two like-charged polar vortices gives rise to the formation of an equatorial vortex necklace, a configuration reminiscent of the polygonal cyclone clusters observed around Jupiter's poles, before its subsequent breakup through a Kelvin--Helmholtz-like instability. Within this framework, the equatorial vortex necklace emerges as the quantized realization of the effective electrostatic line charge that restores global charge neutrality on the sphere. Our results establish spherical superfluids as a versatile platform for realizing and exploring fundamental aspects of Dirac-monopole physics.
\end{abstract}

\maketitle


%
%


\section{Introduction}
Magnetic monopoles occupy a special place in modern physics. Since Dirac's seminal work~\cite{Dirac1931}, they have become a paradigmatic meeting point of gauge fields, topology, and quantum mechanics~\cite{WuYang1975}. A magnetic monopole naturally leads to charge quantization~\cite{Dirac1931}, while its gauge-theoretic description provides one of the most influential examples of gauge geometry in physics and a paradigmatic setting for geometric phases in quantum mechanics~\cite{Simon1983,Berry1984}. Dirac's original formulation, whose gauge potential is necessarily singular along a semi-infinite line known as the Dirac string, was later recast by Wu and Yang in terms of globally consistent gauge patches~\cite{WuYang1975,WuYang1976}, and monopoles subsequently emerged as fundamental objects in gauge theories and grand-unified models through the celebrated solutions of 't Hooft~\cite{tHooft1974} and Polyakov~\cite{Polyakov1974}.

Although no elementary magnetic monopole has yet been observed~\cite{Milton2006}, monopole physics has found remarkable realizations in many-body systems. Examples include emergent magnetic monopoles in spin-ice materials~\cite{Castelnovo2008,Morris2009,Fennell2009}, monopole-like topological defects in superfluid $^3$He~\cite{Volovik1998,Volovik2000,Volovik2020}, and synthetic monopole gauge fields in ultracold atoms~\cite{Dalibard2011,Goldman2014}. In particular, the theoretical prediction~\cite{Pietila2009} and subsequent experimental creation~\cite{Ray2014} of Dirac monopoles in synthetic magnetic fields within spinor Bose--Einstein condensates, together with later studies of their stability and dynamics~\cite{Ollikainen2017}, established ultracold quantum gases as a highly controllable platform for exploring monopole physics within the broader framework of synthetic gauge fields.

Curved quantum fluids offer a natural setting in which effective gauge fields can emerge from geometry itself~\cite{Shi2017} or be engineered through synthetic magnetic monopoles~\cite{Chen2025SyntheticHalf}. A growing body of theoretical work has explored such systems. Bose--Einstein condensation and superfluidity on spherical and shell-shaped manifolds have been investigated theoretically~\cite{Tononi2019,Bereta2019,Moller2020}, while experimentally relevant routes toward shell-shaped condensates have been proposed~\cite{Wolf2022,Veyron2026}. The rapid development of the field is documented in several recent reviews~\cite{Tononi2023,Tononi2024,Dubessy2025,Rhyno2026}.

The possibility of realizing hollow quantum gases was originally anticipated through rf-dressed bubble traps~\cite{Zobay2001,Zobay2004} and later developed theoretically for quantum bubbles~\cite{Tononi2020}. Shell-shaped trapping geometries were subsequently realized experimentally on Earth~\cite{Guo2022,Jia2022}, while fully closed ultracold atomic bubbles have more recently been achieved in microgravity~\cite{Carollo2022}, motivating and enabling investigations of vortex physics on closed surfaces~\cite{Bereta2021,He2023} and of curvature-induced effects in curved quantum fluids~\cite{Biral2024,DeOliveira2025}. 

In these systems, quantized vortices exhibit behaviors with no counterpart in planar geometries, reflecting the coupling between topological defects and curvature~\cite{Bowick2009,Turner2010}. Recent studies have explored vortex-antivortex and vortex-dipole physics in shell-shaped condensates~\cite{Padavic2020,Tononi2026}, vortex dynamics on spherical films~\cite{Bereta2021}, curvature-induced vortex motion on compact manifolds~\cite{Guenther2020,Caracanhas2022}, and nonlinear vortex excitations in spherical quantum fluids~\cite{Cuadra2026}. Notably, the interplay between polar vortices and vortex arrays on a sphere is a recurring motif across scales: long studied in classical point-vortex models~\cite{Polvani1993,Sakajo2004,Sakajo2006}, it has a striking natural counterpart in the persistent polygonal clusters of cyclones encircling Jupiter's poles revealed by the Juno mission~\cite{Adriani2018,Li2020}. 

The sphere also provides a natural arena for monopole physics. In Haldane's spherical geometry, a magnetic monopole placed at the center of a sphere provides a natural description of quantum Hall states on a closed surface~\cite{Haldane1983}. More recently, analogous monopole geometries have been proposed for spherical ultracold-atom systems through synthetic monopole fields, enabling the study of Landau levels and vortex matter~\cite{Zhou2018} as well as of vortex--monopole composite dynamics~\cite{Chen2026IndexTheorem} on curved surfaces.

Despite these advances, a unified and experimentally accessible \emph{dynamical} framework connecting monopole gauge fields, charge quantization, and geometric phases remains lacking. Establishing such a connection would provide a controllable setting in which several hallmark features of Dirac-monopole physics~\cite{Dirac1931,Berry1984,WuYang1975} can be explored simultaneously.

Here we show that massive vortices on a spherical superfluid shell constitute a quantum-fluid realization of the Dirac-monopole paradigm. The monopole charge emerges from the superfluid density and automatically satisfies Dirac's quantization condition, while the resulting monopole field governs the cyclotron-like motion of vortex dipoles. We further show that topological frustration on a closed superfluid surface is resolved through the formation of an equatorial vortex necklace, which constitutes the quantized realization of the effective electrostatic line charge required by global charge neutrality, before undergoing a Kelvin–Helmholtz-like instability.

\begin{figure}[h!]
    \centering
    \includegraphics[width=1.0\linewidth]{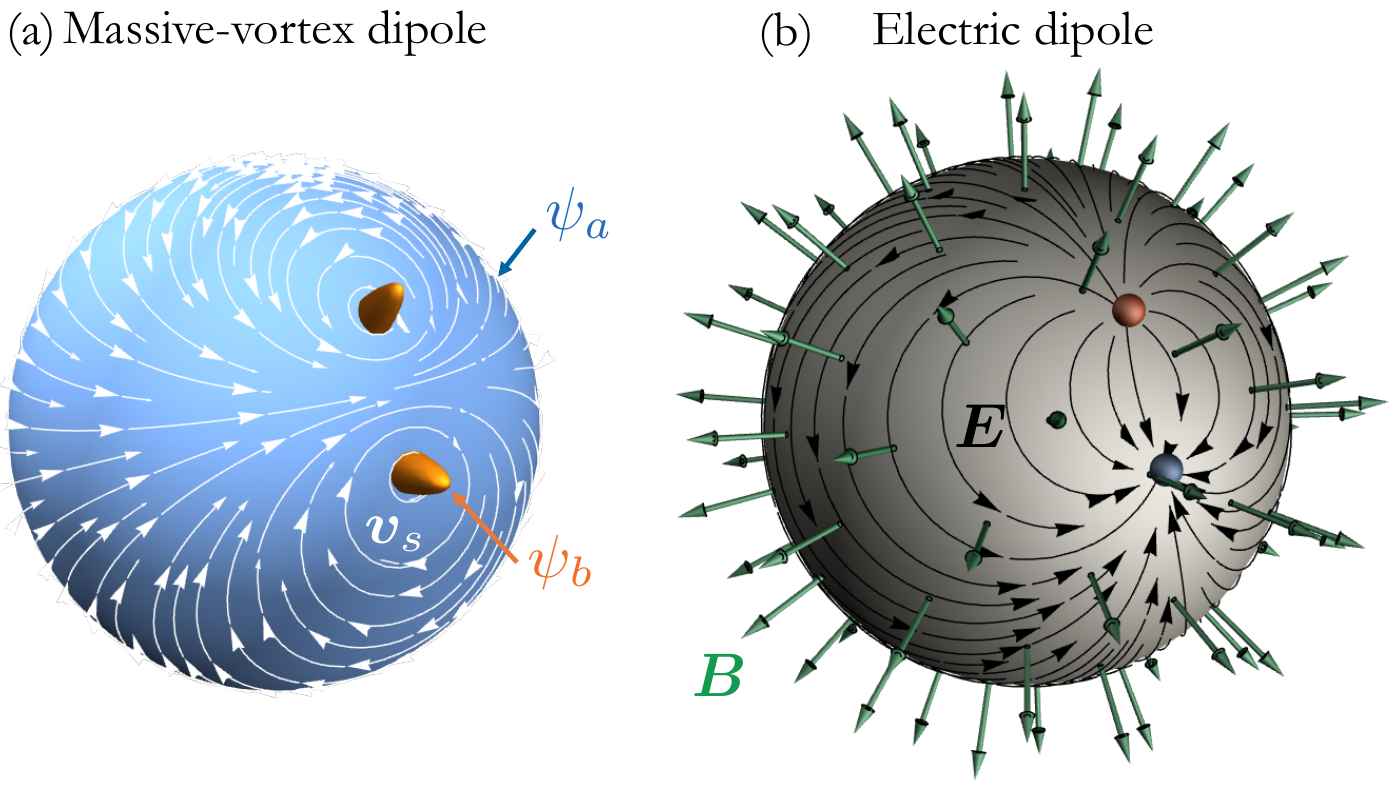}
    \caption{\textbf{Analogy between a massive vortex dipole and an electric dipole in a magnetic monopole field.} (a) Massive vortex dipole in an immiscible binary superfluid confined on a spherical surface. The density of component-$a$, $|\psi_a|^2$, is shown in blue, while the density of component-$b$, $|\psi_b|^2$, localized within the vortex cores, is shown in orange. The white arrows denote the superfluid velocity field $\bm{v}_s$ of component-$a$. (b) Electric dipole constrained to the surface of a sphere in the presence of a magnetic monopole field $\bm{B}$. Black and green arrows denote the electric and magnetic fields, respectively. The dynamics of massive vortices on the sphere is formally analogous to that of charged particles constrained to a spherical surface and subject to a magnetic monopole field.}
    \label{fig:Sketch}
\end{figure}

\section{A Dirac monopole from vortex dynamics} 
The dynamics of $N$ point vortices on the surface of a spherical superfluid film was recently investigated in Ref.~\cite{Bereta2021}. Using a stereographic projection of the sphere onto a tangent plane, the problem was mapped into a complex-potential formulation, allowing the construction of a stream function that completely determines both the vortex interaction energy and the equations of motion.

Denoting by $\mathbf r_j=R\hat{\mathbf r}_j$ the position of the $j$th vortex on a sphere of radius $R$, and by $\Omega_j=(\theta_j,\phi_j)$ its spherical angular coordinates, the vortex interaction energy can be written as
\begin{equation}
E=
\frac{\hbar^2\pi n_a}{m_a}
\left[
\left(\sum_{i=1}^N q_i^2\right)
\ln\left(\frac{R}{\xi}\right)
-
\sum_{i\neq j}
q_iq_j\chi_{ij}
\right],
\label{eq:potential_energy}
\end{equation}
where $N$ is the number of vortices, $n_a$ and $m_a$ denote the surface density and atomic mass of the host condensate, respectively, and $\xi$ is the healing length. Each vortex pair is counted twice in the double sum. Here $\chi_{ij} \equiv \chi(\Omega_i,\,\Omega_j)$ is the pairwise stream function, given by
\begin{equation}
\chi_{ij}
=
\ln\left[
2\sin\left(\frac{\gamma_{ij}}{2}\right)
\right],
\end{equation}
where the angular distance $\gamma_{ij}$ between vortices $i$ and $j$ is defined through $\cos\gamma_{ij} = \cos\theta_i\cos\theta_j + \sin\theta_i\sin\theta_j\cos(\phi_i-\phi_j)$.
The same stream function also determines the equations of motion of massless vortices,
\begin{equation}
\dot{\mathbf r}_k
=
\frac{\hbar}{m_a}
\hat{\mathbf r}_k
\times
\nabla_k
\sum_{j\neq k}
q_j\chi_{kj},
\label{eq:massless_eom}
\end{equation}
where $\nabla_k$ denotes the surface gradient with respect to the position of vortex $k$ on the sphere. Equation~(\ref{eq:massless_eom}) is first order in time and therefore describes purely kinematic vortex dynamics. Unlike the planar case, the compact topology of the sphere requires the total vortex charge to vanish, $\sum_{j=1}^{N} q_j =0$, so that vortices can only occur in charge-neutral configurations, the simplest example being a vortex dipole~\cite{Bereta2021}.

Our starting point is the construction of a Lagrangian whose Euler--Lagrange equations exactly reproduce Eq.~(\ref{eq:massless_eom}) in the absence of inertia. This reverse-engineering procedure provides a natural route toward a massive extension of spherical vortex dynamics~\cite{Richaud2020,Richaud2021PRA,Richaud2022,Caldara2023}. Such an extension is physically relevant whenever vortex cores trap additional matter, including tracer particles~\cite{Bewley2006,Griffin2020,Peretti2023,Tang2023}, quasiparticle bound states~\cite{Kopnin1998,Simula2018,Simonucci2019}, thermal atoms~\cite{Kwon2021,Griffin2009,Jackson2009,Richaud2025}, or a second superfluid component~\cite{Anderson2000,Law2010,Richaud2020,Richaud2021PRA,Williamson2021,Patrick2023,Kanjo2024,Dambroise2025}. As a concrete realization, we consider an immiscible binary Bose--Einstein condensate confined to a thin spherical shell. The wavefunctions $\psi_a$ and $\psi_b$ are normalized to the atom numbers $N_a$ and $N_b$, respectively. Denoting by $m_a$ and $m_b$ the atomic masses of the two components, the corresponding total masses are $M_a=m_aN_a$ and $M_b=m_bN_b$. The majority component $a$ forms a superfluid film with density $n_a=N_a/(4\pi R^2)$ and hosts $N$ quantized vortices, while atoms of the minority component $b$ become localized inside their cores. Assuming that the total mass $M_b$ is evenly distributed among the vortices, each vortex acquires an effective inertial mass $M_b/N$.

The resulting effective Lagrangian reads
\begin{equation}
L=
\sum_{j=1}^{N}
\left[
\frac{M_b}{2N}\dot{\mathbf r}_j^2
+
q_j\mathbf A_j\cdot\dot{\mathbf r}_j
\right]
-
E(\{\mathbf r_j\}),
\label{eq:lagrangian}
\end{equation}
where $E$ is given by Eq.~(\ref{eq:potential_energy}).
Central to this construction is the emergence of the effective gauge potential
\begin{equation}
\mathbf A_j(R,\theta_j)
=
\frac{g}{R}
\frac{1-\cos\theta_j}{\sin\theta_j}
\hat{\boldsymbol\phi}_j, 
\label{eq:vector_potential}
\end{equation}
where the coefficient $g$ is fixed by the superfluid density according to $g=-2\pi\hbar n_aR^2=-\hbar N_a/2$. Equation~(\ref{eq:vector_potential}) is identical to the vector potential introduced by Dirac for a magnetic monopole~\cite{Dirac1931}. The associated magnetic field is
\begin{equation}
\mathbf B_j(R)
=
\boldsymbol{\nabla}\times\mathbf A_j
=
\frac{g}{R^2}\hat{\mathbf r}_j,
\label{eq:magnetic_field}
\end{equation}
which is precisely the radial field generated by a monopole of magnetic charge $g$ located at the center of the sphere (as illustrated in Fig.~\ref{fig:Sketch}(b), the field pierces the spherical surface normally at every point). The monopole is therefore not introduced as an additional ingredient of the theory; rather, it emerges naturally from the effective gauge structure required to reproduce the known vortex dynamics on the sphere. Consequently, the massive-vortex problem admits a direct interpretation in terms of interacting charged particles constrained to a spherical surface and subject to a monopole magnetic field, i.e., the same monopole geometry that underlies Haldane's spherical formulation of the quantum Hall effect~\cite{Haldane1983}, as illustrated schematically in Fig.~\ref{fig:Sketch}. Furthermore, since the effective monopole charge is proportional to the total atom number of the majority component $N_a$, for a vortex of topological charge $q_j$ one finds $q_j g = -\hbar q_jN_a/2$. Since both $q_j$ and $N_a$ are integers, the emergent monopole charge
automatically satisfies
\begin{equation}
q_j g
=
\frac{n\hbar}{2},
\qquad
n=-q_jN_a\in\mathbb Z,
\label{eq:Dirac_quantization_condition}
\end{equation}
which is precisely Dirac's quantization condition~\cite{Dirac1931}.
Remarkably, a condition originally introduced to ensure the consistency of magnetic monopoles emerges here directly from two fundamental properties of the superfluid: quantized vortex circulation and integer atom number.

A complementary perspective on Dirac's quantization condition emerges from the geometric phase~\cite{Anandan1992,Polkinghorne2021} accumulated by a massive vortex core moving on the sphere. As the component-$b$ massive vortex core is transported along a closed trajectory $\mathcal C$ by its component-$a$ host vortex, it acquires a phase factor of purely geometric origin~\cite{Berry1984,Simon1983,Aharonov1987}
\begin{equation}
\Phi_{\rm geom}
=
\frac{q_j}{\hbar}
\oint_{\mathcal C}
\mathbf A\cdot d\mathbf r
=
\frac{q_j g}{\hbar}\,
\Omega_{\mathcal C},
\label{eq:geometric_phase}
\end{equation}
where $\Omega_{\mathcal C}$ is the oriented solid angle enclosed by the trajectory (the geometric phase itself is a property of the host superfluid and is already present in the massless limit~\cite{Haldane1985}; the massive core provides a localized probe through which it becomes directly addressable). For a uniform precession at fixed polar angle $\theta$, choosing the spherical cap containing the north pole, one has $\Omega_{\mathcal C}=2\pi(1-\cos\theta)$ and therefore $\Phi_{\rm geom}=\frac{q_j g}{\hbar}\,2\pi(1-\cos\theta)$. Substituting the effective monopole charge $g=-\hbar N_a/2$ and using $N_{\rm enc}=N_a\Omega_{\mathcal C}/4\pi$ for the number of component-$a$ atoms enclosed by the trajectory, Eq.~(\ref{eq:geometric_phase}) becomes
\begin{equation}
\Phi_{\rm geom}
=
-2\pi q_j N_{\rm enc},
\end{equation}
recovering the celebrated result that the Berry phase accumulated by a vortex is equal (up to the sign convention for the vortex circulation) to $2\pi$ times the number of enclosed superfluid particles~\cite{Haldane1985}. This provides an independent consistency check of the monopole mapping. 
The same contour can also be associated with the complementary spherical surface, whose solid angle differs by $4\pi$. This ambiguity reflects the fact that the monopole field admits different gauge representations related by patch transformations, as in the Wu--Yang formulation of magnetic monopoles~\cite{WuYang1975,WuYang1976}. The geometric phase is therefore defined modulo $\Delta\Phi_{\rm geom} =
4\pi q_j g/\hbar$. Requiring the associated phase factor to be single-valued, $\exp(i\Delta\Phi_{\rm geom})=1$, yields again Dirac's quantization condition~(\ref{eq:Dirac_quantization_condition})~\cite{Dirac1931,WuYang1975}.

The monopole analogy extends beyond the Lagrangian formulation and directly governs the vortex dynamics. The Euler--Lagrange equations associated with Eq.~(\ref{eq:lagrangian}) take the Newton-like form
\begin{equation}
\frac{M_b}{N}\ddot{\mathbf r}_j
=
q_j\,\dot{\mathbf r}_j\times\mathbf B_j
+
q_j\sum_{i\neq j} q_i
\frac{2\pi\hbar^2 n_a}{m_a}
\left[
\frac{\mathbf r_j-\mathbf r_i}
{|\mathbf r_j-\mathbf r_i|^2}
\right]_{\!\parallel},
\label{eq:LorentzCoulomb}
\end{equation}
which manifestly has the form of a Lorentz force acting in the monopole magnetic field (\ref{eq:magnetic_field}) together with a two-dimensional Coulomb interaction arising from the logarithmic vortex potential (\ref{eq:potential_energy}). Here the Coulomb force is understood as its tangential projection onto the spherical surface, while the radial component is balanced by the constraint force that confines the vortices to the sphere. Equivalently, the interaction energy $E$ plays the role of an effective electrostatic potential, whose (negative) surface gradient generates the effective electric field illustrated schematically in Fig.~\ref{fig:Sketch}(b). Equation~(\ref{eq:LorentzCoulomb}) therefore establishes a direct correspondence between massive vortices on a spherical superfluid and interacting charged particles constrained to a sphere and moving in the field of a magnetic monopole, as illustrated schematically in Fig.~\ref{fig:Sketch}.

\section{Cyclotron-like motion of a vortex dipole}
\begin{figure}[h!]
    \centering
    \begin{overpic}[width=1.0\linewidth]{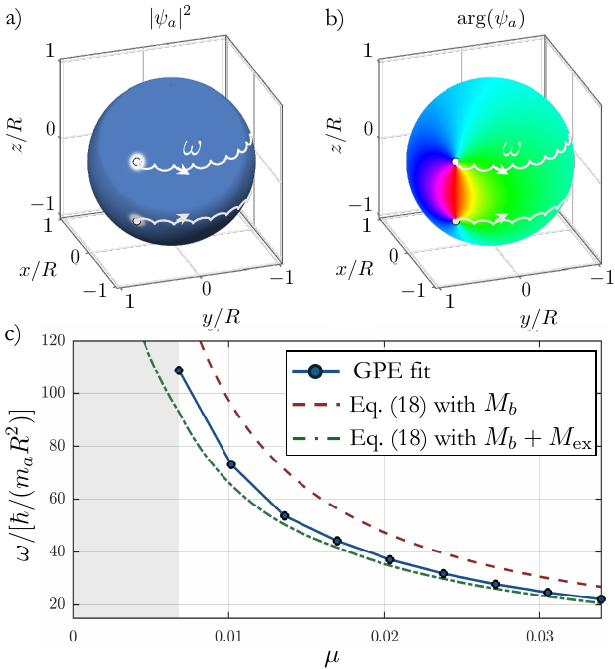}
    \put(52,32.4){\color{white}\rule{34.5mm}{9mm}}
    \put(52.9,38.27){Eq.~(\ref{eq:normal_modes}) with $M_b$}
    \put(52.9,33.3){Eq.~(\ref{eq:normal_modes}) with $M_b+M_{\mathrm{ex}}$}
    \end{overpic}
    \caption{\textbf{Dynamics of a massive vortex dipole on a spherical superfluid.} (a) Density profile $|\psi_a|^2$ of a vortex dipole symmetrically located with respect to the equator. The density is approximately uniform over the sphere (blue) except at the vortex cores, where it vanishes (white). The white curves show the vortex trajectories, which result from the superposition of a uniform azimuthal precession and a small-amplitude oscillation of frequency $\omega$. (b) Corresponding phase field $\arg(\psi_a)$. (c) Dimensionless oscillation frequency $\omega m_a R^2/\hbar$ as a function of the dimensionless vortex mass $\mu=M_b/M_a$. Blue dots: numerical simulations; error bars obtained from the sinusoidal fit are smaller than the symbol size. The red dashed curve corresponds to the prediction of Eq.~(\ref{eq:normal_modes}) obtained from linear-response theory assuming a vortex-core mass equal to the mass of component-$b$ atoms localized within the core. The green dashed curve includes, in addition, the excluded mass associated with the vortex-core density depletion.}
    \label{fig:trajectories_and_omega_vs_mu}
\end{figure}

We now consider the simplest charge-neutral configuration, namely a vortex dipole composed of a vortex and an antivortex with charges $q_\pm=\pm1$. In the massless limit (i.e. for $M_b\to 0^+$), the two vortices move at fixed angular separation and precess uniformly around the sphere~\cite{Bereta2021}. For a dipole symmetric with respect to the equator, $\theta_+=\theta$, $\theta_-=\pi-\theta$, and $\phi_+=\phi_-=\phi$, Ref.~\cite{Bereta2021} gives a purely azimuthal motion, i.e.
\begin{equation}
\dot{\mathbf r}_\pm
=
\frac{\hbar}{2m_aR}\tan\theta\,\hat{\boldsymbol\phi},
\qquad
\dot\phi
=
\frac{\hbar}{2m_aR^2\cos\theta}.
\label{eq:massless_dipole_precession}
\end{equation}
The introduction of vortex inertia ($M_b>0$) qualitatively modifies this purely kinematic motion~\cite{Richaud2020,Richaud2021PRA}. In general, the Lorentz-force term associated with the effective magnetic field produces an additional oscillatory motion superimposed on the uniform precession, as illustrated by the vortex trajectories in Fig.~\ref{fig:trajectories_and_omega_vs_mu}(a)-(b).

A Hamiltonian formulation of the vortex dipole makes the origin of this cyclotron-like dynamics particularly transparent. A detailed analogy with plasma orbit theory has been discussed in Ref.~\cite{Caldara2023}, building on concepts that are standard in plasma physics~\cite{Chen2016}. For $N=2$ vortices, the canonical
angular momenta associated with Eq.~(\ref{eq:lagrangian}) are $l_{\theta_j} = \frac{M_b}{2}R^2\dot\theta_j$ and $l_{\phi_j}=\frac{M_b}{2}R^2\sin^2\theta_j\,\dot\phi_j
+
gq_j(1-\cos\theta_j)$,
with $j=\pm$.
The momentum $l_{\theta_j}$ is purely mechanical, whereas $l_{\phi_j}$ contains both a mechanical contribution and a gauge contribution associated with the monopole vector potential. The relevant Hamiltonian takes the minimal-coupling form
\begin{equation}
H_{\rm dip}
=
\frac{1}{M_b}
\sum_{j=\pm}
\left[
\mathbf p_j-q_j\mathbf A_j(R,\theta_j)
\right]^2
+
E_{\rm dip},
\label{eq:Hdip_minimal}
\end{equation}
where $E_{\rm dip}$ follows directly from Eq.~(\ref{eq:potential_energy}) and $
\mathbf p_j
=
p_{\theta_j}\hat{\boldsymbol\theta}_j
+
p_{\phi_j}\hat{\boldsymbol\phi}_j$, with $
p_{\theta_j}
=l_{\theta_j}/R,$ and
$
p_{\phi_j}
=l_{\phi_j}/(R\sin\theta_j)$.
Since $H_{\rm dip}$ depends only on the relative azimuthal angle
$\phi_+-\phi_-$, the total canonical angular momentum
$
L_z^{\rm can}
=
l_{\phi_+}+l_{\phi_-}
$
is conserved by Noether's theorem.

Equation~(\ref{eq:Hdip_minimal}) is precisely the Hamiltonian of two charged particles of mass $M_b/2$ moving on a sphere and minimally coupled to the monopole vector potential~(\ref{eq:vector_potential}). This correspondence closely parallels the charged-particle problem introduced by Haldane in his spherical formulation of the quantum Hall effect~\cite{Haldane1983}.

A normal-mode analysis of the massive vortex dipole can be performed by linearizing the Hamiltonian dynamics around the uniformly precessing dipole orbit. The resulting eigenfrequencies depend on both the vortex mass and the dipole configuration; technical details are reported in the
Supplemental Material. Introducing the canonical vector
$
\mathbf z
=
\left(
\theta_+,\phi_+,\theta_- ,\phi_-,
l_{\theta_+},l_{\phi_+},l_{\theta_-},l_{\phi_-}
\right)^T,
$
and moving to a frame rotating at angular frequency $\lambda$ through $\phi_j'=\phi_j-\lambda t$ and $\theta_j'=\theta_j$, the Hamiltonian becomes
$
H'_{\rm dip}
=
H_{\rm dip}
-
\lambda
\sum_{j=\pm}l_{\phi_j}.
$
Denoting by $\mathbf z'_{\rm eq}$ the fixed point of the rotating-frame dynamics, small oscillations obey
$\delta\dot{\mathbf z}'
=
\mathbb J\,\mathbb H(\mathbf z'_{\rm eq})\,\delta\mathbf z'$,
where
\begin{equation}
\mathbb J=
\begin{pmatrix}
\mathbf 0_4 & \mathbf I_4\\
-\mathbf I_4 & \mathbf 0_4
\end{pmatrix},
\qquad
\mathbb H(\mathbf z'_{\rm eq})
=
\left.
\frac{\partial^2 H'_{\rm dip}}
{\partial z_\alpha\partial z_\beta}
\right|_{\mathbf z'_{\rm eq}} .
\end{equation}
The normal-mode frequencies are obtained from the eigenvalue problem
\begin{equation}
\mathbb J\,\mathbb H(\mathbf z'_{\rm eq})\,\mathbf u_\nu
=
\lambda_\nu \mathbf u_\nu,
\qquad
\lambda_\nu=\pm i\omega_\nu ,
\label{eq:normal_modes}
\end{equation}
where $\mathbf u_\nu$ denotes the corresponding eigenvector. A particularly transparent limit is provided by the antipodal ($\theta = 0$) dipole configuration. In this configuration, the vortex--vortex interaction force vanishes by symmetry, so that the linearized dynamics is dominated by the monopole field. Because the latter has constant magnitude $|\mathbf B|=|g|/R^2$ over the sphere, the oscillatory motion admits a simple interpretation in terms of cyclotron dynamics. The corresponding characteristic frequency is
\begin{equation}
\omega_{\rm c}
=
\frac{|q_\pm|\,|\mathbf B|}{M_b/2}
=
\frac{2|g|}{M_bR^2}
=
\frac{4\pi\hbar n_a}{M_b},
\label{eq:cyclotron_frequency}
\end{equation}
which is the direct analogue of the cyclotron frequency of a charged particle moving in the monopole magnetic field~\cite{Caldara2023}. Similar cyclotron dynamics underlies a broad range of physical systems, including plasma orbits~\cite{Chen2016}, mass spectrometry~\cite{Maher2015}, Penning traps~\cite{Brown1986}, cyclotron resonance measurements in semiconductors~\cite{Luttinger1956}, and fractional quantum Hall systems~\cite{Munoz_de_las_Heras2020}. In all these cases, the cyclotron frequency is often used to infer the mass or effective mass of the underlying particle or quasiparticle~\cite{Richaud2025}. Likewise, because the monopole charge is fixed by the superfluid density $n_a$, the cyclotron frequency~(\ref{eq:cyclotron_frequency}) is determined entirely by microscopic superfluid parameters.

To benchmark the effective point-vortex theory, we perform real-time Gross--Pitaevskii simulations of a two-component condensate on the sphere, described by
\begin{equation}
i\hbar\partial_t\psi_i
=
\left[
-\frac{\hbar^2}{2m_iR^2}\nabla_{\mathbb{S}}^2
+
\sum_{j=a,b}g_{ij}|\psi_j|^2
\right]\psi_i,
\qquad
i=a,b,
\label{eq:coupled_GPE}
\end{equation}
where $\nabla_{\mathbb{S}}^2$ denotes the Laplace--Beltrami operator on the unit sphere. We focus on the immiscible regime, $g_{ab}>\sqrt{g_{aa}g_{bb}}$, where
$g_{ij}$ denote the effective two-dimensional interaction strengths between components $i$ and $j$. In this regime, the minority component $b$ remains localized inside the vortex cores of the majority component $a$, thereby providing the vortex inertia~\cite{Richaud2020,Richaud2021PRA}. In the simulations, we use parameters representative of an immiscible $^{23}{\rm Na}$--$^{39}{\rm K}$ mixture, as detailed in the Supplemental Material. The equations are solved numerically on a spherical grid, with the Laplace--Beltrami operator evaluated through spherical-harmonic transforms implemented in the SSHT package~\cite{McEwen2011,SSHT}. The vortex trajectories are then obtained by tracking the centers of mass of the localized density peaks of component $b$ [see Fig.~\ref{fig:trajectories_and_omega_vs_mu}(a)-(b)]. The oscillation frequency is extracted through a sinusoidal fit of the tracked trajectories. The fitted frequency is identified with the cyclotron-like normal mode obtained from the linear stability analysis.

The resulting frequencies are shown as blue symbols in Fig.~\ref{fig:trajectories_and_omega_vs_mu}(c) as a function of the dimensionless core mass $\mu=M_b/M_a$. Good agreement is found with the analytical prediction obtained from the cyclotron-like eigenfrequency extracted from the normal-mode analysis of Eq.~(\ref{eq:normal_modes}). The agreement is further improved when the effective vortex mass is taken as $ M_{\rm cores} = M_b+M_{\rm ex}$, where $M_{\rm ex}$ denotes the excluded mass associated with the density depletion inside the vortex core~\cite{Simula2018,Griffin2020,Giuriato2020}. This suggests that both the mass of the minority component trapped in the cores and the depleted mass of the majority component contribute to the inertial response of the vortices. To facilitate reproducibility, the simulation code and analysis routines used in this work are publicly available in Refs.~\cite{GitHubCodeMassiveDipole,GitHubCodeKHI}. Further details on the two-component condensate, the numerical discretization, vortex tracking, frequency extraction, and the robustness of the massive-vortex-dipole dynamics against experimentally relevant residual gravitational accelerations are reported in the Supplemental Material.

\section{Quantized shear layers and effective electrostatics}
In a simply connected planar superfluid, the net topological charge is fixed by the circulation imposed at the boundaries. On a closed spherical surface, however, the situation is fundamentally different: the net vortex charge is constrained to vanish by the topology of the sphere~\cite{Bereta2021}. Imprinting two vortices of identical topological charge $q>0$ at opposite poles attempts to impose a flow configuration that is incompatible with the topological constraints of a closed spherical surface. To explore how the system resolves this topological frustration, we initialize the condensate with a vortex of charge $q>0$ at each pole. The corresponding order parameters may be written locally as $\psi_N=|\psi_N|\,e^{iq\phi}$ and $ \psi_S=|\psi_S|\,e^{-iq\phi}$, where the two expressions describe the northern and southern hemispheres, respectively. The corresponding superfluid velocity fields are
\begin{equation}
\mathbf v_N^{(q)}
=
\frac{q\kappa}{2\pi R\sin\theta}\,
\hat{\boldsymbol\phi},
\qquad
\mathbf v_S^{(q)}
=
-\frac{q\kappa}{2\pi R\sin\theta}\,
\hat{\boldsymbol\phi},
\end{equation}
(where $\kappa=h/m_a$), which become counter-propagating at the equator, $\mathbf v_N^{(q)}\big|_{\theta=\pi/2}=-\mathbf v_S^{(q)}\big|_{\theta=\pi/2}$.
Unlike the previous sections, where the minority component $\psi_b$ endows the vortex cores with inertia, the physics discussed below involves only the host superfluid. Accordingly, the analysis presented below is performed solely in terms of $\psi_a$; this is consistent with the fact that the gauge structure underlying the monopole analogy is already present in the massless vortex dynamics and does not rely on vortex inertia.

\begin{figure*}
    \centering
    \includegraphics[width=1.0\textwidth]{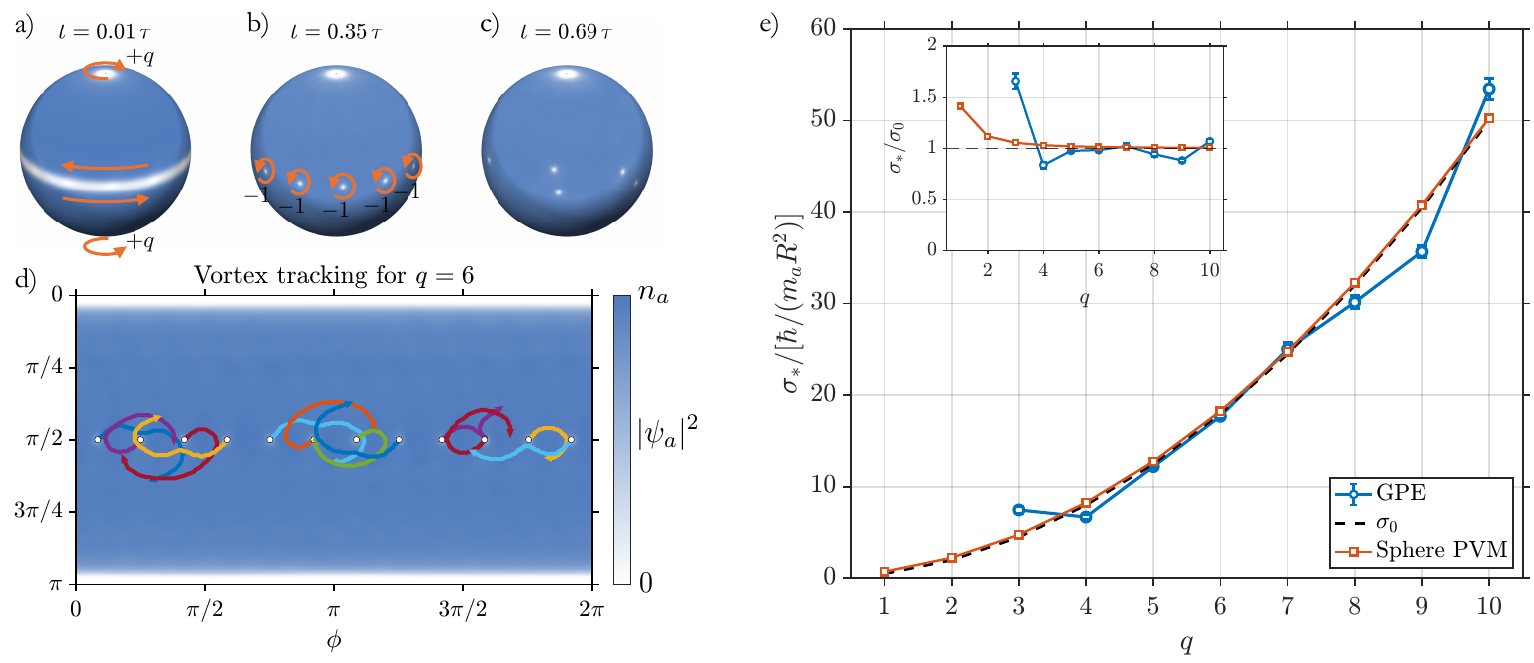}
    \caption{\textbf{Rise and fall of a superfluid shear layer on a sphere.} (a) Initial state with two vortices of charge $+q$ imprinted at the poles and separated by an equatorial barrier. The characteristic time scale of the system is $\tau = m_a R^2/\hbar \simeq 0.145~\mathrm{s}$. (b) Formation of an equatorial vortex necklace composed of $2q$ vortices of charge $-1$. (c) Kelvin--Helmholtz-like breakup of the vortex necklace. (d) Trajectories of the $2q$ vortices on the planisphere obtained from Gross--Pitaevskii simulations following the removal of the equatorial barrier. (e) Maximum instability growth rate $\sigma_\ast$, normalized by the frequency unit $\tau^{-1}=\hbar/(m_aR^2)$, as a function of the polar vortex charge $q$. Blue circles (connected by blue segments) denote values extracted from the vortex-tracking analysis of the Gross--Pitaevskii simulations (see Supplemental Material for further details). The black dashed curve corresponds to the analytical prediction $\sigma_0\sim q^2$ for an infinite vortex chain [see Eq.~(\ref{eq:sigma_0_Aref}) and relevant discussion]~\cite{Aref1995}, while the solid red curve shows the prediction of the spherical point-vortex model (PVM), which accounts for both the spherical geometry and the interaction between the equatorial vortex necklace and the two polar vortices (see Supplemental Material). The inset displays the normalized growth rate $\sigma_\ast/\sigma_0$, highlighting the convergence toward the infinite-chain prediction in the large-$q$ limit.}
    \label{fig:KHI}
\end{figure*}

An equatorial potential barrier, modeled by supplementing Eq.~(\ref{eq:coupled_GPE}) with an additional time-dependent external potential $V(\mathbf r,t)$, initially keeps the two hemispherical superfluids effectively decoupled [Fig.~\ref{fig:KHI}(a)].  Lowering the barrier subsequently brings the two azimuthal flows into contact at the equator, generating a superfluid shear layer whose vorticity satisfies
\begin{equation}
\left(\nabla_{\mathbb{S}}\times\mathbf v\right)\cdot\hat{\mathbf r}
=
\frac{1}{R}\left[
\left(\mathbf v_S^{(q)}-\mathbf v_N^{(q)}\right)
\cdot
\hat{\boldsymbol\phi}
\right]
\delta\!\left(\theta-\frac{\pi}{2}\right),
\label{eq:vortex_sheet}
\end{equation}
where $\nabla_{\mathbb{S}} \times$ denotes the intrinsic surface curl on the sphere and $\mathbf{v}$ is the superfluid velocity field. Unlike a classical shear layer, however, a superfluid cannot sustain an arbitrary continuous distribution of vorticity: circulation is quantized, and the vorticity sheet therefore breaks up into discrete phase singularities. Consequently, the system nucleates an equatorial vortex necklace consisting of $2q$ singly quantized vortices of charge $-1$ [Fig.~\ref{fig:KHI}(b)]. The number of vortices in the necklace is fixed by global charge neutrality, ensuring that the total vortex charge on the sphere remains zero. The necklace therefore provides a natural mechanism through which the system relaxes the topological frustration associated with the imposed polar circulation. The resulting structure (a regular vortex array coexisting with polar vortices on a sphere) is structurally reminiscent of the cyclone polygons encircling Jupiter's poles~\cite{Adriani2018}, whose longevity is understood to rely on anticyclonic shielding of the individual vortices~\cite{Li2020}. More broadly, the vortex necklace plays a role analogous to the defect structures that emerge in frustrated ordered systems on curved manifolds, where global topological constraints are accommodated through defect nucleation~\cite{Bowick2009}.

This process admits a natural electrostatic interpretation in terms of effective electric fields on the sphere. The counter-propagating hemispherical flows are associated with the fields
\begin{equation}
\mathbf E_N^{(q)}
=
\frac{2\pi\hbar^2n_aq}
{m_aR\sin\theta}\,
\hat{\boldsymbol\theta},
\qquad
\mathbf E_S^{(q)}
=
-\frac{2\pi\hbar^2n_aq}
{m_aR\sin\theta}\,
\hat{\boldsymbol\theta},
\label{eq:effective_fields_NS}
\end{equation}
whose explicit construction is provided in the Supplemental Material. They are related [see Eq.~(\ref{eq:LorentzCoulomb})] to the hemispherical superfluid velocities through
\begin{equation}
\mathbf E_\alpha^{(q)}
= \mathbf{B}\times\mathbf{v}_\alpha^{(q)}=
-2\pi\hbar n_a\,
\hat{\mathbf r}\times\mathbf v_\alpha^{(q)},
\qquad
\alpha=N,S,
\label{eq:E_velocity_rotation}
\end{equation}
so that the discontinuity of the azimuthal flow across the equator is mapped onto a discontinuity of the $\hat{\boldsymbol\theta}$ component of the electric field.

The effective electric fields obey Gauss's law on the sphere,
\begin{equation}
\nabla_{\mathbb{S}}\cdot\mathbf E
=
\mathcal G\,\rho,
\qquad
\mathcal G
=
\frac{4\pi^2\hbar^2n_a}{m_a},
\end{equation}
where $\rho$ denotes the effective surface charge density and $\mathcal G$ plays the role of the (inverse) permittivity of this effective two-dimensional electrostatics. Applying the Maxwell boundary condition at the equatorial interface,
\begin{equation}
\label{eq:Maxwell_condition}
(\mathbf E_S^{(q)}-\mathbf E_N^{(q)})\cdot\hat{\boldsymbol\theta}
=
\mathcal G\,\lambda_{\rm eq},
\end{equation}
yields the effective line-charge density at the equator
\begin{equation}
\lambda_{\rm eq}
=
-\frac{q}{\pi R},
\end{equation}
corresponding to the total charge
\begin{equation}
Q_{\rm eq}
=
2\pi R\,\lambda_{\rm eq}
=
-2q.
\end{equation}
Using Eq.~(\ref{eq:E_velocity_rotation}), the Maxwell boundary condition~(\ref{eq:Maxwell_condition}), gives 
\begin{equation}
\left(\mathbf v_S^{(q)}-\mathbf v_N^{(q)}\right)\cdot\hat{\boldsymbol\phi}
=
\kappa\lambda_{\rm eq}.
\label{eq:hydrodynamic_jump}
\end{equation}
Together with Eq.~(\ref{eq:vortex_sheet}), this shows that the effective line-charge density is the electrostatic counterpart of the equatorial vorticity sheet. In a classical electrostatic problem, this charge would form a continuous line distribution along the equator; its hydrodynamic counterpart is the continuous vortex sheet of Eq.~(\ref{eq:vortex_sheet}). Quantization of circulation discretizes the sheet into singly quantized vortices, each carrying effective charge $-1$, so that $Q_{\rm eq}=-2q$ is realized by the equatorial necklace of $2q$ vortices, in agreement with the Gross--Pitaevskii simulations.

The subsequent breakup of the necklace [Fig.~\ref{fig:KHI}(c)], which, unlike its shielded Jovian counterparts~\cite{Li2020}, lacks any stabilizing mechanism, realizes a Kelvin--Helmholtz-like instability of this quantized shear layer~\cite{Takeuchi2010,Baggaley2018,Hernandez2024,Caldara2024}, consistent with classical analyses of vortex sheets and rings on the sphere~\cite{Polvani1993}, including their coupling to pole vortices~\cite{Sakajo2004,Sakajo2006}. The vortex trajectories obtained from the Gross--Pitaevskii simulations [see, e.g., Fig.~\ref{fig:KHI}(d)], reveal the growth of perturbations of the vortex necklace and the eventual departure from the initially regular equatorial configuration $\theta_j=\pi/2$, $\phi_j=\pi j/q$. Following the procedures described in the Supplemental Material, we extract the maximum instability growth rate $\sigma_\ast$ from the early-time exponential growth of the most unstable Fourier mode of the necklace and compute the corresponding prediction of the spherical point-vortex model through a linear stability analysis.

The resulting values of $\sigma_\ast$ are reported in Fig.~\ref{fig:KHI}(e) as a function of the polar vortex charge $q$. As $q$ increases, the number of vortices forming the necklace grows proportionally, while the corresponding equatorial intervortex spacing, $a=\pi R/q$, decreases. In this regime, the necklace increasingly resembles a locally flat vortex chain and the measured growth rate approaches the point-vortex prediction of Aref~\cite{Aref1995},
\begin{equation}
\label{eq:sigma_0_Aref}
\sigma_0=\frac{\pi\kappa}{4a^2},
\end{equation}
derived for an infinite row of identical point vortices. The spherical point-vortex model developed in this work, whose linear stability analysis is presented in the Supplemental Material, captures the finite-curvature corrections arising from the closed geometry and from the interaction between the equatorial necklace and the two polar vortices, providing excellent agreement with the Gross--Pitaevskii simulations over the entire range of $q$ considered. The convergence toward Aref's prediction, highlighted in the inset of Fig.~\ref{fig:KHI}(e), demonstrates how a topologically induced vortex necklace on a closed sphere continuously approaches the Kelvin--Helmholtz instability of an infinite vortex row.

\section{Conclusions}
In summary, we have shown that the Dirac-monopole paradigm, describing an electric charge moving in the field of a magnetic monopole, can be realized using massive quantum vortices on a spherical superfluid film. The resulting monopole framework yields Dirac's quantization condition, admits a geometric-phase interpretation, and predicts cyclotron-like vortex dynamics in quantitative agreement with Gross--Pitaevskii simulations. We further showed that topological frustration on a closed superfluid surface is relieved by the formation of an equatorial vortex necklace, the quantized realization of the effective electrostatic line charge required by global charge neutrality, before its subsequent Kelvin–Helmholtz-like breakup. These results establish spherical superfluids as a versatile platform for exploring monopole physics and related phenomena at the interface of gauge fields, topology, and curved quantum matter.

\begin{acknowledgments}
\section{Acknowledgments}
We thank Matteo Caldara, Pietro Massignan, Miguel Onorato, and Andrea Tononi for stimulating discussions. A.R. received funding from ``La Caixa'' Foundation through the Junior Leader Fellowship LCF/BQ/PR25/12110014. A.R. further acknowledges financial support by the Spanish Ministerio de Ciencia, Innovación y Universidades (grant PID2023-147469NB-C21, financed by MICIU/AEI/10.13039/501100011033 and FEDER-EU).
\end{acknowledgments}

%
\begingroup

\endgroup



\clearpage

\onecolumngrid

%

\setcounter{section}{0}
\setcounter{subsection}{0}
\setcounter{subsubsection}{0}

\setcounter{equation}{0}
\setcounter{figure}{0}
\setcounter{table}{0}
\setcounter{footnote}{0}

\setcounter{secnumdepth}{3}

\setcounter{part}{0}
\setcounter{paragraph}{0}
\setcounter{subparagraph}{0}


\renewcommand{\theequation}{S\arabic{equation}}
\renewcommand{\thefigure}{S\arabic{figure}}
\renewcommand{\thetable}{S\arabic{table}}
\renewcommand{\thesection}{S\arabic{section}}
\renewcommand{\thesubsection}{\thesection.\Alph{subsection}}
\makeatletter
\renewcommand{\p@subsection}{}
\renewcommand{\p@subsubsection}{}
\makeatother
\renewcommand{\theHequation}{SM.\arabic{equation}}
\renewcommand{\theHfigure}{SM.\arabic{figure}}
\renewcommand{\theHtable}{SM.\arabic{table}}
\renewcommand{\theHsection}{SM.\arabic{section}}
\renewcommand{\theHsubsection}{SM.\arabic{section}.\arabic{subsection}}


\begin{center}
\textbf{\large Supplemental Material\\[4mm] 
\Large A magnetic monopole in a superfluid bubble}\\[4mm]
Marianna Sorba,$^{1,\dagger}$ Andrea Richaud,$^{2,\ast}$\\[2mm]
\emph{\small $^1$CNR-INO, Area Science Park, Basovizza, 34149 Trieste, Italy}\\
\emph{\small $^2$Departament de F\'isica, Universitat Polit\`ecnica de Catalunya, Campus Nord B4-B5, E-08034 Barcelona, Spain}\\
\end{center}

\vspace*{-10pt}

%

\section{Linear stability analysis of the massive vortex dipole}
\label{sec:linear_stability}

In this section we provide the details of the normal-mode analysis used in the main text. We consider a vortex dipole with charges $q_\pm=\pm1$ and total core mass $M_b$, so that each vortex carries an inertial mass $M_b/2$. Starting from the Lagrangian given in the main text, the canonical angular momenta are
\begin{equation}
l_{\theta_j}
=
\frac{M_b}{2}R^2\dot\theta_j,
\qquad
l_{\phi_j}
=
\frac{M_b}{2}R^2\sin^2\theta_j\,\dot\phi_j
+
gq_j(1-\cos\theta_j),
\qquad
j=\pm.
\end{equation}
The corresponding Hamiltonian reads
\begin{equation}
H_{\rm dip}
=
\frac{1}{M_bR^2}
\sum_{j=\pm}
\left[
l_{\theta_j}^2
+
\frac{
\left[
l_{\phi_j}
-
gq_j(1-\cos\theta_j)
\right]^2
}
{\sin^2\theta_j}
\right]
+
E_{\rm dip},
\label{eq:SM_Hdip}
\end{equation}
where $E_{\rm dip}$ denotes the interaction energy of the vortex dipole, being
\begin{equation}
E_{\rm dip}
=
-\frac{2\hbar^2\pi n_a}{m_a}\chi_{\rm dip},
\end{equation}
with $\chi_{\rm dip}\equiv q_+q_-\chi_{+-}$.

For the symmetric configuration the equations of motion admit a uniformly precessing solution
\begin{equation}
\theta_+=\theta,
\qquad
\theta_-=\pi-\theta,
\qquad
\phi_+=\phi_-,
\end{equation}
characterized by the angular velocity
\begin{equation}
\dot{\phi}
=
\frac{\hbar}{M_bR\cos\theta}
\left(
2\pi n_aR
-
\sqrt{
\frac{2\pi n_a(2m_an_a\pi R^2-M_b)}
{m_a}
}
\right).
\label{eq:SM_precession}
\end{equation}
In the limit $M_b\rightarrow0$, this expression reduces to the massless point-vortex result~\cite{SM-Bereta2021}
\begin{equation}
\dot{\phi}
=
\frac{\hbar}{2m_aR^2\cos\theta}.
\label{eq:phi_dot_massless}
\end{equation}

To study small oscillations around this uniformly precessing orbit, it is convenient to move to a frame rotating with angular frequency $\lambda=\dot{\phi}$ by introducing $\phi'_j=\phi_j-\lambda t$ and $\theta'_j=\theta_j$. The Hamiltonian in the rotating frame becomes
\begin{equation}
H'_{\rm dip}
=
H_{\rm dip}
-
\lambda
\sum_{j=\pm}
l'_{\phi_j},
\label{eq:SM_Hrot}
\end{equation}
whose equilibrium configuration is
\begin{equation}
\mathbf z'_{\rm eq}
=
\left(
\theta,\phi',
\pi-\theta,\phi',
0,
g(1-\cos\theta)
+\frac{\lambda}{2}M_bR^2\sin^2\theta,
0,
-g(1+\cos\theta)
+\frac{\lambda}{2}M_bR^2\sin^2\theta
\right)^T.
\end{equation}
 
Introducing the canonical vector
\begin{equation}
\mathbf z'
=
\left(
\theta'_+,\phi'_+,\theta'_- ,\phi'_-,
l'_{\theta_+},l'_{\phi_+},
l'_{\theta_-},l'_{\phi_-}
\right)^T,
\end{equation}
Hamilton's equations take the compact form
\begin{equation}
\dot{\mathbf z}'
=
\mathbb J
\nabla_{\mathbf z'}H'_{\rm dip},
\qquad
\mathbb J=
\begin{pmatrix}
\mathbf 0_4 & \mathbf I_4\\
-\mathbf I_4 & \mathbf 0_4
\end{pmatrix}.
\end{equation}
Linearizing around the equilibrium point, $\mathbf z'=\mathbf z'_{\rm eq}+\delta\mathbf z'$, gives
\begin{equation}
\delta\dot{\mathbf z}'
=
\mathbb J
\,
\mathbb H(\mathbf z'_{\rm eq})
\,
\delta\mathbf z',
\label{eq:SM_linearized}
\end{equation}
where
\begin{equation}
\mathbb H_{\alpha\beta}
=
\left.
\frac{\partial^2H'_{\rm dip}}
{\partial z'_\alpha\partial z'_\beta}
\right|_{\mathbf z'=\mathbf z'_{\rm eq}}
\label{eq:Hessian}
\end{equation}
is the Hessian matrix of the rotating-frame Hamiltonian. The normal modes are then obtained from the eigenvalue problem
\begin{equation}
\mathbb J
\,
\mathbb H(\mathbf z'_{\rm eq})
\,
\mathbf u_\nu
=
\lambda_\nu
\mathbf u_\nu.
\label{eq:SM_eigenproblem}
\end{equation}
Stable oscillatory modes correspond to purely imaginary eigenvalues, $\lambda_\nu=\pm i\omega'_\nu$.

The eigenvalue problem~(\ref{eq:SM_eigenproblem}) yields four independent normal-mode frequencies in the rotating frame. Closed analytical expressions are cumbersome and provide little physical insight. In the joint limit $\theta\to0$ and $M_b\to0$, corresponding to a pair of massless antipodal vortices, one finds
\begin{align}
\omega'_1 &= 0,\\
\omega'_2 &=
\frac{4\pi\hbar n_a}{M_b}
-
\frac{\hbar}{2m_aR^2},\\
\omega'_3 &=
\frac{4\pi\hbar n_a}{M_b}
-
\frac{\hbar}{m_aR^2},\\
\omega'_4 &=
\frac{\hbar}{2m_aR^2}.
\end{align}
In the same limit, the precession frequency $\lambda=\dot{\phi}$ satisfies $
\lambda
=
\hbar/(2m_aR^2)$ [see Eq.~(\ref{eq:phi_dot_massless}) ] and the physical interpretation of the different branches becomes particularly simple: the zero mode reflects the rotational symmetry of the problem. The finite frequency $\omega'_4$ coincides with the uniform precession frequency of a massless antipodal vortex dipole on the sphere, while the two large frequencies $\omega'_2$ and $\omega'_3$ correspond to high-frequency inertial (cyclotron-like) modes that diverge as $M_b^{-1}$ when the vortex mass is taken to zero. An example of the evolution of these four branches with the equilibrium polar angle is shown in the right panel of Fig.~\ref{fig:SM_theta_dependence}, where they are compared with the frequencies extracted from the Gross--Pitaevskii simulations.

In the laboratory frame, the dominant high-frequency scale is set by the cyclotron-like motion of a vortex of charge $|q_\pm|=1$ and mass $M_b/2$ in the monopole field. Since $|\mathbf B|=|g|/R^2$, this characteristic frequency is
\begin{equation}
\omega_{\rm c}
=
\frac{|q_\pm||\mathbf B|}{M_b/2}
=
\frac{2|g|}{M_bR^2}
=
\frac{4\pi\hbar n_a}{M_b}.
\end{equation}
As expected, the rotating-frame frequencies are obtained from the corresponding laboratory-frame frequencies through the standard frequency shift associated with the transformation to a uniformly rotating frame.

\section{Gross--Pitaevskii simulations and data analysis}

\subsection{Numerical implementation}
\label{sub:Numerical_implementation}

We solve the coupled Gross--Pitaevskii equations introduced in the main text for an immiscible two-component Bose gas confined to a spherical surface. The simulations reported in Fig.~2 of the main text use a $^{23}{\rm Na}$--$^{39}{\rm K}$ mixture~\cite{SM-Schulze2018}, with $m_a=23\,{\rm amu}$ and $m_b=39\,{\rm amu}$. We take $N_a=5\times10^5$ and vary the number of atoms in component $b$ in the range $N_b\in[10^3,\,10^4]$, thereby tuning the dimensionless vortex mass $\mu=M_b/M_a=N_b m_b/(N_a m_a)$. The radius is fixed to $R=20~\mu\mathrm{m}$, which lies within the range of experimentally relevant bubble-trap geometries proposed for spherical Bose–Einstein condensates~\cite{SM-Dubessy2025}.

The three-dimensional coupling constants are
\begin{equation}
g_{aa}^{\rm 3D}=\frac{4\pi\hbar^2 a_{aa}}{m_a},
\qquad
g_{bb}^{\rm 3D}=\frac{4\pi\hbar^2 a_{bb}}{m_b},
\end{equation}
with $a_{aa}=52.0\,a_0$ and $a_{bb}=7.6\,a_0$, where $a_0$ is the Bohr radius~\cite{SM-Schulze2018}. The intercomponent interaction is chosen as
\begin{equation}
g_{ab}^{\rm 3D}=3\sqrt{g_{aa}^{\rm 3D}g_{bb}^{\rm 3D}},
\end{equation}
which places the system in the immiscible regime. The effective two-dimensional interaction strengths are obtained through the standard quasi-two-dimensional reduction,
\begin{equation}
g_{ij}
=
\frac{g_{ij}^{\rm 3D}}
{\sqrt{2\pi}\,a_z},
\end{equation}
which assumes that the transverse degree of freedom (i.e. the radial thickness of the spherical shell) remains frozen in the harmonic-oscillator ground state. Throughout this work we use $a_z=2~\mu{\rm m}$.

The spherical grid is generated using the SSHT package~\cite{SM-SSHT} with the McEwen--Wiaux sampling scheme~\cite{SM-McEwen2011} and band limit $L=224$. This sampling theorem provides exact spherical-harmonic transforms for band-limited functions on the sphere using a sample-efficient equiangular grid. Surface integrals are evaluated with the weights
\begin{equation}
dS = R^2\sin\theta\,d\theta\,d\phi .
\end{equation}
The Laplace--Beltrami operator is evaluated spectrally: the wavefunction is transformed to spherical-harmonic space, each mode is multiplied by
\begin{equation}
-\frac{\ell(\ell+1)}{R^2},
\end{equation}
and the result is transformed back to real space using SSHT.

The initial state contains a vortex dipole in component $a$, with charges
$q_1=+1$ and $q_2=-1$ located at
\begin{equation}
\theta_1=\frac{\pi}{2}-0.4,\qquad
\theta_2=\frac{\pi}{2}+0.4,\qquad
\phi_1=\phi_2=\frac{3\pi}{4}.
\end{equation}
The phase of component $a$ is initialized from the stereographic-coordinate vortex-dipole phase, while its density is depleted at the two vortex cores using a core size $\xi_{\rm core}=0.8~\mu{\rm m}$. Component $b$ is initialized as a sum of two geodesic Gaussian peaks localized at the same positions, with width $\sigma_b=1~\mu{\rm m}$.

The initial state is relaxed in imaginary time in the presence of Gaussian pinning potentials acting on component $a$ only. The pinning potentials have strength
\begin{equation}
V_0=5\times10^{-31}\,{\rm J}
\simeq
2.5\,\mu_a^{\rm TF},
\end{equation}
where
$\mu_a^{\rm TF}=g_{aa}N_a/(4\pi R^2)$ is the Thomas--Fermi estimate of the chemical potential of component $a$. Their angular width is $\sigma_{\rm pin}=0.04~{\rm rad}$. The imaginary-time propagation is performed with a forward-Euler step, followed by renormalization of both components. We use $10^4$ imaginary-time iterations with time step
\begin{equation}
dt=10^{-6}\,{\rm s}
\simeq
6.9\times10^{-6}\,\tau_a,
\qquad
\tau_a=\frac{m_aR^2}{\hbar}\simeq0.145~{\rm s}.
\end{equation}

After the imaginary-time preparation, the pinning potentials are removed and the coupled Gross--Pitaevskii equations are evolved in real time. The real-time propagation is performed using a fourth-order Runge--Kutta scheme, again with $dt=10^{-6}\,{\rm s}$. To improve numerical stability and suppress spurious high-wavenumber components generated by the nonlinear evolution, after each time step the spherical-harmonic coefficients are multiplied by the exponential spectral filter $f_\ell=\exp\!\left[-\left(\ell/(L-1)\right)^{18} \right]$. The filter leaves the physically resolved scales essentially unaffected, while smoothly damping only the highest harmonic modes close to the spectral cutoff. The real-time simulations are propagated for $2\times10^5$ time steps, corresponding to a total evolution time of $0.2~{\rm s}\simeq1.38\,\tau_a$. Observables and vortex positions are saved every $10^3$ time steps. During the real-time evolution, the total energy and angular momentum are monitored as numerical diagnostics. Across all simulations, the relative variation of the total energy remains below $2.5\times10^{-4}\%$, while that of the total angular momentum remains below $2.3\times10^{-1}\%$, confirming the excellent conservation properties of the numerical scheme. The vortex trajectories used for Fig.~2 of the main text are obtained by tracking the localized density peaks of component $b$. More precisely, the center-of-mass position of $|\psi_b|^2$ is computed separately in the northern and southern hemispheres, yielding the two time-dependent vortex-core positions.

\subsection{Extraction of the dipole oscillation frequency}
\label{sub:Extraction_of_the_dipole_frequency}

The oscillation frequency reported in Fig.~2(c) of the main text is extracted directly from the vortex trajectories obtained from the Gross--Pitaevskii simulations. Starting from the time-dependent vortex-core positions, we compute the instantaneous angular (geodesic) separation between the two vortices,
\begin{equation}
\cos\gamma(t)
=
\cos\theta_N(t)\cos\theta_S(t)
+
\sin\theta_N(t)\sin\theta_S(t)
\cos\!\left[\phi_N(t)-\phi_S(t)\right],
\end{equation}
where $(\theta_N,\phi_N)$ and $(\theta_S,\phi_S)$ denote the vortex positions in the northern and southern hemispheres, respectively. Since the cyclotron-like oscillations of the two vortices occur in antiphase [see Fig.~2(a) of the main text], the angular separation $\gamma(t)$ provides a differential readout of the motion, enhancing the oscillatory signal with respect to the displacement of each individual vortex.

For all values of $N_b$ considered in this work, the signal $\gamma(t)$ consists of small oscillations around its equilibrium value over the simulated time interval. We therefore fit it with the sinusoidal model
\begin{equation}
\gamma(t)
=
\gamma_0
+
A\sin(\omega t+\varphi),
\label{eq:gamma_fit_model}
\end{equation}
where $\gamma_0$, $A$, $\omega$, and $\varphi$ are free fitting parameters. To improve the robustness of the nonlinear optimization, the initial guess for $\omega$ is obtained from the dominant Fourier component of $\gamma(t)-\langle\gamma\rangle$, while the remaining parameters are initialized from the mean value and oscillation amplitude of the signal. The final fit is obtained through least-squares minimization, and the resulting angular frequency $\omega$ is used in Fig.~2(c) of the main text.

\begin{figure}[h!]
    \centering
\includegraphics[width=\textwidth]{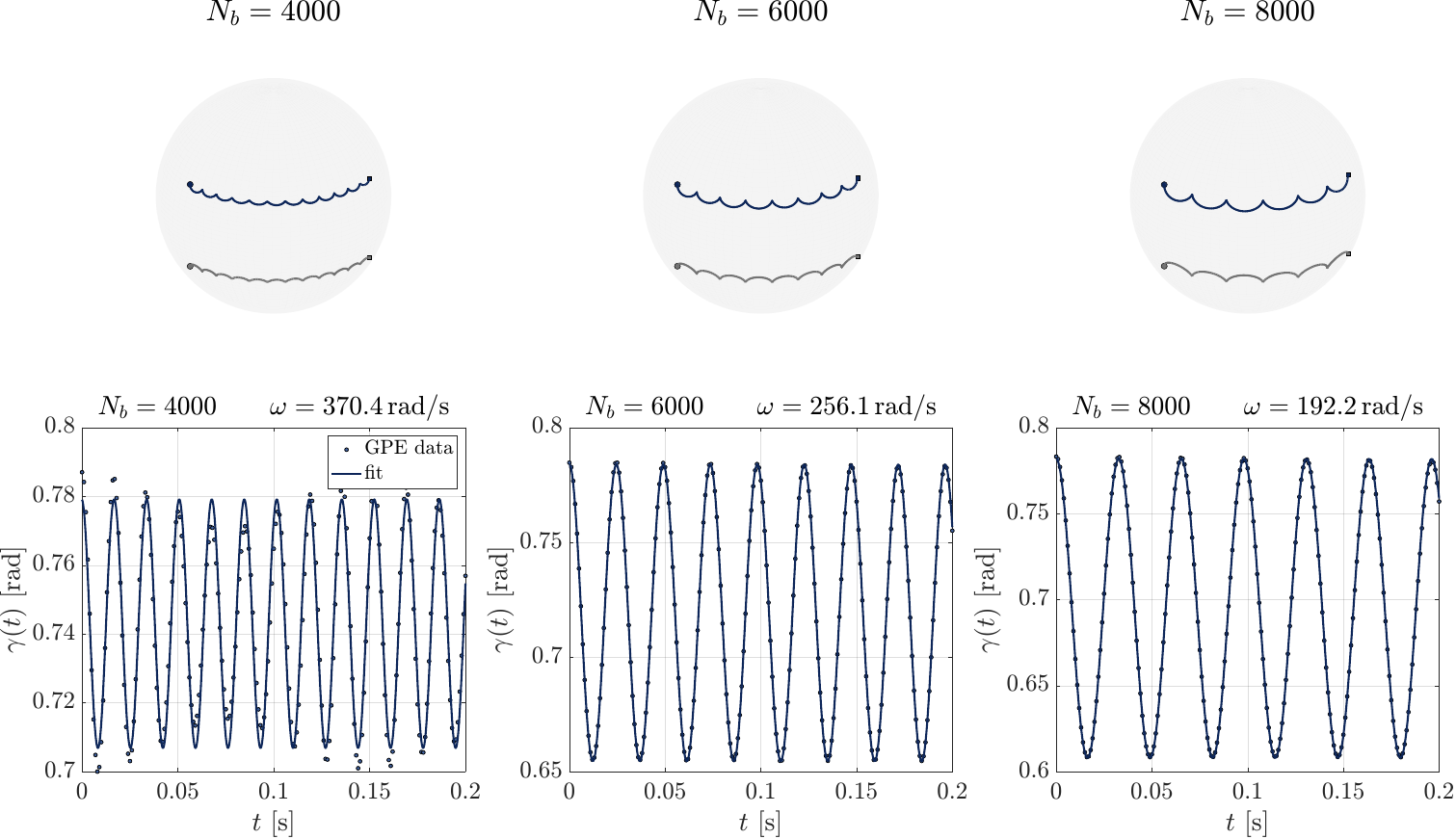}
\caption{\textbf{Extraction of the dipole oscillation frequency.} Top panels: trajectories of the two vortex cores on the sphere for three representative values of the minority population $N_b$. Bottom panels: corresponding evolution of the angular separation $\gamma(t)$ extracted from the Gross--Pitaevskii simulations (symbols), together with the best-fit curves obtained from Eq.~(\ref{eq:gamma_fit_model}) (solid lines). The fitted angular frequency $\omega$ is compared with the analytical normal-mode prediction in Fig.~2(c) of the main text. For the three representative simulations shown here ($N_b=4000$, $6000$, and $8000$), the relative variations of the total energy are below $2.2\times10^{-4}\%$, while those of the total angular momentum remain below $2.1\times10^{-1}\%$, confirming the excellent numerical conservation properties of the integration scheme.}
    \label{fig:SM_gamma_fit}
\end{figure}

Selected examples of the fitted oscillations are shown in Fig.~\ref{fig:SM_gamma_fit}. The excellent agreement between the numerical data and the sinusoidal fits indicates that, over the simulated time window, the dynamics is dominated by a single oscillatory mode. This justifies identifying the fitted frequency with the cyclotron-like eigenfrequency predicted by the linear stability analysis.

\subsection{Dependence on the initial dipole position}
\label{sub:Dependence_on_initial_position}

We also investigate how the cyclotron-like oscillation frequency depends on the initial position of the vortex dipole on the sphere. To this end, we repeat the Gross--Pitaevskii simulations for symmetric initial conditions
\begin{equation}
\theta_+(0)=\theta_0,
\qquad
\theta_-(0)=\pi-\theta_0,
\qquad
\phi_+(0)=\phi_-(0),
\end{equation}
while keeping all other physical and numerical parameters fixed. For each value of $\theta_0$, the vortex trajectories are tracked as described above, and the oscillation frequency is extracted from the sinusoidal fit of the angular separation $\gamma(t)$.

\begin{figure}[h!]
    \centering
    \includegraphics[width=\textwidth]{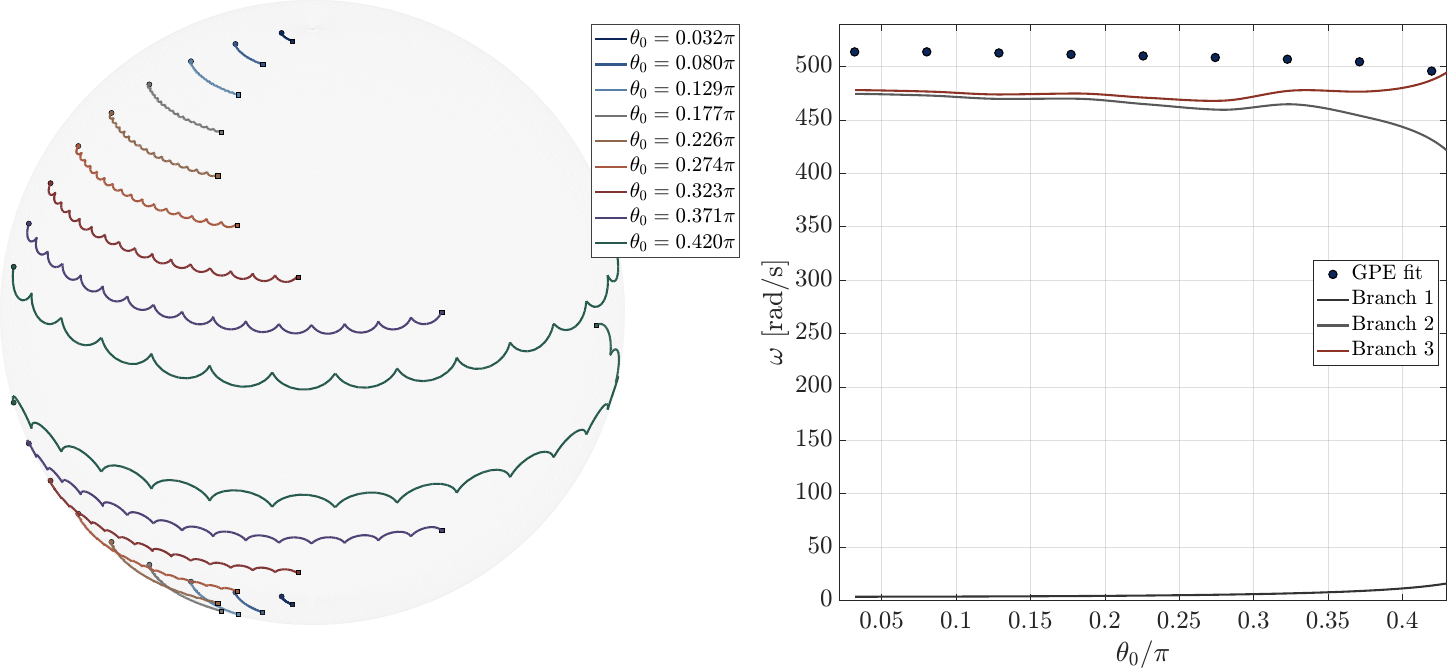}
    \caption{\textbf{Dependence of the dipole oscillation frequency on the initial polar angle.} Left panel: real-time vortex trajectories on the sphere for different symmetric initial configurations, $\theta_+(0)=\theta_0$ and $\theta_-(0)=\pi-\theta_0$, obtained from Gross--Pitaevskii simulations with dimensionless mass ratio $\mu=0.01$. Right panel: oscillation frequency extracted from the Gross--Pitaevskii simulations as a function of $\theta_0/\pi$, compared with the positive-frequency branches of the linearized Hamiltonian dynamics of Sec.~\ref{sec:linear_stability}, computed using the effective vortex mass $M_b+M_{\mathrm{ex}}$ [see Eq.~(\ref{eq:Mex})]. The numerical frequencies qualitatively follow the second positive branch of the linearized spectrum over the explored range of initial conditions. The zero-frequency branch, not shown, reflects the rotational invariance of the system.}
\label{fig:SM_theta_dependence}
\end{figure}

Selected trajectories are shown in the left panel of Fig.~\ref{fig:SM_theta_dependence}. As $\theta_0$ is varied, the vortices precess along different latitudes while undergoing the small oscillatory motion associated with the massive-core dynamics. For each simulation, the excluded mass is estimated from the Gross--Pitaevskii density profile as
\begin{equation}
M_{\mathrm{ex}}
=
m_a
\int
\max\!\left[n_{\rm bulk}-n_a(\mathbf r),\,0\right]\,dS,
\label{eq:Mex}
\end{equation}
where $n_a(\mathbf r)=|\psi_a(\mathbf r)|^2$ is the condensate density of component $a$ and $n_{\rm bulk}$ is the bulk density far from the vortex cores. The corresponding fitted frequencies are shown in the right panel of Fig.~\ref{fig:SM_theta_dependence} as a function of $\theta_0/\pi$ and are compared with the eigenfrequencies obtained from the linearized Hamiltonian analysis of Sec.~\ref{sec:linear_stability}, using the effective vortex mass $M_b+M_{\mathrm{ex}}$. The numerical frequencies qualitatively follow the second positive branch of the linearized spectrum over the entire explored range of initial conditions. As expected from the rotational symmetry discussed in Sec.~\ref{sec:linear_stability}, the spectrum also contains a zero-frequency mode for all values of $\theta_0$.

\subsection{Robustness against residual gravity}
\label{sub:Robustness_against_residual_gravity}

We investigate the robustness of the massive-vortex-dipole dynamics against a possible residual gravitational acceleration. This test is relevant to experiments with shell-shaped condensates, since gravity breaks the north–south symmetry of the atomic distribution and, if sufficiently strong, can substantially deform or even open the shell~\cite{SM-Wolf2022}. For the microgravity bubble-trap experiments considered in Ref.~\cite{SM-Tononi2020}, residual microgravitational corrections were estimated to be of order $10^{-6}g$, where $g=9.81{\rm ms^{-2}}$ denotes the gravitational acceleration at the Earth’s surface. Moreover, for the mixture-based shell configuration and parameters analyzed in Ref.~\cite{SM-Wolf2022}, residual accelerations $g_{\rm res}\lesssim10^{-5}g$, characteristic of platforms such as the International Space Station and drop-tower facilities, were found to be compatible with an almost ideal closed shell. The precise acceleration threshold is not universal, but depends on quantities such as the particle numbers and trapping frequencies.

To determine whether residual accelerations in this range affect the dynamics discussed in the main text, we repeat the real-time Gross–Pitaevskii simulation of the massive vortex dipole for $N_b=6000$, keeping all other physical and numerical parameters unchanged. We take the residual acceleration to point along the negative $z$ direction and include the corresponding gravitational potentials
\begin{equation}
V_{g,a}(\theta) = m_a g_{\rm res}R\cos\theta,
\qquad
V_{g,b}(\theta) = m_b g_{\rm res}R\cos\theta,
\label{eq:residual_gravity_potentials}
\end{equation}
for components $a$ and $b$, respectively. Since $z=R\cos\theta$, these potentials decrease toward the south pole, $\theta=\pi$, and therefore produce a force directed along $-\hat{\mathbf z}$.

The real-time propagation is initialized from the same vortex-dipole state previously relaxed in the absence of gravity. The residual gravitational potential is then switched on at the beginning of the real-time evolution. We consider
\begin{equation}
\frac{g_{\rm res}}{g} = 10^{-7},\ 10^{-6},\ 10^{-5},\ 10^{-4}.
\end{equation}
The first three values span the microgravity regime relevant to the experimental conditions discussed above. The largest value, $g_{\rm res}=10^{-4}g$, is included as a conservative upper bound rather than as a nominal value for present orbital microgravity platforms.

\begin{figure}[h!]
\centering
\includegraphics[width=\textwidth]
{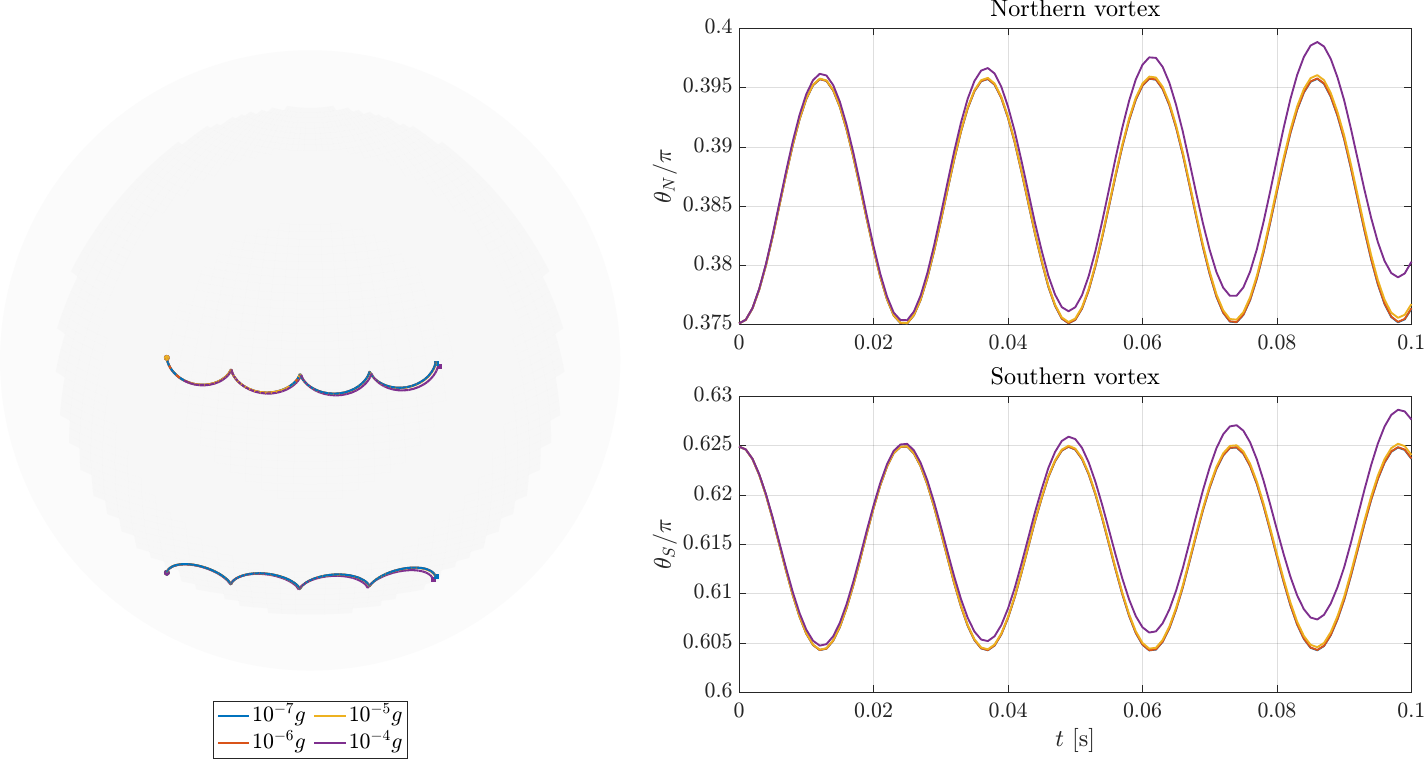} 
\caption{\textbf{Effect of residual gravity on the massive vortex dipole.} \textbf{Left:} Trajectories of the northern and southern vortex cores obtained from two-component Gross–Pitaevskii simulations with $N_b=6000$ and residual accelerations $g_{\rm res}/g=10^{-7},10^{-6},10^{-5},10^{-4}$. The two trajectories associated with a given value of $g_{\rm res}$ are shown using the same color. Circles and squares indicate the initial and final positions, respectively. \textbf{Right:} Time evolution of the polar coordinates of the northern (top) and southern (bottom) vortex cores. For $g_{\rm res}\leq10^{-5}g$, the dynamics is indistinguishable, within the numerical resolution, from that obtained for $g_{\rm res}=0$. At the conservative upper bound $g_{\rm res}=10^{-4}g$, both vortices display a small drift toward the south pole.}
\label{fig:SM_residual_gravity}
\end{figure}

The resulting vortex trajectories are shown in the left panel of Fig.~\ref{fig:SM_residual_gravity}. For $g_{\rm res}=10^{-7}g$, $10^{-6}g$, and $10^{-5}g$, the trajectories and the corresponding polar-coordinate evolution are indistinguishable, within the resolution of our simulations, from those obtained in the ideal gravity-free case. In particular, neither the cyclotron like oscillations nor the mean polar positions of the vortices display a visible systematic change over the simulated time interval. Only for the conservative value $g_{\rm res}=10^{-4}g$ does a small systematic effect become apparent: both vortex cores exhibit a weak drift toward the south pole (see right panels of Fig.~\ref{fig:SM_residual_gravity}). These results show that the massive-vortex-dipole dynamics reported in the main text is robust against residual gravitational accelerations in the experimentally relevant microgravity range considered here.

\subsection{Numerical protocol for the Kelvin--Helmholtz simulations}
\label{sub:Numerical_protocol_for_KHI}

The simulations reported in Fig.~3 of the main text are performed within a \emph{single-component} Gross--Pitaevskii model, in contrast to the two-component simulations discussed in Secs.~\ref{sub:Numerical_implementation}-\ref{sub:Dependence_on_initial_position} for the massive-vortex dynamics. Since the purpose of these calculations is to investigate the formation and subsequent breakup of the equatorial vortex necklace in the host condensate, the component-$b$ used to make the vortices massive is omitted. The condensate dynamics is therefore governed by
\begin{equation}
i\hbar\partial_t\psi
=
\left[
-\frac{\hbar^2}{2m_aR^2}\nabla_\Omega^2
+
g_{aa}|\psi|^2
+
V(\mathbf r,t)
\right]\psi,
\label{eq:GPE_single_component}
\end{equation}
where $V(\mathbf r,t)$ denotes the external potential used to prepare the
initial state and tune the coupling between the two hemispheres. The
equatorial barrier is modeled as
\begin{equation}
V(\theta,t)
=
V_{\rm bar}(t)
\exp\!\left[
-\frac{(\theta-\pi/2)^2}{2\sigma_{\rm bar}^2}
\right],
\end{equation}
with initial height $V_{\rm bar}(0)=1.2\,\mu_a^{\rm TF}$ and angular width $\sigma_{\rm bar}=0.05~\mathrm{rad}$. The barrier height is held constant during the imaginary-time preparation and subsequently ramped down linearly in real time, as described below. Unless otherwise specified below, all numerical parameters and the spatial discretization are identical to those described in Sec.~\ref{sub:Numerical_implementation}.

The initial configuration is obtained by imaginary-time propagation in the presence of the equatorial barrier defined above, which effectively decouples the two hemispheres. Two additional localized repulsive pinning potentials are applied at the north and south poles in order to stabilize multiply quantized polar vortices during the state preparation. The condensate phase is initialized using a two-patch construction, with $\arg(\psi)\simeq +q\phi$ in the northern hemisphere and $\arg(\psi)\simeq -q\phi$ in the southern hemisphere, smoothly interpolated across a narrow equatorial band lying within the barrier region, where the condensate density is strongly suppressed. This phase profile represents a pair of like-charged polar vortices of winding number $q$, one located at each pole. The corresponding phase mismatch across the equatorial barrier is $\Delta\arg(\psi)=2q\phi$, corresponding to a circulation discontinuity of $2q\kappa$ across the equator. Once the barrier is removed, this discontinuity provides the initial condition for the formation of the equatorial vortex necklace discussed in the main text.

After the imaginary-time preparation, the polar pinning potentials are removed and the real-time dynamics is started from the relaxed state. During the real-time evolution, the height $V_{\rm bar}(t)$ of the equatorial barrier is ramped down linearly from its initial value to zero over $T_{\rm ramp}=50~\mathrm{ms} \simeq 0.35\,\tau_a$, after which the condensate evolves in the absence of the barrier. As the barrier is lowered, singly quantized vortices nucleate along the equator and organize into the vortex necklace discussed in the main text. The whole dynamics is computed propagating Eq.~(\ref{eq:GPE_single_component}) in real time. 

Unlike the two-component simulations of Sec.~\ref{sub:Extraction_of_the_dipole_frequency}, where vortex positions are extracted from the center of mass of the localized component-$b$ density, here vortex cores are identified directly from the minima of the single-component condensate density.

\subsection{Extraction of the instability growth rate}
To quantify the development of the Kelvin--Helmholtz instability, we post-process the vortex trajectories extracted from the Gross--Pitaevskii simulations. The vortex cores are identified from the density minima of component $a$, yielding the time-dependent coordinates
\begin{equation}
(\theta_j(t),\phi_j(t)),
\qquad
j=0,\ldots,N_v-1,
\end{equation}
where $N_v=2q$ is the number of vortices forming the necklace after the removal of the equatorial barrier.

As a reference state, we choose the first saved configuration after the barrier has been completely removed. Denoting the corresponding time by $t_0$, we define
\begin{equation}
\theta_j^{(0)}=\theta_j(t_0),
\qquad
\phi_j^{(0)}=\phi_j(t_0).
\end{equation}
The vortices are ordered according to their azimuthal positions in this reference configuration, and their azimuthal trajectories are subsequently unwrapped to remove artificial discontinuities associated with the periodicity of the angular coordinate $\phi$.

We then introduce the shifted time variable
\begin{equation}
t' = t-t_0,
\end{equation}
and combine the polar and azimuthal displacements into the complex displacement
\begin{equation}
u_j(t')
=
\delta\theta_j(t')
+i\,\delta\phi_j(t'),
\end{equation}
where
\begin{equation}
\delta\theta_j(t')
=
\theta_j(t')-\theta_j^{(0)},
\qquad
\delta\phi_j(t')
=
\phi_j(t')-\phi_j^{(0)}.
\end{equation}

The displacement is then decomposed into discrete Fourier modes,
\begin{equation}
\label{eq:c_m}
c_m(t')
=
\frac{1}{N_v}
\sum_{j=0}^{N_v-1}
u_j(t')
\exp\!\left(
-i\frac{2\pi m j}{N_v}
\right).
\end{equation}

Inspired by the linear stability analysis of vortex chains~\cite{SM-Aref1995} and necklaces~\cite{SM-Hernandez2024,SM-Caldara2024}, and corroborated by the spherical point-vortex model presented in the main text, we focus on the mode expected to display the largest instability growth rate. For a necklace containing $N_v=2q$ vortices, the highest-wavenumber mode supported by the discrete ring is
\begin{equation}
m=\frac{N_v}{2}=q,
\end{equation}
and we therefore monitor the amplitude
\begin{equation}
A(t')=|c_q(t')|.
\end{equation}

During the linear stage of the instability, this quantity is expected to grow exponentially,
\begin{equation}
A(t')\simeq A_0 e^{\sigma_\ast t'},
\end{equation}
or equivalently,
\begin{equation}
\log A(t')
\simeq
\log A_0+\sigma_\ast t'.
\end{equation}

The instability growth rate $\sigma_\ast$ is obtained from a linear fit of $\log |c_q|$ as a function of $t'$. To isolate the exponential growth regime automatically, only data points satisfying 
\begin{equation}
\label{eq:fitting_criteria}
A_{\min}\le |c_q(t')| \le A_{\max}
\end{equation}
are considered. Among all contiguous fitting intervals containing at least a prescribed minimum number of points, we retain the segment with positive slope and coefficient of determination larger than a threshold $R^2_{\min}$. The selected interval maximizes a score that mildly favors longer fitting windows while avoiding excessively late-time data, where nonlinear effects become important. The uncertainty on $\sigma_\ast$ is estimated from the standard error of the fitted slope.

\begin{figure}[h!]
    \centering
    \includegraphics[width=\textwidth]{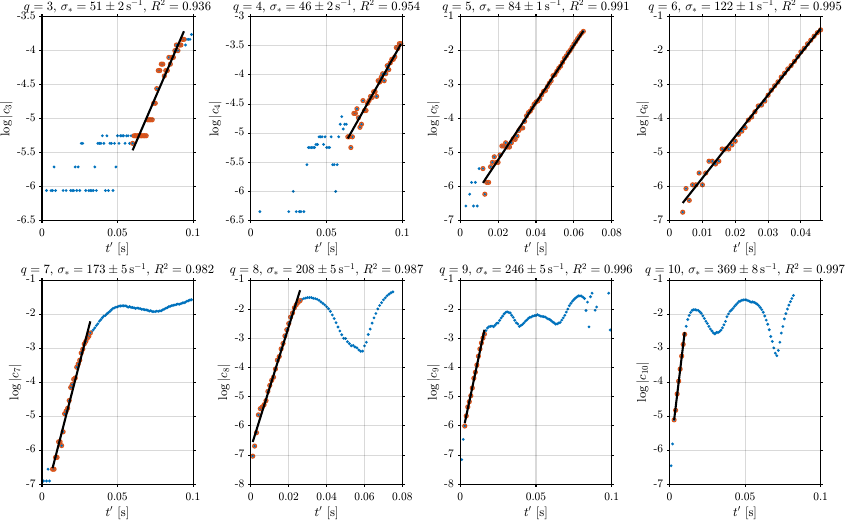}
    \caption{\textbf{Extraction of the instability growth rate from the Gross--Pitaevskii simulations.} For each value of $q$, the quantity $\log|c_q|$ [defined in Eq.~(\ref{eq:c_m})] is plotted as a function of $t'=t-t_0$, where $t_0$ denotes the instant at which the equatorial barrier has been completely removed. Here $c_q$ denotes the Fourier amplitude of the mode $m=q$, corresponding to the highest-wavenumber mode supported by a vortex necklace containing $N_v=2q$ vortices. Small blue dots denote all candidate points satisfying the fitting criterion of Eq.~(\ref{eq:fitting_criteria}), while red circles indicate the subset selected by the automatic fitting procedure. The solid black line shows the corresponding least-squares fit of $\log|c_q|$. The fitted slope yields the instability growth rate $\sigma_\ast$.}
    \label{fig:Sup_Mat_Log_fit}
\end{figure}

Representative examples of the fitting procedure are shown in Fig.~\ref{fig:Sup_Mat_Log_fit}. For each value of $q$, the figure reports the evolution of $\log |c_q|$ as a function of $t'=t-t_0$, together with the automatically selected fitting interval and the corresponding least-squares linear regression. The excellent linearity observed over the selected windows supports the interpretation of the early-time dynamics as an exponential instability and provides a robust estimate of the growth rate $\sigma_\ast$.

\section{Effective electrostatic description of the polar-vortex configuration}
\label{sec:effective_electrostatics}

As discussed in the main text, a globally defined electrostatic field on a closed spherical surface requires overall charge neutrality. Consequently, the effective electric field associated with an isolated polar vortex cannot be constructed directly. Instead, we introduce an auxiliary neutral dipole consisting of two opposite effective charges placed at the north and south poles, derive the corresponding electric field, from which the effective northern and southern fields used in the main text are obtained. This construction yields the effective fields entering the Maxwell boundary condition discussed in the main text.

According to the effective Coulomb interaction appearing in the Newton-like equation of motion, the electric field generated by an effective charge $q_i$ located at $\mathbf r_i$ is
\begin{equation}
\mathbf E_i(\mathbf r)
=
\frac{2\pi\hbar^2 n_a}{m_a}
q_i
\left[
\frac{\mathbf r-\mathbf r_i}
{|\mathbf r-\mathbf r_i|^2}
\right]_{\parallel},
\label{eq:SM_single_source_field}
\end{equation}
where $[\cdots]_{\parallel}$ denotes projection onto the tangent plane at $\mathbf r$, and $|\mathbf r-\mathbf r_i|$ is the chordal distance in the embedding three-dimensional space, consistently with the effective force
appearing in the main text.

For a source charge $+q$ located at the north pole, $\mathbf r_N=R\hat{\mathbf z}$, the projection onto the tangent plane is
\begin{equation}
\left[
\frac{\mathbf r-\mathbf r_N}
{|\mathbf r-\mathbf r_N|^2}
\right]_{\parallel}
=
\frac{1}{2R}
\cot\!\left(\frac{\theta}{2}\right)
\hat{\boldsymbol\theta},
\label{eq:SM_north_projection}
\end{equation}
whereas for a source charge $+q$ located at the south pole, $\mathbf r_S=-R\hat{\mathbf z}$,
\begin{equation}
\left[
\frac{\mathbf r-\mathbf r_S}
{|\mathbf r-\mathbf r_S|^2}
\right]_{\parallel}
=
-
\frac{1}{2R}
\tan\!\left(\frac{\theta}{2}\right)
\hat{\boldsymbol\theta}.
\label{eq:SM_south_projection}
\end{equation}

Consider first the auxiliary dipole consisting of a charge $+q$ at the north pole and a charge $-q$ at the south pole. The resulting field in the northern hemisphere is therefore
\begin{equation}
\mathbf E_N^{(q)}=
\frac{2\pi\hbar^2 n_a q}{m_a}
\left[
\frac{1}{2R}
\cot\!\left(\frac{\theta}{2}\right)
+
\frac{1}{2R}
\tan\!\left(\frac{\theta}{2}\right)
\right]
\hat{\boldsymbol\theta}
=
\frac{2\pi\hbar^2 n_a q}
{m_aR\sin\theta}
\hat{\boldsymbol\theta},
\label{eq:SM_effective_field_N}
\end{equation}
where the trigonometric identity $ \cot\!\left(\frac{\theta}{2}\right)+\tan\!\left(\frac{\theta}{2}\right)=\frac{2}{\sin\theta}$ has been used.

Likewise, interchanging the signs of the two charges gives the field in the southern hemisphere,
\begin{equation}
\mathbf E_S^{(q)}
=
-
\frac{2\pi\hbar^2 n_a q}
{m_aR\sin\theta}
\hat{\boldsymbol\theta}.
\label{eq:SM_effective_field_S}
\end{equation}
These effective fields are those employed in the main text to derive the Maxwell boundary condition and the corresponding equatorial line charge.

\section{Instability growth rate of an equatorial vortex necklace}

The instability growth rate shown in Fig.~3(e) of the main text is obtained from a linear stability analysis of the spherical point-vortex model. Starting from the effective equations of motion  of the main text, we consider the stationary configuration consisting of two vortices of charge $+q$ located at the north and south poles and an equatorial necklace formed by $2q$ equally spaced vortices of charge $-1$. This configuration satisfies the global neutrality condition and constitutes an equilibrium of the effective Hamiltonian dynamics.

As in the case of the massive vortex dipole discussed in Sec.~\ref{sec:linear_stability}, we formulate the problem in terms of the canonical coordinates and momenta of the equatorial vortices and construct the corresponding Hamiltonian, which includes the monopole-induced kinetic contribution together with the logarithmic vortex-vortex interaction. The Hamiltonian is then expanded to quadratic order around the equilibrium configuration,
\begin{equation}
\theta_j=\frac{\pi}{2},
\qquad
\phi_j=\frac{2\pi(j-1)}{2q},
\qquad
l_{\theta_j}=0,
\qquad
l_{\phi_j}=-g,
\end{equation}
which corresponds to vanishing mechanical velocities. The nonzero value of
$l_{\phi_j}$ originates from the gauge contribution associated with the monopole vector potential~\cite{SM-Richaud2021PRA}. The stability analysis is performed within the massive formulation introduced in the main text, using a vanishingly small core mass. We have verified that the resulting growth rates are indistinguishable from those obtained with a strictly massless (first-order) point-vortex scheme, confirming that the instability is governed by the kinematic vortex dynamics.

The normal modes are obtained from the eigenvalue problem 
\begin{equation}
\mathbb J
\,\mathbb H(\mathbf z_{\rm eq})
\,\mathbf u_\nu
=
\lambda_\nu
\mathbf u_\nu,
\end{equation}
[see Eq.~(\ref{eq:SM_eigenproblem}) of Sec.~\ref{sec:linear_stability}] where $\mathbb H(\mathbf z_{\rm eq})$ is the Hessian of the Hamiltonian evaluated at the equilibrium configuration [see Eq.(~\ref{eq:Hessian})], following exactly the procedure described in Sec.~\ref{sec:linear_stability}. Unlike the vortex-dipole case, the equilibrium is dynamically unstable, so that the spectrum contains eigenvalues with nonzero real part. The instability growth rate is therefore identified with the largest positive real part,
\begin{equation}
\sigma_\ast
=
\max_\nu
\left[
\mathrm{Re}(\lambda_\nu)
\right].
\end{equation}

The resulting values of $\sigma_\ast$ are reported in Fig.~3(e) of the main text (solid red curve), where they are compared with the growth rates extracted from the Gross--Pitaevskii simulations.




\begingroup

\setcounter{enumiv}{0}

\endgroup


\begin{thebibliography}{95}%
\makeatletter
\providecommand \@ifxundefined [1]{%
 \@ifx{#1\undefined}
}%
\providecommand \@ifnum [1]{%
 \ifnum #1\expandafter \@firstoftwo
 \else \expandafter \@secondoftwo
 \fi
}%
\providecommand \@ifx [1]{%
 \ifx #1\expandafter \@firstoftwo
 \else \expandafter \@secondoftwo
 \fi
}%
\providecommand \natexlab [1]{#1}%
\providecommand \enquote  [1]{``#1''}%
\providecommand \bibnamefont  [1]{#1}%
\providecommand \bibfnamefont [1]{#1}%
\providecommand \citenamefont [1]{#1}%
\providecommand \href@noop [0]{\@secondoftwo}%
\providecommand \href [0]{\begingroup \@sanitize@url \@href}%
\providecommand \@href[1]{\@@startlink{#1}\@@href}%
\providecommand \@@href[1]{\endgroup#1\@@endlink}%
\providecommand \@sanitize@url [0]{\catcode `\\12\catcode `\$12\catcode `\&12\catcode `\#12\catcode `\^12\catcode `\_12\catcode `\%12\relax}%
\providecommand \@@startlink[1]{}%
\providecommand \@@endlink[0]{}%
\providecommand \url  [0]{\begingroup\@sanitize@url \@url }%
\providecommand \@url [1]{\endgroup\@href {#1}{\urlprefix }}%
\providecommand \urlprefix  [0]{URL }%
\providecommand \Eprint [0]{\href }%
\providecommand \doibase [0]{https://doi.org/}%
\providecommand \selectlanguage [0]{\@gobble}%
\providecommand \bibinfo  [0]{\@secondoftwo}%
\providecommand \bibfield  [0]{\@secondoftwo}%
\providecommand \translation [1]{[#1]}%
\providecommand \BibitemOpen [0]{}%
\providecommand \bibitemStop [0]{}%
\providecommand \bibitemNoStop [0]{.\EOS\space}%
\providecommand \EOS [0]{\spacefactor3000\relax}%
\providecommand \BibitemShut  [1]{\csname bibitem#1\endcsname}%
\let\auto@bib@innerbib\@empty
\bibitem [{\citenamefont {Dirac}(1931)}]{Dirac1931}%
  \BibitemOpen
  \bibfield  {author} {\bibinfo {author} {\bibfnamefont {P.~A.~M.}\ \bibnamefont {Dirac}},\ }\bibfield  {title} {\bibinfo {title} {Quantised singularities in the electromagnetic field},\ }\href {https://doi.org/10.1098/rspa.1931.0130} {\bibfield  {journal} {\bibinfo  {journal} {Proc. R. Soc. A}\ }\textbf {\bibinfo {volume} {133}},\ \bibinfo {pages} {60} (\bibinfo {year} {1931})}\BibitemShut {NoStop}%
\bibitem [{\citenamefont {Wu}\ and\ \citenamefont {Yang}(1975)}]{WuYang1975}%
  \BibitemOpen
  \bibfield  {author} {\bibinfo {author} {\bibfnamefont {T.~T.}\ \bibnamefont {Wu}}\ and\ \bibinfo {author} {\bibfnamefont {C.~N.}\ \bibnamefont {Yang}},\ }\bibfield  {title} {\bibinfo {title} {Concept of nonintegrable phase factors and global formulation of gauge fields},\ }\href {https://doi.org/10.1103/PhysRevD.12.3845} {\bibfield  {journal} {\bibinfo  {journal} {Phys. Rev. D}\ }\textbf {\bibinfo {volume} {12}},\ \bibinfo {pages} {3845} (\bibinfo {year} {1975})}\BibitemShut {NoStop}%
\bibitem [{\citenamefont {Simon}(1983)}]{Simon1983}%
  \BibitemOpen
  \bibfield  {author} {\bibinfo {author} {\bibfnamefont {B.}~\bibnamefont {Simon}},\ }\bibfield  {title} {\bibinfo {title} {Holonomy, the quantum adiabatic theorem, and {B}erry's phase},\ }\href {https://doi.org/10.1103/PhysRevLett.51.2167} {\bibfield  {journal} {\bibinfo  {journal} {Phys. Rev. Lett.}\ }\textbf {\bibinfo {volume} {51}},\ \bibinfo {pages} {2167} (\bibinfo {year} {1983})}\BibitemShut {NoStop}%
\bibitem [{\citenamefont {Berry}(1984)}]{Berry1984}%
  \BibitemOpen
  \bibfield  {author} {\bibinfo {author} {\bibfnamefont {M.~V.}\ \bibnamefont {Berry}},\ }\bibfield  {title} {\bibinfo {title} {Quantal phase factors accompanying adiabatic changes},\ }\href {https://doi.org/10.1098/rspa.1984.0023} {\bibfield  {journal} {\bibinfo  {journal} {Proc. R. Soc. A}\ }\textbf {\bibinfo {volume} {392}},\ \bibinfo {pages} {45} (\bibinfo {year} {1984})}\BibitemShut {NoStop}%
\bibitem [{\citenamefont {Wu}\ and\ \citenamefont {Yang}(1976)}]{WuYang1976}%
  \BibitemOpen
  \bibfield  {author} {\bibinfo {author} {\bibfnamefont {T.~T.}\ \bibnamefont {Wu}}\ and\ \bibinfo {author} {\bibfnamefont {C.~N.}\ \bibnamefont {Yang}},\ }\bibfield  {title} {\bibinfo {title} {{Dirac} monopole without strings: Monopole harmonics},\ }\href {https://doi.org/10.1016/0550-3213(76)90143-7} {\bibfield  {journal} {\bibinfo  {journal} {Nucl. Phys. B}\ }\textbf {\bibinfo {volume} {107}},\ \bibinfo {pages} {365} (\bibinfo {year} {1976})}\BibitemShut {NoStop}%
\bibitem [{\citenamefont {'t~Hooft}(1974)}]{tHooft1974}%
  \BibitemOpen
  \bibfield  {author} {\bibinfo {author} {\bibfnamefont {G.}~\bibnamefont {'t~Hooft}},\ }\bibfield  {title} {\bibinfo {title} {Magnetic monopoles in unified gauge theories},\ }\href {https://doi.org/10.1016/0550-3213(74)90486-6} {\bibfield  {journal} {\bibinfo  {journal} {Nucl. Phys. B}\ }\textbf {\bibinfo {volume} {79}},\ \bibinfo {pages} {276} (\bibinfo {year} {1974})}\BibitemShut {NoStop}%
\bibitem [{\citenamefont {Polyakov}(1974)}]{Polyakov1974}%
  \BibitemOpen
  \bibfield  {author} {\bibinfo {author} {\bibfnamefont {A.~M.}\ \bibnamefont {Polyakov}},\ }\bibfield  {title} {\bibinfo {title} {Particle spectrum in quantum field theory},\ }\href@noop {} {\bibfield  {journal} {\bibinfo  {journal} {JETP Lett.}\ }\textbf {\bibinfo {volume} {20}},\ \bibinfo {pages} {194} (\bibinfo {year} {1974})}\BibitemShut {NoStop}%
\bibitem [{\citenamefont {Milton}(2006)}]{Milton2006}%
  \BibitemOpen
  \bibfield  {author} {\bibinfo {author} {\bibfnamefont {K.~A.}\ \bibnamefont {Milton}},\ }\bibfield  {title} {\bibinfo {title} {Theoretical and experimental status of magnetic monopoles},\ }\href {https://doi.org/10.1088/0034-4885/69/6/R02} {\bibfield  {journal} {\bibinfo  {journal} {Rep. Prog. Phys.}\ }\textbf {\bibinfo {volume} {69}},\ \bibinfo {pages} {1637} (\bibinfo {year} {2006})}\BibitemShut {NoStop}%
\bibitem [{\citenamefont {Castelnovo}\ \emph {et~al.}(2008)\citenamefont {Castelnovo}, \citenamefont {Moessner},\ and\ \citenamefont {Sondhi}}]{Castelnovo2008}%
  \BibitemOpen
  \bibfield  {author} {\bibinfo {author} {\bibfnamefont {C.}~\bibnamefont {Castelnovo}}, \bibinfo {author} {\bibfnamefont {R.}~\bibnamefont {Moessner}},\ and\ \bibinfo {author} {\bibfnamefont {S.~L.}\ \bibnamefont {Sondhi}},\ }\bibfield  {title} {\bibinfo {title} {Magnetic monopoles in spin ice},\ }\href {https://doi.org/10.1038/nature06433} {\bibfield  {journal} {\bibinfo  {journal} {Nature}\ }\textbf {\bibinfo {volume} {451}},\ \bibinfo {pages} {42} (\bibinfo {year} {2008})}\BibitemShut {NoStop}%
\bibitem [{\citenamefont {Morris}\ \emph {et~al.}(2009)\citenamefont {Morris}, \citenamefont {Tennant}, \citenamefont {Grigera}, \citenamefont {Klemke}, \citenamefont {Castelnovo}, \citenamefont {Moessner}, \citenamefont {Czternasty}, \citenamefont {Meissner}, \citenamefont {Rule}, \citenamefont {Hoffmann} \emph {et~al.}}]{Morris2009}%
  \BibitemOpen
  \bibfield  {author} {\bibinfo {author} {\bibfnamefont {D.~J.~P.}\ \bibnamefont {Morris}}, \bibinfo {author} {\bibfnamefont {D.~A.}\ \bibnamefont {Tennant}}, \bibinfo {author} {\bibfnamefont {S.~A.}\ \bibnamefont {Grigera}}, \bibinfo {author} {\bibfnamefont {B.}~\bibnamefont {Klemke}}, \bibinfo {author} {\bibfnamefont {C.}~\bibnamefont {Castelnovo}}, \bibinfo {author} {\bibfnamefont {R.}~\bibnamefont {Moessner}}, \bibinfo {author} {\bibfnamefont {C.}~\bibnamefont {Czternasty}}, \bibinfo {author} {\bibfnamefont {M.}~\bibnamefont {Meissner}}, \bibinfo {author} {\bibfnamefont {K.}~\bibnamefont {Rule}}, \bibinfo {author} {\bibfnamefont {J.-U.}\ \bibnamefont {Hoffmann}}, \emph {et~al.},\ }\bibfield  {title} {\bibinfo {title} {{Dirac} strings and magnetic monopoles in the spin ice {Dy2Ti2O7}},\ }\href {https://doi.org/10.1126/science.1178868} {\bibfield  {journal} {\bibinfo  {journal} {Science}\ }\textbf {\bibinfo {volume} {326}},\ \bibinfo {pages} {411} (\bibinfo {year} {2009})}\BibitemShut {NoStop}%
\bibitem [{\citenamefont {Fennell}\ \emph {et~al.}(2009)\citenamefont {Fennell}, \citenamefont {Deen}, \citenamefont {Wildes}, \citenamefont {Schmalzl}, \citenamefont {Prabhakaran}, \citenamefont {Boothroyd}, \citenamefont {Aldus}, \citenamefont {McMorrow},\ and\ \citenamefont {Bramwell}}]{Fennell2009}%
  \BibitemOpen
  \bibfield  {author} {\bibinfo {author} {\bibfnamefont {T.}~\bibnamefont {Fennell}}, \bibinfo {author} {\bibfnamefont {P.}~\bibnamefont {Deen}}, \bibinfo {author} {\bibfnamefont {A.}~\bibnamefont {Wildes}}, \bibinfo {author} {\bibfnamefont {K.}~\bibnamefont {Schmalzl}}, \bibinfo {author} {\bibfnamefont {D.}~\bibnamefont {Prabhakaran}}, \bibinfo {author} {\bibfnamefont {A.}~\bibnamefont {Boothroyd}}, \bibinfo {author} {\bibfnamefont {R.}~\bibnamefont {Aldus}}, \bibinfo {author} {\bibfnamefont {D.}~\bibnamefont {McMorrow}},\ and\ \bibinfo {author} {\bibfnamefont {S.}~\bibnamefont {Bramwell}},\ }\bibfield  {title} {\bibinfo {title} {Magnetic coulomb phase in the spin ice {Ho2Ti2O7}},\ }\href {https://doi.org/10.1126/science.1177582} {\bibfield  {journal} {\bibinfo  {journal} {Science}\ }\textbf {\bibinfo {volume} {326}},\ \bibinfo {pages} {415} (\bibinfo {year} {2009})}\BibitemShut {NoStop}%
\bibitem [{\citenamefont {Volovik}(1998)}]{Volovik1998}%
  \BibitemOpen
  \bibfield  {author} {\bibinfo {author} {\bibfnamefont {G.~E.}\ \bibnamefont {Volovik}},\ }\bibfield  {title} {\bibinfo {title} {Gravity of monopole and string and the gravitational constant in {$^3$He-A}},\ }\href {https://doi.org/10.1134/1.567704} {\bibfield  {journal} {\bibinfo  {journal} {JETP Letters}\ }\textbf {\bibinfo {volume} {67}},\ \bibinfo {pages} {698} (\bibinfo {year} {1998})}\BibitemShut {NoStop}%
\bibitem [{\citenamefont {Volovik}(2000)}]{Volovik2000}%
  \BibitemOpen
  \bibfield  {author} {\bibinfo {author} {\bibfnamefont {G.~E.}\ \bibnamefont {Volovik}},\ }\bibfield  {title} {\bibinfo {title} {Monopoles and fractional vortices in chiral superconductors},\ }\href {https://doi.org/10.1073/pnas.97.6.2431} {\bibfield  {journal} {\bibinfo  {journal} {Proceedings of the National Academy of Sciences}\ }\textbf {\bibinfo {volume} {97}},\ \bibinfo {pages} {2431} (\bibinfo {year} {2000})}\BibitemShut {NoStop}%
\bibitem [{\citenamefont {Volovik}\ and\ \citenamefont {Zhang}(2020)}]{Volovik2020}%
  \BibitemOpen
  \bibfield  {author} {\bibinfo {author} {\bibfnamefont {G.~E.}\ \bibnamefont {Volovik}}\ and\ \bibinfo {author} {\bibfnamefont {K.}~\bibnamefont {Zhang}},\ }\bibfield  {title} {\bibinfo {title} {String monopoles, string walls, vortex skyrmions, and nexus objects in the polar distorted $b$ phase of $^{3}\mathrm{He}$},\ }\href {https://doi.org/10.1103/PhysRevResearch.2.023263} {\bibfield  {journal} {\bibinfo  {journal} {Phys. Rev. Res.}\ }\textbf {\bibinfo {volume} {2}},\ \bibinfo {pages} {023263} (\bibinfo {year} {2020})}\BibitemShut {NoStop}%
\bibitem [{\citenamefont {Dalibard}\ \emph {et~al.}(2011)\citenamefont {Dalibard}, \citenamefont {Gerbier}, \citenamefont {Juzeli\ifmmode~\bar{u}\else \={u}\fi{}nas},\ and\ \citenamefont {\"Ohberg}}]{Dalibard2011}%
  \BibitemOpen
  \bibfield  {author} {\bibinfo {author} {\bibfnamefont {J.}~\bibnamefont {Dalibard}}, \bibinfo {author} {\bibfnamefont {F.}~\bibnamefont {Gerbier}}, \bibinfo {author} {\bibfnamefont {G.}~\bibnamefont {Juzeli\ifmmode~\bar{u}\else \={u}\fi{}nas}},\ and\ \bibinfo {author} {\bibfnamefont {P.}~\bibnamefont {\"Ohberg}},\ }\bibfield  {title} {\bibinfo {title} {Colloquium: Artificial gauge potentials for neutral atoms},\ }\href {https://doi.org/10.1103/RevModPhys.83.1523} {\bibfield  {journal} {\bibinfo  {journal} {Rev. Mod. Phys.}\ }\textbf {\bibinfo {volume} {83}},\ \bibinfo {pages} {1523} (\bibinfo {year} {2011})}\BibitemShut {NoStop}%
\bibitem [{\citenamefont {Goldman}\ \emph {et~al.}(2014)\citenamefont {Goldman}, \citenamefont {Juzeliunas}, \citenamefont {{\"O}hberg},\ and\ \citenamefont {Spielman}}]{Goldman2014}%
  \BibitemOpen
  \bibfield  {author} {\bibinfo {author} {\bibfnamefont {N.}~\bibnamefont {Goldman}}, \bibinfo {author} {\bibfnamefont {G.}~\bibnamefont {Juzeliunas}}, \bibinfo {author} {\bibfnamefont {P.}~\bibnamefont {{\"O}hberg}},\ and\ \bibinfo {author} {\bibfnamefont {I.~B.}\ \bibnamefont {Spielman}},\ }\bibfield  {title} {\bibinfo {title} {Light-induced gauge fields for ultracold atoms},\ }\href {https://doi.org/10.1088/0034-4885/77/12/126401} {\bibfield  {journal} {\bibinfo  {journal} {Rep. Prog. Phys.}\ }\textbf {\bibinfo {volume} {77}},\ \bibinfo {pages} {126401} (\bibinfo {year} {2014})}\BibitemShut {NoStop}%
\bibitem [{\citenamefont {Pietil\"a}\ and\ \citenamefont {M\"ott\"onen}(2009)}]{Pietila2009}%
  \BibitemOpen
  \bibfield  {author} {\bibinfo {author} {\bibfnamefont {V.}~\bibnamefont {Pietil\"a}}\ and\ \bibinfo {author} {\bibfnamefont {M.}~\bibnamefont {M\"ott\"onen}},\ }\bibfield  {title} {\bibinfo {title} {Creation of {Dirac} monopoles in spinor {Bose-Einstein} condensates},\ }\href {https://doi.org/10.1103/PhysRevLett.103.030401} {\bibfield  {journal} {\bibinfo  {journal} {Phys. Rev. Lett.}\ }\textbf {\bibinfo {volume} {103}},\ \bibinfo {pages} {030401} (\bibinfo {year} {2009})}\BibitemShut {NoStop}%
\bibitem [{\citenamefont {Ray}\ \emph {et~al.}(2014)\citenamefont {Ray}, \citenamefont {Ruokokoski}, \citenamefont {Kandel}, \citenamefont {M{\"o}tt{\"o}nen},\ and\ \citenamefont {Hall}}]{Ray2014}%
  \BibitemOpen
  \bibfield  {author} {\bibinfo {author} {\bibfnamefont {M.~W.}\ \bibnamefont {Ray}}, \bibinfo {author} {\bibfnamefont {E.}~\bibnamefont {Ruokokoski}}, \bibinfo {author} {\bibfnamefont {S.}~\bibnamefont {Kandel}}, \bibinfo {author} {\bibfnamefont {M.}~\bibnamefont {M{\"o}tt{\"o}nen}},\ and\ \bibinfo {author} {\bibfnamefont {D.~S.}\ \bibnamefont {Hall}},\ }\bibfield  {title} {\bibinfo {title} {Observation of {Dirac} monopoles in a synthetic magnetic field},\ }\href {https://doi.org/10.1038/nature12954} {\bibfield  {journal} {\bibinfo  {journal} {Nature}\ }\textbf {\bibinfo {volume} {505}},\ \bibinfo {pages} {657} (\bibinfo {year} {2014})}\BibitemShut {NoStop}%
\bibitem [{\citenamefont {Ollikainen}\ \emph {et~al.}(2017)\citenamefont {Ollikainen}, \citenamefont {Tiurev}, \citenamefont {Blinova}, \citenamefont {Lee}, \citenamefont {Hall},\ and\ \citenamefont {M\"ott\"onen}}]{Ollikainen2017}%
  \BibitemOpen
  \bibfield  {author} {\bibinfo {author} {\bibfnamefont {T.}~\bibnamefont {Ollikainen}}, \bibinfo {author} {\bibfnamefont {K.}~\bibnamefont {Tiurev}}, \bibinfo {author} {\bibfnamefont {A.}~\bibnamefont {Blinova}}, \bibinfo {author} {\bibfnamefont {W.}~\bibnamefont {Lee}}, \bibinfo {author} {\bibfnamefont {D.~S.}\ \bibnamefont {Hall}},\ and\ \bibinfo {author} {\bibfnamefont {M.}~\bibnamefont {M\"ott\"onen}},\ }\bibfield  {title} {\bibinfo {title} {Experimental realization of a {Dirac} monopole through the decay of an isolated monopole},\ }\href {https://doi.org/10.1103/PhysRevX.7.021023} {\bibfield  {journal} {\bibinfo  {journal} {Phys. Rev. X}\ }\textbf {\bibinfo {volume} {7}},\ \bibinfo {pages} {021023} (\bibinfo {year} {2017})}\BibitemShut {NoStop}%
\bibitem [{\citenamefont {Shi}\ and\ \citenamefont {Zhai}(2017)}]{Shi2017}%
  \BibitemOpen
  \bibfield  {author} {\bibinfo {author} {\bibfnamefont {Z.-Y.}\ \bibnamefont {Shi}}\ and\ \bibinfo {author} {\bibfnamefont {H.}~\bibnamefont {Zhai}},\ }\bibfield  {title} {\bibinfo {title} {Emergent gauge field for a chiral bound state on curved surface},\ }\href {https://doi.org/10.1088/1361-6455/aa84fa} {\bibfield  {journal} {\bibinfo  {journal} {J. Phys. B}\ }\textbf {\bibinfo {volume} {50}},\ \bibinfo {pages} {184006} (\bibinfo {year} {2017})}\BibitemShut {NoStop}%
\bibitem [{\citenamefont {Chen}\ \emph {et~al.}(2025)\citenamefont {Chen}, \citenamefont {Jiang}, \citenamefont {Bai}, \citenamefont {Yang},\ and\ \citenamefont {Zheng}}]{Chen2025SyntheticHalf}%
  \BibitemOpen
  \bibfield  {author} {\bibinfo {author} {\bibfnamefont {X.-Y.}\ \bibnamefont {Chen}}, \bibinfo {author} {\bibfnamefont {L.}~\bibnamefont {Jiang}}, \bibinfo {author} {\bibfnamefont {W.-K.}\ \bibnamefont {Bai}}, \bibinfo {author} {\bibfnamefont {T.}~\bibnamefont {Yang}},\ and\ \bibinfo {author} {\bibfnamefont {J.-H.}\ \bibnamefont {Zheng}},\ }\bibfield  {title} {\bibinfo {title} {Synthetic half-integer magnetic monopole and single-vortex dynamics in spherical {Bose-Einstein} condensates},\ }\href {https://doi.org/10.1103/PhysRevA.111.033322} {\bibfield  {journal} {\bibinfo  {journal} {Phys. Rev. A}\ }\textbf {\bibinfo {volume} {111}},\ \bibinfo {pages} {033322} (\bibinfo {year} {2025})}\BibitemShut {NoStop}%
\bibitem [{\citenamefont {Tononi}\ and\ \citenamefont {Salasnich}(2019)}]{Tononi2019}%
  \BibitemOpen
  \bibfield  {author} {\bibinfo {author} {\bibfnamefont {A.}~\bibnamefont {Tononi}}\ and\ \bibinfo {author} {\bibfnamefont {L.}~\bibnamefont {Salasnich}},\ }\bibfield  {title} {\bibinfo {title} {{Bose-Einstein} condensation on the surface of a sphere},\ }\href {https://doi.org/10.1103/PhysRevLett.123.160403} {\bibfield  {journal} {\bibinfo  {journal} {Phys. Rev. Lett.}\ }\textbf {\bibinfo {volume} {123}},\ \bibinfo {pages} {160403} (\bibinfo {year} {2019})}\BibitemShut {NoStop}%
\bibitem [{\citenamefont {Bereta}\ \emph {et~al.}(2019)\citenamefont {Bereta}, \citenamefont {Madeira}, \citenamefont {Bagnato},\ and\ \citenamefont {Caracanhas}}]{Bereta2019}%
  \BibitemOpen
  \bibfield  {author} {\bibinfo {author} {\bibfnamefont {S.~J.}\ \bibnamefont {Bereta}}, \bibinfo {author} {\bibfnamefont {L.}~\bibnamefont {Madeira}}, \bibinfo {author} {\bibfnamefont {V.~S.}\ \bibnamefont {Bagnato}},\ and\ \bibinfo {author} {\bibfnamefont {M.~A.}\ \bibnamefont {Caracanhas}},\ }\bibfield  {title} {\bibinfo {title} {{Bose–Einstein} condensation in spherically symmetric traps},\ }\href {https://doi.org/10.1119/1.5125092} {\bibfield  {journal} {\bibinfo  {journal} {Am. J. Phys.}\ }\textbf {\bibinfo {volume} {87}},\ \bibinfo {pages} {924} (\bibinfo {year} {2019})}\BibitemShut {NoStop}%
\bibitem [{\citenamefont {M{\'o}ller}\ \emph {et~al.}(2020)\citenamefont {M{\'o}ller}, \citenamefont {dos Santos}, \citenamefont {Bagnato},\ and\ \citenamefont {Pelster}}]{Moller2020}%
  \BibitemOpen
  \bibfield  {author} {\bibinfo {author} {\bibfnamefont {N.~S.}\ \bibnamefont {M{\'o}ller}}, \bibinfo {author} {\bibfnamefont {F.~E.~A.}\ \bibnamefont {dos Santos}}, \bibinfo {author} {\bibfnamefont {V.~S.}\ \bibnamefont {Bagnato}},\ and\ \bibinfo {author} {\bibfnamefont {A.}~\bibnamefont {Pelster}},\ }\bibfield  {title} {\bibinfo {title} {{Bose-Einstein} condensation on curved manifolds},\ }\href {https://doi.org/10.1088/1367-2630/ab91fb} {\bibfield  {journal} {\bibinfo  {journal} {New J. Phys.}\ }\textbf {\bibinfo {volume} {22}},\ \bibinfo {pages} {063059} (\bibinfo {year} {2020})}\BibitemShut {NoStop}%
\bibitem [{\citenamefont {Wolf}\ \emph {et~al.}(2022)\citenamefont {Wolf}, \citenamefont {Boegel}, \citenamefont {Meister}, \citenamefont {Bala\ifmmode~\check{z}\else \v{z}\fi{}}, \citenamefont {Gaaloul},\ and\ \citenamefont {Efremov}}]{Wolf2022}%
  \BibitemOpen
  \bibfield  {author} {\bibinfo {author} {\bibfnamefont {A.}~\bibnamefont {Wolf}}, \bibinfo {author} {\bibfnamefont {P.}~\bibnamefont {Boegel}}, \bibinfo {author} {\bibfnamefont {M.}~\bibnamefont {Meister}}, \bibinfo {author} {\bibfnamefont {A.}~\bibnamefont {Bala\ifmmode~\check{z}\else \v{z}\fi{}}}, \bibinfo {author} {\bibfnamefont {N.}~\bibnamefont {Gaaloul}},\ and\ \bibinfo {author} {\bibfnamefont {M.~A.}\ \bibnamefont {Efremov}},\ }\bibfield  {title} {\bibinfo {title} {Shell-shaped {Bose-Einstein} condensates based on dual-species mixtures},\ }\href {https://doi.org/10.1103/PhysRevA.106.013309} {\bibfield  {journal} {\bibinfo  {journal} {Phys. Rev. A}\ }\textbf {\bibinfo {volume} {106}},\ \bibinfo {pages} {013309} (\bibinfo {year} {2022})}\BibitemShut {NoStop}%
\bibitem [{\citenamefont {Veyron}\ \emph {et~al.}(2026)\citenamefont {Veyron}, \citenamefont {M{\'e}tayer}, \citenamefont {Gerent}, \citenamefont {Huang}, \citenamefont {Beraud}, \citenamefont {Garraway}, \citenamefont {Bernon},\ and\ \citenamefont {Battelier}}]{Veyron2026}%
  \BibitemOpen
  \bibfield  {author} {\bibinfo {author} {\bibfnamefont {R.}~\bibnamefont {Veyron}}, \bibinfo {author} {\bibfnamefont {C.}~\bibnamefont {M{\'e}tayer}}, \bibinfo {author} {\bibfnamefont {J.-B.}\ \bibnamefont {Gerent}}, \bibinfo {author} {\bibfnamefont {R.}~\bibnamefont {Huang}}, \bibinfo {author} {\bibfnamefont {E.}~\bibnamefont {Beraud}}, \bibinfo {author} {\bibfnamefont {B.~M.}\ \bibnamefont {Garraway}}, \bibinfo {author} {\bibfnamefont {S.}~\bibnamefont {Bernon}},\ and\ \bibinfo {author} {\bibfnamefont {B.}~\bibnamefont {Battelier}},\ }\bibfield  {title} {\bibinfo {title} {All-optical bubble trap for ultracold atoms in microgravity},\ }\bibfield  {journal} {\bibinfo  {journal} {AVS Quantum Sci.}\ }\textbf {\bibinfo {volume} {8}},\ \href {https://doi.org/10.1116/5.0305448} {10.1116/5.0305448} (\bibinfo {year} {2026})\BibitemShut {NoStop}%
\bibitem [{\citenamefont {Tononi}\ and\ \citenamefont {Salasnich}(2023)}]{Tononi2023}%
  \BibitemOpen
  \bibfield  {author} {\bibinfo {author} {\bibfnamefont {A.}~\bibnamefont {Tononi}}\ and\ \bibinfo {author} {\bibfnamefont {L.}~\bibnamefont {Salasnich}},\ }\bibfield  {title} {\bibinfo {title} {Low-dimensional quantum gases in curved geometries},\ }\href {https://doi.org/10.1038/s42254-023-00591-2} {\bibfield  {journal} {\bibinfo  {journal} {Nat. Rev. Phys.}\ }\textbf {\bibinfo {volume} {5}},\ \bibinfo {pages} {398} (\bibinfo {year} {2023})}\BibitemShut {NoStop}%
\bibitem [{\citenamefont {Tononi}\ and\ \citenamefont {Salasnich}(2024)}]{Tononi2024}%
  \BibitemOpen
  \bibfield  {author} {\bibinfo {author} {\bibfnamefont {A.}~\bibnamefont {Tononi}}\ and\ \bibinfo {author} {\bibfnamefont {L.}~\bibnamefont {Salasnich}},\ }\bibfield  {title} {\bibinfo {title} {Shell-shaped atomic gases},\ }\href {https://doi.org/10.1016/j.physrep.2024.04.004} {\bibfield  {journal} {\bibinfo  {journal} {Phys. Rep.}\ }\textbf {\bibinfo {volume} {1072}},\ \bibinfo {pages} {1} (\bibinfo {year} {2024})}\BibitemShut {NoStop}%
\bibitem [{\citenamefont {Dubessy}\ and\ \citenamefont {Perrin}(2025)}]{Dubessy2025}%
  \BibitemOpen
  \bibfield  {author} {\bibinfo {author} {\bibfnamefont {R.}~\bibnamefont {Dubessy}}\ and\ \bibinfo {author} {\bibfnamefont {H.}~\bibnamefont {Perrin}},\ }\bibfield  {title} {\bibinfo {title} {Quantum gases in bubble traps},\ }\href {https://doi.org/10.1116/5.0242948} {\bibfield  {journal} {\bibinfo  {journal} {AVS Quantum Sci.}\ }\textbf {\bibinfo {volume} {7}},\ \bibinfo {pages} {010501} (\bibinfo {year} {2025})}\BibitemShut {NoStop}%
\bibitem [{\citenamefont {Rhyno}\ \emph {et~al.}(2026)\citenamefont {Rhyno}, \citenamefont {Sun}, \citenamefont {Bedessem}, \citenamefont {Gaaloul}, \citenamefont {Lundblad},\ and\ \citenamefont {Vishveshwara}}]{Rhyno2026}%
  \BibitemOpen
  \bibfield  {author} {\bibinfo {author} {\bibfnamefont {B.}~\bibnamefont {Rhyno}}, \bibinfo {author} {\bibfnamefont {K.}~\bibnamefont {Sun}}, \bibinfo {author} {\bibfnamefont {J.}~\bibnamefont {Bedessem}}, \bibinfo {author} {\bibfnamefont {N.}~\bibnamefont {Gaaloul}}, \bibinfo {author} {\bibfnamefont {N.}~\bibnamefont {Lundblad}},\ and\ \bibinfo {author} {\bibfnamefont {S.}~\bibnamefont {Vishveshwara}},\ }\bibfield  {title} {\bibinfo {title} {Shell-shaped {Bose-Einstein} condensates: Dynamics, excitations, and thermodynamics},\ }\href {https://doi.org/10.1116/5.0320794} {\bibfield  {journal} {\bibinfo  {journal} {AVS Quantum Sci.}\ }\textbf {\bibinfo {volume} {8}},\ \bibinfo {pages} {010501} (\bibinfo {year} {2026})}\BibitemShut {NoStop}%
\bibitem [{\citenamefont {Zobay}\ and\ \citenamefont {Garraway}(2001)}]{Zobay2001}%
  \BibitemOpen
  \bibfield  {author} {\bibinfo {author} {\bibfnamefont {O.}~\bibnamefont {Zobay}}\ and\ \bibinfo {author} {\bibfnamefont {B.~M.}\ \bibnamefont {Garraway}},\ }\bibfield  {title} {\bibinfo {title} {Two-dimensional atom trapping in field-induced adiabatic potentials},\ }\href {https://doi.org/10.1103/PhysRevLett.86.1195} {\bibfield  {journal} {\bibinfo  {journal} {Phys. Rev. Lett.}\ }\textbf {\bibinfo {volume} {86}},\ \bibinfo {pages} {1195} (\bibinfo {year} {2001})}\BibitemShut {NoStop}%
\bibitem [{\citenamefont {Zobay}\ and\ \citenamefont {Garraway}(2004)}]{Zobay2004}%
  \BibitemOpen
  \bibfield  {author} {\bibinfo {author} {\bibfnamefont {O.}~\bibnamefont {Zobay}}\ and\ \bibinfo {author} {\bibfnamefont {B.~M.}\ \bibnamefont {Garraway}},\ }\bibfield  {title} {\bibinfo {title} {Atom trapping and two-dimensional {Bose-Einstein} condensates in field-induced adiabatic potentials},\ }\href {https://doi.org/10.1103/PhysRevA.69.023605} {\bibfield  {journal} {\bibinfo  {journal} {Phys. Rev. A}\ }\textbf {\bibinfo {volume} {69}},\ \bibinfo {pages} {023605} (\bibinfo {year} {2004})}\BibitemShut {NoStop}%
\bibitem [{\citenamefont {Tononi}\ \emph {et~al.}(2020)\citenamefont {Tononi}, \citenamefont {Cinti},\ and\ \citenamefont {Salasnich}}]{Tononi2020}%
  \BibitemOpen
  \bibfield  {author} {\bibinfo {author} {\bibfnamefont {A.}~\bibnamefont {Tononi}}, \bibinfo {author} {\bibfnamefont {F.}~\bibnamefont {Cinti}},\ and\ \bibinfo {author} {\bibfnamefont {L.}~\bibnamefont {Salasnich}},\ }\bibfield  {title} {\bibinfo {title} {Quantum bubbles in microgravity},\ }\href {https://doi.org/10.1103/PhysRevLett.125.010402} {\bibfield  {journal} {\bibinfo  {journal} {Phys. Rev. Lett.}\ }\textbf {\bibinfo {volume} {125}},\ \bibinfo {pages} {010402} (\bibinfo {year} {2020})}\BibitemShut {NoStop}%
\bibitem [{\citenamefont {Guo}\ \emph {et~al.}(2022)\citenamefont {Guo}, \citenamefont {Mercado~Gutierrez}, \citenamefont {Rey}, \citenamefont {Badr}, \citenamefont {Perrin}, \citenamefont {Longchambon}, \citenamefont {Bagnato}, \citenamefont {Perrin},\ and\ \citenamefont {Dubessy}}]{Guo2022}%
  \BibitemOpen
  \bibfield  {author} {\bibinfo {author} {\bibfnamefont {Y.}~\bibnamefont {Guo}}, \bibinfo {author} {\bibfnamefont {E.}~\bibnamefont {Mercado~Gutierrez}}, \bibinfo {author} {\bibfnamefont {D.}~\bibnamefont {Rey}}, \bibinfo {author} {\bibfnamefont {T.}~\bibnamefont {Badr}}, \bibinfo {author} {\bibfnamefont {A.}~\bibnamefont {Perrin}}, \bibinfo {author} {\bibfnamefont {L.}~\bibnamefont {Longchambon}}, \bibinfo {author} {\bibfnamefont {V.~S.}\ \bibnamefont {Bagnato}}, \bibinfo {author} {\bibfnamefont {H.}~\bibnamefont {Perrin}},\ and\ \bibinfo {author} {\bibfnamefont {R.}~\bibnamefont {Dubessy}},\ }\bibfield  {title} {\bibinfo {title} {Expansion of a quantum gas in a shell trap},\ }\href {https://doi.org/10.1088/1367-2630/ac919f} {\bibfield  {journal} {\bibinfo  {journal} {New J. Phys.}\ }\textbf {\bibinfo {volume} {24}},\ \bibinfo {pages} {093040} (\bibinfo {year} {2022})}\BibitemShut {NoStop}%
\bibitem [{\citenamefont {Jia}\ \emph {et~al.}(2022)\citenamefont {Jia}, \citenamefont {Huang}, \citenamefont {Qiu}, \citenamefont {Zhou}, \citenamefont {Yan},\ and\ \citenamefont {Wang}}]{Jia2022}%
  \BibitemOpen
  \bibfield  {author} {\bibinfo {author} {\bibfnamefont {F.}~\bibnamefont {Jia}}, \bibinfo {author} {\bibfnamefont {Z.}~\bibnamefont {Huang}}, \bibinfo {author} {\bibfnamefont {L.}~\bibnamefont {Qiu}}, \bibinfo {author} {\bibfnamefont {R.}~\bibnamefont {Zhou}}, \bibinfo {author} {\bibfnamefont {Y.}~\bibnamefont {Yan}},\ and\ \bibinfo {author} {\bibfnamefont {D.}~\bibnamefont {Wang}},\ }\bibfield  {title} {\bibinfo {title} {Expansion dynamics of a shell-shaped bose-einstein condensate},\ }\href {https://doi.org/10.1103/PhysRevLett.129.243402} {\bibfield  {journal} {\bibinfo  {journal} {Phys. Rev. Lett.}\ }\textbf {\bibinfo {volume} {129}},\ \bibinfo {pages} {243402} (\bibinfo {year} {2022})}\BibitemShut {NoStop}%
\bibitem [{\citenamefont {Carollo}\ \emph {et~al.}(2022)\citenamefont {Carollo}, \citenamefont {Aveline}, \citenamefont {Rhyno}, \citenamefont {Vishveshwara}, \citenamefont {Lannert}, \citenamefont {Murphree}, \citenamefont {Elliott}, \citenamefont {Williams}, \citenamefont {Thompson},\ and\ \citenamefont {Lundblad}}]{Carollo2022}%
  \BibitemOpen
  \bibfield  {author} {\bibinfo {author} {\bibfnamefont {R.~A.}\ \bibnamefont {Carollo}}, \bibinfo {author} {\bibfnamefont {D.~C.}\ \bibnamefont {Aveline}}, \bibinfo {author} {\bibfnamefont {B.}~\bibnamefont {Rhyno}}, \bibinfo {author} {\bibfnamefont {S.}~\bibnamefont {Vishveshwara}}, \bibinfo {author} {\bibfnamefont {C.}~\bibnamefont {Lannert}}, \bibinfo {author} {\bibfnamefont {J.~D.}\ \bibnamefont {Murphree}}, \bibinfo {author} {\bibfnamefont {E.~R.}\ \bibnamefont {Elliott}}, \bibinfo {author} {\bibfnamefont {J.~R.}\ \bibnamefont {Williams}}, \bibinfo {author} {\bibfnamefont {R.~J.}\ \bibnamefont {Thompson}},\ and\ \bibinfo {author} {\bibfnamefont {N.}~\bibnamefont {Lundblad}},\ }\bibfield  {title} {\bibinfo {title} {Observation of ultracold atomic bubbles in orbital microgravity},\ }\href {https://doi.org/10.1038/s41586-022-04639-8} {\bibfield  {journal} {\bibinfo  {journal} {Nature}\ }\textbf {\bibinfo {volume} {606}},\ \bibinfo {pages} {281} (\bibinfo {year} {2022})}\BibitemShut {NoStop}%
\bibitem [{\citenamefont {Bereta}\ \emph {et~al.}(2021)\citenamefont {Bereta}, \citenamefont {Caracanhas},\ and\ \citenamefont {Fetter}}]{Bereta2021}%
  \BibitemOpen
  \bibfield  {author} {\bibinfo {author} {\bibfnamefont {S.~J.}\ \bibnamefont {Bereta}}, \bibinfo {author} {\bibfnamefont {M.~A.}\ \bibnamefont {Caracanhas}},\ and\ \bibinfo {author} {\bibfnamefont {A.~L.}\ \bibnamefont {Fetter}},\ }\bibfield  {title} {\bibinfo {title} {Superfluid vortex dynamics on a spherical film},\ }\href {https://doi.org/10.1103/PhysRevA.103.053306} {\bibfield  {journal} {\bibinfo  {journal} {Phys. Rev. A}\ }\textbf {\bibinfo {volume} {103}},\ \bibinfo {pages} {053306} (\bibinfo {year} {2021})}\BibitemShut {NoStop}%
\bibitem [{\citenamefont {He}\ and\ \citenamefont {Chien}(2023)}]{He2023}%
  \BibitemOpen
  \bibfield  {author} {\bibinfo {author} {\bibfnamefont {Y.}~\bibnamefont {He}}\ and\ \bibinfo {author} {\bibfnamefont {C.-C.}\ \bibnamefont {Chien}},\ }\bibfield  {title} {\bibinfo {title} {Vortex structure and spectrum of an atomic fermi superfluid in a spherical bubble trap},\ }\href {https://doi.org/10.1103/PhysRevA.108.053303} {\bibfield  {journal} {\bibinfo  {journal} {Phys. Rev. A}\ }\textbf {\bibinfo {volume} {108}},\ \bibinfo {pages} {053303} (\bibinfo {year} {2023})}\BibitemShut {NoStop}%
\bibitem [{\citenamefont {Biral}\ \emph {et~al.}(2024)\citenamefont {Biral}, \citenamefont {M{\'o}ller}, \citenamefont {Pelster},\ and\ \citenamefont {dos Santos}}]{Biral2024}%
  \BibitemOpen
  \bibfield  {author} {\bibinfo {author} {\bibfnamefont {E.~J.}\ \bibnamefont {Biral}}, \bibinfo {author} {\bibfnamefont {N.~S.}\ \bibnamefont {M{\'o}ller}}, \bibinfo {author} {\bibfnamefont {A.}~\bibnamefont {Pelster}},\ and\ \bibinfo {author} {\bibfnamefont {F.~E.~A.}\ \bibnamefont {dos Santos}},\ }\bibfield  {title} {\bibinfo {title} {{Bose--Einstein} condensates and the thin-shell limit in anisotropic bubble traps},\ }\href {https://doi.org/10.1088/1367-2630/ad1a29} {\bibfield  {journal} {\bibinfo  {journal} {New J. Phys.}\ }\textbf {\bibinfo {volume} {26}},\ \bibinfo {pages} {013035} (\bibinfo {year} {2024})}\BibitemShut {NoStop}%
\bibitem [{\citenamefont {de~Oliveira}\ and\ \citenamefont {Salom{\'e}~M{\'o}ller}(2025)}]{DeOliveira2025}%
  \BibitemOpen
  \bibfield  {author} {\bibinfo {author} {\bibfnamefont {S.~M.}\ \bibnamefont {de~Oliveira}}\ and\ \bibinfo {author} {\bibfnamefont {N.}~\bibnamefont {Salom{\'e}~M{\'o}ller}},\ }\bibfield  {title} {\bibinfo {title} {Geometric potential for a {Bose-Einstein} condensate on a curved surface},\ }\href {https://doi.org/10.1116/5.0281991} {\bibfield  {journal} {\bibinfo  {journal} {AVS Quantum Sci.}\ }\textbf {\bibinfo {volume} {7}},\ \bibinfo {pages} {033203} (\bibinfo {year} {2025})}\BibitemShut {NoStop}%
\bibitem [{\citenamefont {Bowick}\ and\ \citenamefont {Giomi}(2009)}]{Bowick2009}%
  \BibitemOpen
  \bibfield  {author} {\bibinfo {author} {\bibfnamefont {M.~J.}\ \bibnamefont {Bowick}}\ and\ \bibinfo {author} {\bibfnamefont {L.}~\bibnamefont {Giomi}},\ }\bibfield  {title} {\bibinfo {title} {Two-dimensional matter: order, curvature and defects},\ }\href {https://doi.org/10.1080/00018730903043166} {\bibfield  {journal} {\bibinfo  {journal} {Adv. Phys.}\ }\textbf {\bibinfo {volume} {58}},\ \bibinfo {pages} {449} (\bibinfo {year} {2009})}\BibitemShut {NoStop}%
\bibitem [{\citenamefont {Turner}\ \emph {et~al.}(2010)\citenamefont {Turner}, \citenamefont {Vitelli},\ and\ \citenamefont {Nelson}}]{Turner2010}%
  \BibitemOpen
  \bibfield  {author} {\bibinfo {author} {\bibfnamefont {A.~M.}\ \bibnamefont {Turner}}, \bibinfo {author} {\bibfnamefont {V.}~\bibnamefont {Vitelli}},\ and\ \bibinfo {author} {\bibfnamefont {D.~R.}\ \bibnamefont {Nelson}},\ }\bibfield  {title} {\bibinfo {title} {Vortices on curved surfaces},\ }\href {https://doi.org/10.1103/RevModPhys.82.1301} {\bibfield  {journal} {\bibinfo  {journal} {Rev. Mod. Phys.}\ }\textbf {\bibinfo {volume} {82}},\ \bibinfo {pages} {1301} (\bibinfo {year} {2010})}\BibitemShut {NoStop}%
\bibitem [{\citenamefont {Padavi\ifmmode~\acute{c}\else \'{c}\fi{}}\ \emph {et~al.}(2020)\citenamefont {Padavi\ifmmode~\acute{c}\else \'{c}\fi{}}, \citenamefont {Sun}, \citenamefont {Lannert},\ and\ \citenamefont {Vishveshwara}}]{Padavic2020}%
  \BibitemOpen
  \bibfield  {author} {\bibinfo {author} {\bibfnamefont {K.}~\bibnamefont {Padavi\ifmmode~\acute{c}\else \'{c}\fi{}}}, \bibinfo {author} {\bibfnamefont {K.}~\bibnamefont {Sun}}, \bibinfo {author} {\bibfnamefont {C.}~\bibnamefont {Lannert}},\ and\ \bibinfo {author} {\bibfnamefont {S.}~\bibnamefont {Vishveshwara}},\ }\bibfield  {title} {\bibinfo {title} {Vortex-antivortex physics in shell-shaped {Bose-Einstein} condensates},\ }\href {https://doi.org/10.1103/PhysRevA.102.043305} {\bibfield  {journal} {\bibinfo  {journal} {Phys. Rev. A}\ }\textbf {\bibinfo {volume} {102}},\ \bibinfo {pages} {043305} (\bibinfo {year} {2020})}\BibitemShut {NoStop}%
\bibitem [{\citenamefont {Tononi}(2026)}]{Tononi2026}%
  \BibitemOpen
  \bibfield  {author} {\bibinfo {author} {\bibfnamefont {A.}~\bibnamefont {Tononi}},\ }\bibfield  {title} {\bibinfo {title} {Vortex dipoles in expanding shell-shaped {Bose-Einstein} condensates},\ }\bibfield  {journal} {\bibinfo  {journal} {arXiv preprint arXiv:2604.20407}\ }\href {https://doi.org/10.48550/arXiv.2604.20407} {10.48550/arXiv.2604.20407} (\bibinfo {year} {2026})\BibitemShut {NoStop}%
\bibitem [{\citenamefont {Guenther}\ \emph {et~al.}(2020)\citenamefont {Guenther}, \citenamefont {Massignan},\ and\ \citenamefont {Fetter}}]{Guenther2020}%
  \BibitemOpen
  \bibfield  {author} {\bibinfo {author} {\bibfnamefont {N.-E.}\ \bibnamefont {Guenther}}, \bibinfo {author} {\bibfnamefont {P.}~\bibnamefont {Massignan}},\ and\ \bibinfo {author} {\bibfnamefont {A.~L.}\ \bibnamefont {Fetter}},\ }\bibfield  {title} {\bibinfo {title} {Superfluid vortex dynamics on a torus and other toroidal surfaces of revolution},\ }\href {https://doi.org/10.1103/PhysRevA.101.053606} {\bibfield  {journal} {\bibinfo  {journal} {Phys. Rev. A}\ }\textbf {\bibinfo {volume} {101}},\ \bibinfo {pages} {053606} (\bibinfo {year} {2020})}\BibitemShut {NoStop}%
\bibitem [{\citenamefont {Caracanhas}\ \emph {et~al.}(2022)\citenamefont {Caracanhas}, \citenamefont {Massignan},\ and\ \citenamefont {Fetter}}]{Caracanhas2022}%
  \BibitemOpen
  \bibfield  {author} {\bibinfo {author} {\bibfnamefont {M.~A.}\ \bibnamefont {Caracanhas}}, \bibinfo {author} {\bibfnamefont {P.}~\bibnamefont {Massignan}},\ and\ \bibinfo {author} {\bibfnamefont {A.~L.}\ \bibnamefont {Fetter}},\ }\bibfield  {title} {\bibinfo {title} {Superfluid vortex dynamics on an ellipsoid and other surfaces of revolution},\ }\href {https://doi.org/10.1103/PhysRevA.105.023307} {\bibfield  {journal} {\bibinfo  {journal} {Phys. Rev. A}\ }\textbf {\bibinfo {volume} {105}},\ \bibinfo {pages} {023307} (\bibinfo {year} {2022})}\BibitemShut {NoStop}%
\bibitem [{\citenamefont {Cuadra}\ \emph {et~al.}(2026)\citenamefont {Cuadra}, \citenamefont {Gasenzer}, \citenamefont {Proment},\ and\ \citenamefont {Villois}}]{Cuadra2026}%
  \BibitemOpen
  \bibfield  {author} {\bibinfo {author} {\bibfnamefont {N.}~\bibnamefont {Cuadra}}, \bibinfo {author} {\bibfnamefont {T.}~\bibnamefont {Gasenzer}}, \bibinfo {author} {\bibfnamefont {D.}~\bibnamefont {Proment}},\ and\ \bibinfo {author} {\bibfnamefont {A.}~\bibnamefont {Villois}},\ }\bibfield  {title} {\bibinfo {title} {Jones-roberts solitary waves and the onset of rotation in a spherical surface condensate},\ }\bibfield  {journal} {\bibinfo  {journal} {arXiv preprint arXiv:2605.18297}\ }\href {https://doi.org/10.48550/arXiv.2605.18297} {10.48550/arXiv.2605.18297} (\bibinfo {year} {2026})\BibitemShut {NoStop}%
\bibitem [{\citenamefont {Polvani}\ and\ \citenamefont {Dritschel}(1993)}]{Polvani1993}%
  \BibitemOpen
  \bibfield  {author} {\bibinfo {author} {\bibfnamefont {L.~M.}\ \bibnamefont {Polvani}}\ and\ \bibinfo {author} {\bibfnamefont {D.~G.}\ \bibnamefont {Dritschel}},\ }\bibfield  {title} {\bibinfo {title} {Wave and vortex dynamics on the surface of a sphere},\ }\href {https://doi.org/10.1017/S0022112093002381} {\bibfield  {journal} {\bibinfo  {journal} {J. Fluid Mech.}\ }\textbf {\bibinfo {volume} {255}},\ \bibinfo {pages} {35} (\bibinfo {year} {1993})}\BibitemShut {NoStop}%
\bibitem [{\citenamefont {Sakajo}(2004)}]{Sakajo2004}%
  \BibitemOpen
  \bibfield  {author} {\bibinfo {author} {\bibfnamefont {T.}~\bibnamefont {Sakajo}},\ }\bibfield  {title} {\bibinfo {title} {Motion of a vortex sheet on a sphere with pole vortices},\ }\href {https://doi.org/10.1063/1.1644148} {\bibfield  {journal} {\bibinfo  {journal} {Phys. Fluids}\ }\textbf {\bibinfo {volume} {16}},\ \bibinfo {pages} {717} (\bibinfo {year} {2004})}\BibitemShut {NoStop}%
\bibitem [{\citenamefont {Sakajo}(2006)}]{Sakajo2006}%
  \BibitemOpen
  \bibfield  {author} {\bibinfo {author} {\bibfnamefont {T.}~\bibnamefont {Sakajo}},\ }\bibfield  {title} {\bibinfo {title} {Invariant dynamical systems embedded in the n-vortex problem on a sphere with pole vortices},\ }\href {https://doi.org/10.1016/j.physd.2006.04.002} {\bibfield  {journal} {\bibinfo  {journal} {Phys. D: Nonlinear Phenom.}\ }\textbf {\bibinfo {volume} {217}},\ \bibinfo {pages} {142} (\bibinfo {year} {2006})}\BibitemShut {NoStop}%
\bibitem [{\citenamefont {Adriani}\ \emph {et~al.}(2018)\citenamefont {Adriani}, \citenamefont {Mura}, \citenamefont {Orton}, \citenamefont {Hansen}, \citenamefont {Altieri}, \citenamefont {Moriconi}, \citenamefont {Rogers}, \citenamefont {Eichst{\"a}dt}, \citenamefont {Momary}, \citenamefont {Ingersoll} \emph {et~al.}}]{Adriani2018}%
  \BibitemOpen
  \bibfield  {author} {\bibinfo {author} {\bibfnamefont {A.}~\bibnamefont {Adriani}}, \bibinfo {author} {\bibfnamefont {A.}~\bibnamefont {Mura}}, \bibinfo {author} {\bibfnamefont {G.}~\bibnamefont {Orton}}, \bibinfo {author} {\bibfnamefont {C.}~\bibnamefont {Hansen}}, \bibinfo {author} {\bibfnamefont {F.}~\bibnamefont {Altieri}}, \bibinfo {author} {\bibfnamefont {M.}~\bibnamefont {Moriconi}}, \bibinfo {author} {\bibfnamefont {J.}~\bibnamefont {Rogers}}, \bibinfo {author} {\bibfnamefont {G.}~\bibnamefont {Eichst{\"a}dt}}, \bibinfo {author} {\bibfnamefont {T.}~\bibnamefont {Momary}}, \bibinfo {author} {\bibfnamefont {A.~P.}\ \bibnamefont {Ingersoll}}, \emph {et~al.},\ }\bibfield  {title} {\bibinfo {title} {Clusters of cyclones encircling jupiter’s poles},\ }\href {https://doi.org/10.1038/nature25491} {\bibfield  {journal} {\bibinfo  {journal} {Nature}\ }\textbf {\bibinfo {volume} {555}},\ \bibinfo {pages} {216} (\bibinfo {year} {2018})}\BibitemShut {NoStop}%
\bibitem [{\citenamefont {Li}\ \emph {et~al.}(2020)\citenamefont {Li}, \citenamefont {Ingersoll}, \citenamefont {Klipfel},\ and\ \citenamefont {Brettle}}]{Li2020}%
  \BibitemOpen
  \bibfield  {author} {\bibinfo {author} {\bibfnamefont {C.}~\bibnamefont {Li}}, \bibinfo {author} {\bibfnamefont {A.~P.}\ \bibnamefont {Ingersoll}}, \bibinfo {author} {\bibfnamefont {A.~P.}\ \bibnamefont {Klipfel}},\ and\ \bibinfo {author} {\bibfnamefont {H.}~\bibnamefont {Brettle}},\ }\bibfield  {title} {\bibinfo {title} {Modeling the stability of polygonal patterns of vortices at the poles of jupiter as revealed by the juno spacecraft},\ }\href {https://doi.org/10.1073/pnas.2008440117} {\bibfield  {journal} {\bibinfo  {journal} {Proc. Natl. Acad. Sci. U. S. A.}\ }\textbf {\bibinfo {volume} {117}},\ \bibinfo {pages} {24082} (\bibinfo {year} {2020})}\BibitemShut {NoStop}%
\bibitem [{\citenamefont {Haldane}(1983)}]{Haldane1983}%
  \BibitemOpen
  \bibfield  {author} {\bibinfo {author} {\bibfnamefont {F.~D.~M.}\ \bibnamefont {Haldane}},\ }\bibfield  {title} {\bibinfo {title} {Fractional quantization of the hall effect: A hierarchy of incompressible quantum fluid states},\ }\href {https://doi.org/10.1103/PhysRevLett.51.605} {\bibfield  {journal} {\bibinfo  {journal} {Phys. Rev. Lett.}\ }\textbf {\bibinfo {volume} {51}},\ \bibinfo {pages} {605} (\bibinfo {year} {1983})}\BibitemShut {NoStop}%
\bibitem [{\citenamefont {Zhou}\ \emph {et~al.}(2018)\citenamefont {Zhou}, \citenamefont {Wu}, \citenamefont {Guo}, \citenamefont {Wang}, \citenamefont {Pu},\ and\ \citenamefont {Zhou}}]{Zhou2018}%
  \BibitemOpen
  \bibfield  {author} {\bibinfo {author} {\bibfnamefont {X.-F.}\ \bibnamefont {Zhou}}, \bibinfo {author} {\bibfnamefont {C.}~\bibnamefont {Wu}}, \bibinfo {author} {\bibfnamefont {G.-C.}\ \bibnamefont {Guo}}, \bibinfo {author} {\bibfnamefont {R.}~\bibnamefont {Wang}}, \bibinfo {author} {\bibfnamefont {H.}~\bibnamefont {Pu}},\ and\ \bibinfo {author} {\bibfnamefont {Z.-W.}\ \bibnamefont {Zhou}},\ }\bibfield  {title} {\bibinfo {title} {Synthetic {Landau} levels and spinor vortex matter on a {Haldane} spherical surface with a magnetic monopole},\ }\href {https://doi.org/10.1103/PhysRevLett.120.130402} {\bibfield  {journal} {\bibinfo  {journal} {Phys. Rev. Lett.}\ }\textbf {\bibinfo {volume} {120}},\ \bibinfo {pages} {130402} (\bibinfo {year} {2018})}\BibitemShut {NoStop}%
\bibitem [{\citenamefont {Chen}\ \emph {et~al.}(2026)\citenamefont {Chen}, \citenamefont {Jiang}, \citenamefont {Yang},\ and\ \citenamefont {Zheng}}]{Chen2026IndexTheorem}%
  \BibitemOpen
  \bibfield  {author} {\bibinfo {author} {\bibfnamefont {X.-Y.}\ \bibnamefont {Chen}}, \bibinfo {author} {\bibfnamefont {L.}~\bibnamefont {Jiang}}, \bibinfo {author} {\bibfnamefont {T.}~\bibnamefont {Yang}},\ and\ \bibinfo {author} {\bibfnamefont {J.-H.}\ \bibnamefont {Zheng}},\ }\bibfield  {title} {\bibinfo {title} {Index theorem and vortex kinetics in bose-einstein condensates on a haldane sphere with a magnetic monopole},\ }\href {https://doi.org/10.1103/2msv-lk1m} {\bibfield  {journal} {\bibinfo  {journal} {Phys. Rev. A}\ }\textbf {\bibinfo {volume} {113}},\ \bibinfo {pages} {L041304} (\bibinfo {year} {2026})}\BibitemShut {NoStop}%
\bibitem [{\citenamefont {Richaud}\ \emph {et~al.}(2020)\citenamefont {Richaud}, \citenamefont {Penna}, \citenamefont {Mayol},\ and\ \citenamefont {Guilleumas}}]{Richaud2020}%
  \BibitemOpen
  \bibfield  {author} {\bibinfo {author} {\bibfnamefont {A.}~\bibnamefont {Richaud}}, \bibinfo {author} {\bibfnamefont {V.}~\bibnamefont {Penna}}, \bibinfo {author} {\bibfnamefont {R.}~\bibnamefont {Mayol}},\ and\ \bibinfo {author} {\bibfnamefont {M.}~\bibnamefont {Guilleumas}},\ }\bibfield  {title} {\bibinfo {title} {{Vortices with massive cores in a binary mixture of {Bose-Einstein} condensates}},\ }\href {https://doi.org/10.1103/PhysRevA.101.013630} {\bibfield  {journal} {\bibinfo  {journal} {Phys. Rev. A}\ }\textbf {\bibinfo {volume} {101}},\ \bibinfo {pages} {013630} (\bibinfo {year} {2020})}\BibitemShut {NoStop}%
\bibitem [{\citenamefont {Richaud}\ \emph {et~al.}(2021)\citenamefont {Richaud}, \citenamefont {Penna},\ and\ \citenamefont {Fetter}}]{Richaud2021PRA}%
  \BibitemOpen
  \bibfield  {author} {\bibinfo {author} {\bibfnamefont {A.}~\bibnamefont {Richaud}}, \bibinfo {author} {\bibfnamefont {V.}~\bibnamefont {Penna}},\ and\ \bibinfo {author} {\bibfnamefont {A.~L.}\ \bibnamefont {Fetter}},\ }\bibfield  {title} {\bibinfo {title} {{Dynamics of massive point vortices in a binary mixture of {Bose-Einstein} condensates}},\ }\href {https://doi.org/10.1103/PhysRevA.103.023311} {\bibfield  {journal} {\bibinfo  {journal} {Phys. Rev. A}\ }\textbf {\bibinfo {volume} {103}},\ \bibinfo {pages} {023311} (\bibinfo {year} {2021})}\BibitemShut {NoStop}%
\bibitem [{\citenamefont {Richaud}\ \emph {et~al.}(2022)\citenamefont {Richaud}, \citenamefont {Massignan}, \citenamefont {Penna},\ and\ \citenamefont {Fetter}}]{Richaud2022}%
  \BibitemOpen
  \bibfield  {author} {\bibinfo {author} {\bibfnamefont {A.}~\bibnamefont {Richaud}}, \bibinfo {author} {\bibfnamefont {P.}~\bibnamefont {Massignan}}, \bibinfo {author} {\bibfnamefont {V.}~\bibnamefont {Penna}},\ and\ \bibinfo {author} {\bibfnamefont {A.~L.}\ \bibnamefont {Fetter}},\ }\bibfield  {title} {\bibinfo {title} {{Dynamics of a massive superfluid vortex in ${r}^{k}$ confining potentials}},\ }\href {https://doi.org/10.1103/PhysRevA.106.063307} {\bibfield  {journal} {\bibinfo  {journal} {Phys. Rev. A}\ }\textbf {\bibinfo {volume} {106}},\ \bibinfo {pages} {063307} (\bibinfo {year} {2022})}\BibitemShut {NoStop}%
\bibitem [{\citenamefont {Caldara}\ \emph {et~al.}(2023)\citenamefont {Caldara}, \citenamefont {Richaud}, \citenamefont {Capone},\ and\ \citenamefont {Massignan}}]{Caldara2023}%
  \BibitemOpen
  \bibfield  {author} {\bibinfo {author} {\bibfnamefont {M.}~\bibnamefont {Caldara}}, \bibinfo {author} {\bibfnamefont {A.}~\bibnamefont {Richaud}}, \bibinfo {author} {\bibfnamefont {M.}~\bibnamefont {Capone}},\ and\ \bibinfo {author} {\bibfnamefont {P.}~\bibnamefont {Massignan}},\ }\bibfield  {title} {\bibinfo {title} {{Massive superfluid vortices and vortex necklaces on a planar annulus}},\ }\href {https://doi.org/10.21468/SciPostPhys.15.2.057} {\bibfield  {journal} {\bibinfo  {journal} {SciPost Phys.}\ }\textbf {\bibinfo {volume} {15}},\ \bibinfo {pages} {057} (\bibinfo {year} {2023})}\BibitemShut {NoStop}%
\bibitem [{\citenamefont {Bewley}\ \emph {et~al.}(2006)\citenamefont {Bewley}, \citenamefont {Lathrop},\ and\ \citenamefont {Sreenivasan}}]{Bewley2006}%
  \BibitemOpen
  \bibfield  {author} {\bibinfo {author} {\bibfnamefont {G.~P.}\ \bibnamefont {Bewley}}, \bibinfo {author} {\bibfnamefont {D.~P.}\ \bibnamefont {Lathrop}},\ and\ \bibinfo {author} {\bibfnamefont {K.~R.}\ \bibnamefont {Sreenivasan}},\ }\bibfield  {title} {\bibinfo {title} {{Visualization of quantized vortices}},\ }\href {https://doi.org/10.1038/441588a} {\bibfield  {journal} {\bibinfo  {journal} {Nature}\ }\textbf {\bibinfo {volume} {441}},\ \bibinfo {pages} {588} (\bibinfo {year} {2006})}\BibitemShut {NoStop}%
\bibitem [{\citenamefont {Griffin}\ \emph {et~al.}(2020)\citenamefont {Griffin}, \citenamefont {Shukla}, \citenamefont {Brachet},\ and\ \citenamefont {Nazarenko}}]{Griffin2020}%
  \BibitemOpen
  \bibfield  {author} {\bibinfo {author} {\bibfnamefont {A.}~\bibnamefont {Griffin}}, \bibinfo {author} {\bibfnamefont {V.}~\bibnamefont {Shukla}}, \bibinfo {author} {\bibfnamefont {M.-E.}\ \bibnamefont {Brachet}},\ and\ \bibinfo {author} {\bibfnamefont {S.}~\bibnamefont {Nazarenko}},\ }\bibfield  {title} {\bibinfo {title} {{Magnus-force model for active particles trapped on superfluid vortices}},\ }\href {https://doi.org/10.1103/PhysRevA.101.053601} {\bibfield  {journal} {\bibinfo  {journal} {Phys. Rev. A}\ }\textbf {\bibinfo {volume} {101}},\ \bibinfo {pages} {053601} (\bibinfo {year} {2020})}\BibitemShut {NoStop}%
\bibitem [{\citenamefont {Peretti}\ \emph {et~al.}(2023)\citenamefont {Peretti}, \citenamefont {Vessaire}, \citenamefont {Durozoy},\ and\ \citenamefont {Gibert}}]{Peretti2023}%
  \BibitemOpen
  \bibfield  {author} {\bibinfo {author} {\bibfnamefont {C.}~\bibnamefont {Peretti}}, \bibinfo {author} {\bibfnamefont {J.}~\bibnamefont {Vessaire}}, \bibinfo {author} {\bibfnamefont {{\'E}.}~\bibnamefont {Durozoy}},\ and\ \bibinfo {author} {\bibfnamefont {M.}~\bibnamefont {Gibert}},\ }\bibfield  {title} {\bibinfo {title} {Direct visualization of the quantum vortex lattice structure, oscillations, and destabilization in rotating $^4\mathrm{He}$},\ }\href {https://doi.org/https://doi.org/10.1126/sciadv.adh2899} {\bibfield  {journal} {\bibinfo  {journal} {Sci. Adv.}\ }\textbf {\bibinfo {volume} {9}},\ \bibinfo {pages} {eadh2899} (\bibinfo {year} {2023})}\BibitemShut {NoStop}%
\bibitem [{\citenamefont {Tang}\ \emph {et~al.}(2023)\citenamefont {Tang}, \citenamefont {Guo}, \citenamefont {Kobayashi}, \citenamefont {Yui}, \citenamefont {Tsubota},\ and\ \citenamefont {Kanai}}]{Tang2023}%
  \BibitemOpen
  \bibfield  {author} {\bibinfo {author} {\bibfnamefont {Y.}~\bibnamefont {Tang}}, \bibinfo {author} {\bibfnamefont {W.}~\bibnamefont {Guo}}, \bibinfo {author} {\bibfnamefont {H.}~\bibnamefont {Kobayashi}}, \bibinfo {author} {\bibfnamefont {S.}~\bibnamefont {Yui}}, \bibinfo {author} {\bibfnamefont {M.}~\bibnamefont {Tsubota}},\ and\ \bibinfo {author} {\bibfnamefont {T.}~\bibnamefont {Kanai}},\ }\bibfield  {title} {\bibinfo {title} {Imaging quantized vortex rings in superfluid helium to evaluate quantum dissipation},\ }\href {https://doi.org/10.1038/s41467-023-38787-w} {\bibfield  {journal} {\bibinfo  {journal} {Nat. Commun.}\ }\textbf {\bibinfo {volume} {14}},\ \bibinfo {pages} {2941} (\bibinfo {year} {2023})}\BibitemShut {NoStop}%
\bibitem [{\citenamefont {Kopnin}\ and\ \citenamefont {Vinokur}(1998)}]{Kopnin1998}%
  \BibitemOpen
  \bibfield  {author} {\bibinfo {author} {\bibfnamefont {N.~B.}\ \bibnamefont {Kopnin}}\ and\ \bibinfo {author} {\bibfnamefont {V.~M.}\ \bibnamefont {Vinokur}},\ }\bibfield  {title} {\bibinfo {title} {Dynamic vortex mass in clean fermi superfluids and superconductors},\ }\href {https://doi.org/10.1103/PhysRevLett.81.3952} {\bibfield  {journal} {\bibinfo  {journal} {Phys. Rev. Lett.}\ }\textbf {\bibinfo {volume} {81}},\ \bibinfo {pages} {3952} (\bibinfo {year} {1998})}\BibitemShut {NoStop}%
\bibitem [{\citenamefont {Simula}(2018)}]{Simula2018}%
  \BibitemOpen
  \bibfield  {author} {\bibinfo {author} {\bibfnamefont {T.}~\bibnamefont {Simula}},\ }\bibfield  {title} {\bibinfo {title} {{Vortex mass in a superfluid}},\ }\href {https://doi.org/10.1103/PhysRevA.97.023609} {\bibfield  {journal} {\bibinfo  {journal} {Phys. Rev. A}\ }\textbf {\bibinfo {volume} {97}},\ \bibinfo {pages} {023609} (\bibinfo {year} {2018})}\BibitemShut {NoStop}%
\bibitem [{\citenamefont {Simonucci}\ \emph {et~al.}(2019)\citenamefont {Simonucci}, \citenamefont {Pieri},\ and\ \citenamefont {Strinati}}]{Simonucci2019}%
  \BibitemOpen
  \bibfield  {author} {\bibinfo {author} {\bibfnamefont {S.}~\bibnamefont {Simonucci}}, \bibinfo {author} {\bibfnamefont {P.}~\bibnamefont {Pieri}},\ and\ \bibinfo {author} {\bibfnamefont {G.~C.}\ \bibnamefont {Strinati}},\ }\bibfield  {title} {\bibinfo {title} {{Bound states in a superfluid vortex: A detailed study along the BCS-BEC crossover}},\ }\href {https://doi.org/10.1103/PhysRevB.99.134506} {\bibfield  {journal} {\bibinfo  {journal} {Phys. Rev. B}\ }\textbf {\bibinfo {volume} {99}},\ \bibinfo {pages} {134506} (\bibinfo {year} {2019})}\BibitemShut {NoStop}%
\bibitem [{\citenamefont {Kwon}\ \emph {et~al.}(2021)\citenamefont {Kwon}, \citenamefont {Del~Pace}, \citenamefont {Xhani}, \citenamefont {Galantucci}, \citenamefont {Muzi~Falconi}, \citenamefont {Inguscio}, \citenamefont {Scazza},\ and\ \citenamefont {Roati}}]{Kwon2021}%
  \BibitemOpen
  \bibfield  {author} {\bibinfo {author} {\bibfnamefont {W.~J.}\ \bibnamefont {Kwon}}, \bibinfo {author} {\bibfnamefont {G.}~\bibnamefont {Del~Pace}}, \bibinfo {author} {\bibfnamefont {K.}~\bibnamefont {Xhani}}, \bibinfo {author} {\bibfnamefont {L.}~\bibnamefont {Galantucci}}, \bibinfo {author} {\bibfnamefont {A.}~\bibnamefont {Muzi~Falconi}}, \bibinfo {author} {\bibfnamefont {M.}~\bibnamefont {Inguscio}}, \bibinfo {author} {\bibfnamefont {F.}~\bibnamefont {Scazza}},\ and\ \bibinfo {author} {\bibfnamefont {G.}~\bibnamefont {Roati}},\ }\bibfield  {title} {\bibinfo {title} {{Sound emission and annihilations in a programmable quantum vortex collider}},\ }\href {https://doi.org/10.1038/s41586-021-04047-4} {\bibfield  {journal} {\bibinfo  {journal} {Nature}\ }\textbf {\bibinfo {volume} {600}},\ \bibinfo {pages} {64} (\bibinfo {year} {2021})}\BibitemShut {NoStop}%
\bibitem [{\citenamefont {Griffin}\ \emph {et~al.}(2009)\citenamefont {Griffin}, \citenamefont {Nikuni},\ and\ \citenamefont {Zaremba}}]{Griffin2009}%
  \BibitemOpen
  \bibfield  {author} {\bibinfo {author} {\bibfnamefont {A.}~\bibnamefont {Griffin}}, \bibinfo {author} {\bibfnamefont {T.}~\bibnamefont {Nikuni}},\ and\ \bibinfo {author} {\bibfnamefont {E.}~\bibnamefont {Zaremba}},\ }\href {https://doi.org/10.1017/CBO9780511575150} {\emph {\bibinfo {title} {{Bose-condensed gases at finite temperatures}}}}\ (\bibinfo  {publisher} {Cambridge University Press},\ \bibinfo {year} {2009})\BibitemShut {NoStop}%
\bibitem [{\citenamefont {Jackson}\ \emph {et~al.}(2009)\citenamefont {Jackson}, \citenamefont {Proukakis}, \citenamefont {Barenghi},\ and\ \citenamefont {Zaremba}}]{Jackson2009}%
  \BibitemOpen
  \bibfield  {author} {\bibinfo {author} {\bibfnamefont {B.}~\bibnamefont {Jackson}}, \bibinfo {author} {\bibfnamefont {N.~P.}\ \bibnamefont {Proukakis}}, \bibinfo {author} {\bibfnamefont {C.~F.}\ \bibnamefont {Barenghi}},\ and\ \bibinfo {author} {\bibfnamefont {E.}~\bibnamefont {Zaremba}},\ }\bibfield  {title} {\bibinfo {title} {Finite-temperature vortex dynamics in bose-einstein condensates},\ }\href {https://doi.org/10.1103/PhysRevA.79.053615} {\bibfield  {journal} {\bibinfo  {journal} {Phys. Rev. A}\ }\textbf {\bibinfo {volume} {79}},\ \bibinfo {pages} {053615} (\bibinfo {year} {2009})}\BibitemShut {NoStop}%
\bibitem [{\citenamefont {Richaud}\ \emph {et~al.}(2025)\citenamefont {Richaud}, \citenamefont {Caldara}, \citenamefont {Capone}, \citenamefont {Massignan},\ and\ \citenamefont {Wlaz\l{}owski}}]{Richaud2025}%
  \BibitemOpen
  \bibfield  {author} {\bibinfo {author} {\bibfnamefont {A.}~\bibnamefont {Richaud}}, \bibinfo {author} {\bibfnamefont {M.}~\bibnamefont {Caldara}}, \bibinfo {author} {\bibfnamefont {M.}~\bibnamefont {Capone}}, \bibinfo {author} {\bibfnamefont {P.}~\bibnamefont {Massignan}},\ and\ \bibinfo {author} {\bibfnamefont {G.}~\bibnamefont {Wlaz\l{}owski}},\ }\bibfield  {title} {\bibinfo {title} {Dynamical signature of vortex mass in {Fermi} superfluids},\ }\href {https://doi.org/10.1103/zcb4-dldf} {\bibfield  {journal} {\bibinfo  {journal} {Phys. Rev. A}\ }\textbf {\bibinfo {volume} {112}},\ \bibinfo {pages} {L051306} (\bibinfo {year} {2025})}\BibitemShut {NoStop}%
\bibitem [{\citenamefont {Anderson}\ \emph {et~al.}(2000)\citenamefont {Anderson}, \citenamefont {Haljan}, \citenamefont {Wieman},\ and\ \citenamefont {Cornell}}]{Anderson2000}%
  \BibitemOpen
  \bibfield  {author} {\bibinfo {author} {\bibfnamefont {B.~P.}\ \bibnamefont {Anderson}}, \bibinfo {author} {\bibfnamefont {P.~C.}\ \bibnamefont {Haljan}}, \bibinfo {author} {\bibfnamefont {C.~E.}\ \bibnamefont {Wieman}},\ and\ \bibinfo {author} {\bibfnamefont {E.~A.}\ \bibnamefont {Cornell}},\ }\bibfield  {title} {\bibinfo {title} {{Vortex Precession in {Bose-Einstein} Condensates: Observations with Filled and Empty Cores}},\ }\href {https://doi.org/10.1103/PhysRevLett.85.2857} {\bibfield  {journal} {\bibinfo  {journal} {Phys. Rev. Lett.}\ }\textbf {\bibinfo {volume} {85}},\ \bibinfo {pages} {2857} (\bibinfo {year} {2000})}\BibitemShut {NoStop}%
\bibitem [{\citenamefont {Law}\ \emph {et~al.}(2010)\citenamefont {Law}, \citenamefont {Kevrekidis},\ and\ \citenamefont {Tuckerman}}]{Law2010}%
  \BibitemOpen
  \bibfield  {author} {\bibinfo {author} {\bibfnamefont {K.~J.~H.}\ \bibnamefont {Law}}, \bibinfo {author} {\bibfnamefont {P.~G.}\ \bibnamefont {Kevrekidis}},\ and\ \bibinfo {author} {\bibfnamefont {L.~S.}\ \bibnamefont {Tuckerman}},\ }\bibfield  {title} {\bibinfo {title} {{Stable Vortex--Bright-Soliton Structures in Two-Component {Bose-Einstein} Condensates}},\ }\href {https://doi.org/10.1103/PhysRevLett.105.160405} {\bibfield  {journal} {\bibinfo  {journal} {Phys. Rev. Lett.}\ }\textbf {\bibinfo {volume} {105}},\ \bibinfo {pages} {160405} (\bibinfo {year} {2010})}\BibitemShut {NoStop}%
\bibitem [{\citenamefont {Williamson}\ and\ \citenamefont {Blakie}(2021)}]{Williamson2021}%
  \BibitemOpen
  \bibfield  {author} {\bibinfo {author} {\bibfnamefont {L.~A.}\ \bibnamefont {Williamson}}\ and\ \bibinfo {author} {\bibfnamefont {P.~B.}\ \bibnamefont {Blakie}},\ }\bibfield  {title} {\bibinfo {title} {Damped point-vortex model for polar-core spin vortices in a ferromagnetic spin-1 bose-einstein condensate},\ }\href {https://doi.org/10.1103/PhysRevResearch.3.013154} {\bibfield  {journal} {\bibinfo  {journal} {Phys. Rev. Res.}\ }\textbf {\bibinfo {volume} {3}},\ \bibinfo {pages} {013154} (\bibinfo {year} {2021})}\BibitemShut {NoStop}%
\bibitem [{\citenamefont {Patrick}\ \emph {et~al.}(2023)\citenamefont {Patrick}, \citenamefont {Gupta}, \citenamefont {Gregory},\ and\ \citenamefont {Barenghi}}]{Patrick2023}%
  \BibitemOpen
  \bibfield  {author} {\bibinfo {author} {\bibfnamefont {S.}~\bibnamefont {Patrick}}, \bibinfo {author} {\bibfnamefont {A.}~\bibnamefont {Gupta}}, \bibinfo {author} {\bibfnamefont {R.}~\bibnamefont {Gregory}},\ and\ \bibinfo {author} {\bibfnamefont {C.~F.}\ \bibnamefont {Barenghi}},\ }\bibfield  {title} {\bibinfo {title} {{Stability of quantized vortices in two-component condensates}},\ }\href {https://doi.org/10.1103/PhysRevResearch.5.033201} {\bibfield  {journal} {\bibinfo  {journal} {Phys. Rev. Res.}\ }\textbf {\bibinfo {volume} {5}},\ \bibinfo {pages} {033201} (\bibinfo {year} {2023})}\BibitemShut {NoStop}%
\bibitem [{\citenamefont {Kanjo}\ and\ \citenamefont {Takeuchi}(2024)}]{Kanjo2024}%
  \BibitemOpen
  \bibfield  {author} {\bibinfo {author} {\bibfnamefont {A.}~\bibnamefont {Kanjo}}\ and\ \bibinfo {author} {\bibfnamefont {H.}~\bibnamefont {Takeuchi}},\ }\bibfield  {title} {\bibinfo {title} {Universal description of massive point vortices and verification methods of vortex inertia in superfluids},\ }\href {https://doi.org/10.1103/PhysRevA.110.063311} {\bibfield  {journal} {\bibinfo  {journal} {Phys. Rev. A}\ }\textbf {\bibinfo {volume} {110}},\ \bibinfo {pages} {063311} (\bibinfo {year} {2024})}\BibitemShut {NoStop}%
\bibitem [{\citenamefont {D'Ambroise}\ \emph {et~al.}(2025)\citenamefont {D'Ambroise}, \citenamefont {Wang}, \citenamefont {Ticknor}, \citenamefont {Carretero-Gonz\'alez},\ and\ \citenamefont {Kevrekidis}}]{Dambroise2025}%
  \BibitemOpen
  \bibfield  {author} {\bibinfo {author} {\bibfnamefont {J.}~\bibnamefont {D'Ambroise}}, \bibinfo {author} {\bibfnamefont {W.}~\bibnamefont {Wang}}, \bibinfo {author} {\bibfnamefont {C.}~\bibnamefont {Ticknor}}, \bibinfo {author} {\bibfnamefont {R.}~\bibnamefont {Carretero-Gonz\'alez}},\ and\ \bibinfo {author} {\bibfnamefont {P.~G.}\ \bibnamefont {Kevrekidis}},\ }\bibfield  {title} {\bibinfo {title} {Stability and dynamics of massive vortices in two-component {Bose-Einstein} condensates},\ }\href {https://doi.org/10.1103/PhysRevE.111.034216} {\bibfield  {journal} {\bibinfo  {journal} {Phys. Rev. E}\ }\textbf {\bibinfo {volume} {111}},\ \bibinfo {pages} {034216} (\bibinfo {year} {2025})}\BibitemShut {NoStop}%
\bibitem [{\citenamefont {Anandan}(1992)}]{Anandan1992}%
  \BibitemOpen
  \bibfield  {author} {\bibinfo {author} {\bibfnamefont {J.}~\bibnamefont {Anandan}},\ }\bibfield  {title} {\bibinfo {title} {The geometric phase},\ }\href {https://doi.org/10.1038/360307a0} {\bibfield  {journal} {\bibinfo  {journal} {Nature}\ }\textbf {\bibinfo {volume} {360}},\ \bibinfo {pages} {307} (\bibinfo {year} {1992})}\BibitemShut {NoStop}%
\bibitem [{\citenamefont {Polkinghorne}\ \emph {et~al.}(2021)\citenamefont {Polkinghorne}, \citenamefont {Groszek},\ and\ \citenamefont {Simula}}]{Polkinghorne2021}%
  \BibitemOpen
  \bibfield  {author} {\bibinfo {author} {\bibfnamefont {R.~E.~S.}\ \bibnamefont {Polkinghorne}}, \bibinfo {author} {\bibfnamefont {A.~J.}\ \bibnamefont {Groszek}},\ and\ \bibinfo {author} {\bibfnamefont {T.~P.}\ \bibnamefont {Simula}},\ }\bibfield  {title} {\bibinfo {title} {Geometric phases of a vortex in a superfluid},\ }\href {https://doi.org/10.1103/PhysRevA.104.L041305} {\bibfield  {journal} {\bibinfo  {journal} {Phys. Rev. A}\ }\textbf {\bibinfo {volume} {104}},\ \bibinfo {pages} {L041305} (\bibinfo {year} {2021})}\BibitemShut {NoStop}%
\bibitem [{\citenamefont {Aharonov}\ and\ \citenamefont {Anandan}(1987)}]{Aharonov1987}%
  \BibitemOpen
  \bibfield  {author} {\bibinfo {author} {\bibfnamefont {Y.}~\bibnamefont {Aharonov}}\ and\ \bibinfo {author} {\bibfnamefont {J.}~\bibnamefont {Anandan}},\ }\bibfield  {title} {\bibinfo {title} {Phase change during a cyclic quantum evolution},\ }\href {https://doi.org/10.1103/PhysRevLett.58.1593} {\bibfield  {journal} {\bibinfo  {journal} {Phys. Rev. Lett.}\ }\textbf {\bibinfo {volume} {58}},\ \bibinfo {pages} {1593} (\bibinfo {year} {1987})}\BibitemShut {NoStop}%
\bibitem [{\citenamefont {Haldane}\ and\ \citenamefont {Wu}(1985)}]{Haldane1985}%
  \BibitemOpen
  \bibfield  {author} {\bibinfo {author} {\bibfnamefont {F.~D.~M.}\ \bibnamefont {Haldane}}\ and\ \bibinfo {author} {\bibfnamefont {Y.-S.}\ \bibnamefont {Wu}},\ }\bibfield  {title} {\bibinfo {title} {Quantum dynamics and statistics of vortices in two-dimensional superfluids},\ }\href {https://doi.org/10.1103/PhysRevLett.55.2887} {\bibfield  {journal} {\bibinfo  {journal} {Phys. Rev. Lett.}\ }\textbf {\bibinfo {volume} {55}},\ \bibinfo {pages} {2887} (\bibinfo {year} {1985})}\BibitemShut {NoStop}%
\bibitem [{\citenamefont {Chen}(2016)}]{Chen2016}%
  \BibitemOpen
  \bibfield  {author} {\bibinfo {author} {\bibfnamefont {F.~F.}\ \bibnamefont {Chen}},\ }\bibinfo {title} {{Introduction to Plasma Physics and Controlled Fusion}}\ (\bibinfo  {publisher} {Springer Cham},\ \bibinfo {year} {2016})\ Chap.~\bibinfo {chapter} {1},\ \bibinfo {edition} {3rd}\ ed.\BibitemShut {Stop}%
\bibitem [{\citenamefont {Maher}\ \emph {et~al.}(2015)\citenamefont {Maher}, \citenamefont {Jjunju},\ and\ \citenamefont {Taylor}}]{Maher2015}%
  \BibitemOpen
  \bibfield  {author} {\bibinfo {author} {\bibfnamefont {S.}~\bibnamefont {Maher}}, \bibinfo {author} {\bibfnamefont {F.~P.~M.}\ \bibnamefont {Jjunju}},\ and\ \bibinfo {author} {\bibfnamefont {S.}~\bibnamefont {Taylor}},\ }\bibfield  {title} {\bibinfo {title} {Colloquium: 100 years of mass spectrometry: Perspectives and future trends},\ }\href {https://doi.org/10.1103/RevModPhys.87.113} {\bibfield  {journal} {\bibinfo  {journal} {Rev. Mod. Phys.}\ }\textbf {\bibinfo {volume} {87}},\ \bibinfo {pages} {113} (\bibinfo {year} {2015})}\BibitemShut {NoStop}%
\bibitem [{\citenamefont {Brown}\ and\ \citenamefont {Gabrielse}(1986)}]{Brown1986}%
  \BibitemOpen
  \bibfield  {author} {\bibinfo {author} {\bibfnamefont {L.~S.}\ \bibnamefont {Brown}}\ and\ \bibinfo {author} {\bibfnamefont {G.}~\bibnamefont {Gabrielse}},\ }\bibfield  {title} {\bibinfo {title} {{Geonium theory: Physics of a single electron or ion in a Penning trap}},\ }\href {https://doi.org/10.1103/RevModPhys.58.233} {\bibfield  {journal} {\bibinfo  {journal} {Rev. Mod. Phys.}\ }\textbf {\bibinfo {volume} {58}},\ \bibinfo {pages} {233} (\bibinfo {year} {1986})}\BibitemShut {NoStop}%
\bibitem [{\citenamefont {Luttinger}(1956)}]{Luttinger1956}%
  \BibitemOpen
  \bibfield  {author} {\bibinfo {author} {\bibfnamefont {J.~M.}\ \bibnamefont {Luttinger}},\ }\bibfield  {title} {\bibinfo {title} {Quantum theory of cyclotron resonance in semiconductors: General theory},\ }\href {https://doi.org/10.1103/PhysRev.102.1030} {\bibfield  {journal} {\bibinfo  {journal} {Phys. Rev.}\ }\textbf {\bibinfo {volume} {102}},\ \bibinfo {pages} {1030} (\bibinfo {year} {1956})}\BibitemShut {NoStop}%
\bibitem [{\citenamefont {Mu\~noz de~las Heras}\ \emph {et~al.}(2020)\citenamefont {Mu\~noz de~las Heras}, \citenamefont {Macaluso},\ and\ \citenamefont {Carusotto}}]{Munoz_de_las_Heras2020}%
  \BibitemOpen
  \bibfield  {author} {\bibinfo {author} {\bibfnamefont {A.}~\bibnamefont {Mu\~noz de~las Heras}}, \bibinfo {author} {\bibfnamefont {E.}~\bibnamefont {Macaluso}},\ and\ \bibinfo {author} {\bibfnamefont {I.}~\bibnamefont {Carusotto}},\ }\bibfield  {title} {\bibinfo {title} {Anyonic molecules in atomic fractional quantum hall liquids: A quantitative probe of fractional charge and anyonic statistics},\ }\href {https://doi.org/10.1103/PhysRevX.10.041058} {\bibfield  {journal} {\bibinfo  {journal} {Phys. Rev. X}\ }\textbf {\bibinfo {volume} {10}},\ \bibinfo {pages} {041058} (\bibinfo {year} {2020})}\BibitemShut {NoStop}%
\bibitem [{\citenamefont {McEwen}\ and\ \citenamefont {Wiaux}(2011)}]{McEwen2011}%
  \BibitemOpen
  \bibfield  {author} {\bibinfo {author} {\bibfnamefont {J.~D.}\ \bibnamefont {McEwen}}\ and\ \bibinfo {author} {\bibfnamefont {Y.}~\bibnamefont {Wiaux}},\ }\bibfield  {title} {\bibinfo {title} {A novel sampling theorem on the sphere},\ }\href {https://doi.org/10.1109/TSP.2011.2166394} {\bibfield  {journal} {\bibinfo  {journal} {IEEE Trans. Signal Process.}\ }\textbf {\bibinfo {volume} {59}},\ \bibinfo {pages} {5876} (\bibinfo {year} {2011})}\BibitemShut {NoStop}%
\bibitem [{\citenamefont {McEwen}\ and\ \citenamefont {collaborators}(2025)}]{SSHT}%
  \BibitemOpen
  \bibfield  {author} {\bibinfo {author} {\bibfnamefont {J.~D.}\ \bibnamefont {McEwen}}\ and\ \bibinfo {author} {\bibnamefont {collaborators}},\ }\href {https://astro-informatics.github.io/ssht/} {\bibinfo {title} {Ssht: Spin spherical harmonic transforms}} (\bibinfo {year} {2025}),\ \bibinfo {note} {software package}\BibitemShut {NoStop}%
\bibitem [{\citenamefont {Giuriato}\ \emph {et~al.}(2020)\citenamefont {Giuriato}, \citenamefont {Krstulovic},\ and\ \citenamefont {Nazarenko}}]{Giuriato2020}%
  \BibitemOpen
  \bibfield  {author} {\bibinfo {author} {\bibfnamefont {U.}~\bibnamefont {Giuriato}}, \bibinfo {author} {\bibfnamefont {G.}~\bibnamefont {Krstulovic}},\ and\ \bibinfo {author} {\bibfnamefont {S.}~\bibnamefont {Nazarenko}},\ }\bibfield  {title} {\bibinfo {title} {How trapped particles interact with and sample superfluid vortex excitations},\ }\href {https://doi.org/10.1103/PhysRevResearch.2.023149} {\bibfield  {journal} {\bibinfo  {journal} {Phys. Rev. Res.}\ }\textbf {\bibinfo {volume} {2}},\ \bibinfo {pages} {023149} (\bibinfo {year} {2020})}\BibitemShut {NoStop}%
\bibitem [{\citenamefont {Richaud}\ and\ \citenamefont {Sorba}(2026{\natexlab{a}})}]{GitHubCodeMassiveDipole}%
  \BibitemOpen
  \bibfield  {author} {\bibinfo {author} {\bibfnamefont {A.}~\bibnamefont {Richaud}}\ and\ \bibinfo {author} {\bibfnamefont {M.}~\bibnamefont {Sorba}},\ }\href {https://doi.org/10.5281/zenodo.21351319} {\bibinfo {title} {Massive vortex dipole on a spherical superfluid shell}} (\bibinfo {year} {2026}{\natexlab{a}})\BibitemShut {NoStop}%
\bibitem [{\citenamefont {Richaud}\ and\ \citenamefont {Sorba}(2026{\natexlab{b}})}]{GitHubCodeKHI}%
  \BibitemOpen
  \bibfield  {author} {\bibinfo {author} {\bibfnamefont {A.}~\bibnamefont {Richaud}}\ and\ \bibinfo {author} {\bibfnamefont {M.}~\bibnamefont {Sorba}},\ }\href {https://doi.org/10.5281/zenodo.21367835} {\bibinfo {title} {Superfluid kelvin--helmholtz instability on a spherical shell}} (\bibinfo {year} {2026}{\natexlab{b}})\BibitemShut {NoStop}%
\bibitem [{\citenamefont {Aref}(1995)}]{Aref1995}%
  \BibitemOpen
  \bibfield  {author} {\bibinfo {author} {\bibfnamefont {H.}~\bibnamefont {Aref}},\ }\bibfield  {title} {\bibinfo {title} {On the equilibrium and stability of a row of point vortices},\ }\href {https://doi.org/10.1017/s002211209500245x} {\bibfield  {journal} {\bibinfo  {journal} {J. Fluid Mech.}\ }\textbf {\bibinfo {volume} {290}},\ \bibinfo {pages} {167–181} (\bibinfo {year} {1995})}\BibitemShut {NoStop}%
\bibitem [{\citenamefont {Takeuchi}\ \emph {et~al.}(2010)\citenamefont {Takeuchi}, \citenamefont {Suzuki}, \citenamefont {Kasamatsu}, \citenamefont {Saito},\ and\ \citenamefont {Tsubota}}]{Takeuchi2010}%
  \BibitemOpen
  \bibfield  {author} {\bibinfo {author} {\bibfnamefont {H.}~\bibnamefont {Takeuchi}}, \bibinfo {author} {\bibfnamefont {N.}~\bibnamefont {Suzuki}}, \bibinfo {author} {\bibfnamefont {K.}~\bibnamefont {Kasamatsu}}, \bibinfo {author} {\bibfnamefont {H.}~\bibnamefont {Saito}},\ and\ \bibinfo {author} {\bibfnamefont {M.}~\bibnamefont {Tsubota}},\ }\bibfield  {title} {\bibinfo {title} {{Quantum Kelvin-Helmholtz instability in phase-separated two-component {Bose-Einstein} condensates}},\ }\href {https://doi.org/10.1103/PhysRevB.81.094517} {\bibfield  {journal} {\bibinfo  {journal} {Phys. Rev. B}\ }\textbf {\bibinfo {volume} {81}},\ \bibinfo {pages} {094517} (\bibinfo {year} {2010})}\BibitemShut {NoStop}%
\bibitem [{\citenamefont {Baggaley}\ and\ \citenamefont {Parker}(2018)}]{Baggaley2018}%
  \BibitemOpen
  \bibfield  {author} {\bibinfo {author} {\bibfnamefont {A.~W.}\ \bibnamefont {Baggaley}}\ and\ \bibinfo {author} {\bibfnamefont {N.~G.}\ \bibnamefont {Parker}},\ }\bibfield  {title} {\bibinfo {title} {{Kelvin-Helmholtz instability in a single-component atomic superfluid}},\ }\href {https://doi.org/10.1103/PhysRevA.97.053608} {\bibfield  {journal} {\bibinfo  {journal} {Phys. Rev. A}\ }\textbf {\bibinfo {volume} {97}},\ \bibinfo {pages} {053608} (\bibinfo {year} {2018})}\BibitemShut {NoStop}%
\bibitem [{\citenamefont {Hernandez-Rajkov}\ \emph {et~al.}(2024)\citenamefont {Hernandez-Rajkov}, \citenamefont {Grani}, \citenamefont {Scazza}, \citenamefont {Pace}, \citenamefont {Kwon}, \citenamefont {Inguscio}, \citenamefont {Xhani}, \citenamefont {Fort}, \citenamefont {Modugno}, \citenamefont {Marino},\ and\ \citenamefont {Roati}}]{Hernandez2024}%
  \BibitemOpen
  \bibfield  {author} {\bibinfo {author} {\bibfnamefont {D.}~\bibnamefont {Hernandez-Rajkov}}, \bibinfo {author} {\bibfnamefont {N.}~\bibnamefont {Grani}}, \bibinfo {author} {\bibfnamefont {F.}~\bibnamefont {Scazza}}, \bibinfo {author} {\bibfnamefont {G.~D.}\ \bibnamefont {Pace}}, \bibinfo {author} {\bibfnamefont {W.~J.}\ \bibnamefont {Kwon}}, \bibinfo {author} {\bibfnamefont {M.}~\bibnamefont {Inguscio}}, \bibinfo {author} {\bibfnamefont {K.}~\bibnamefont {Xhani}}, \bibinfo {author} {\bibfnamefont {C.}~\bibnamefont {Fort}}, \bibinfo {author} {\bibfnamefont {M.}~\bibnamefont {Modugno}}, \bibinfo {author} {\bibfnamefont {F.}~\bibnamefont {Marino}},\ and\ \bibinfo {author} {\bibfnamefont {G.}~\bibnamefont {Roati}},\ }\bibfield  {title} {\bibinfo {title} {{Connecting shear flow and vortex array instabilities in annular atomic superfluids}},\ }\href {https://doi.org/10.1038/s41567-024-02466-4} {\bibfield  {journal} {\bibinfo  {journal} {Nat. Phys.}\ }\textbf {\bibinfo {volume} {20}},\ \bibinfo {pages} {939}
  (\bibinfo {year} {2024})}\BibitemShut {NoStop}%
\bibitem [{\citenamefont {Caldara}\ \emph {et~al.}(2024)\citenamefont {Caldara}, \citenamefont {Richaud}, \citenamefont {Capone},\ and\ \citenamefont {Massignan}}]{Caldara2024}%
  \BibitemOpen
  \bibfield  {author} {\bibinfo {author} {\bibfnamefont {M.}~\bibnamefont {Caldara}}, \bibinfo {author} {\bibfnamefont {A.}~\bibnamefont {Richaud}}, \bibinfo {author} {\bibfnamefont {M.}~\bibnamefont {Capone}},\ and\ \bibinfo {author} {\bibfnamefont {P.}~\bibnamefont {Massignan}},\ }\bibfield  {title} {\bibinfo {title} {{Suppression of the superfluid Kelvin-Helmholtz instability due to massive vortex cores, friction and confinement}},\ }\href {https://doi.org/10.21468/SciPostPhys.17.3.076} {\bibfield  {journal} {\bibinfo  {journal} {SciPost Phys.}\ }\textbf {\bibinfo {volume} {17}},\ \bibinfo {pages} {076} (\bibinfo {year} {2024})}\BibitemShut {NoStop}%
\end{thebibliography}

\begin{thebibliography}{11}%
\makeatletter
\providecommand \@ifxundefined [1]{%
 \@ifx{#1\undefined}
}%
\providecommand \@ifnum [1]{%
 \ifnum #1\expandafter \@firstoftwo
 \else \expandafter \@secondoftwo
 \fi
}%
\providecommand \@ifx [1]{%
 \ifx #1\expandafter \@firstoftwo
 \else \expandafter \@secondoftwo
 \fi
}%
\providecommand \natexlab [1]{#1}%
\providecommand \enquote  [1]{``#1''}%
\providecommand \bibnamefont  [1]{#1}%
\providecommand \bibfnamefont [1]{#1}%
\providecommand \citenamefont [1]{#1}%
\providecommand \href@noop [0]{\@secondoftwo}%
\providecommand \href [0]{\begingroup \@sanitize@url \@href}%
\providecommand \@href[1]{\@@startlink{#1}\@@href}%
\providecommand \@@href[1]{\endgroup#1\@@endlink}%
\providecommand \@sanitize@url [0]{\catcode `\\12\catcode `\$12\catcode `\&12\catcode `\#12\catcode `\^12\catcode `\_12\catcode `\%12\relax}%
\providecommand \@@startlink[1]{}%
\providecommand \@@endlink[0]{}%
\providecommand \url  [0]{\begingroup\@sanitize@url \@url }%
\providecommand \@url [1]{\endgroup\@href {#1}{\urlprefix }}%
\providecommand \urlprefix  [0]{URL }%
\providecommand \Eprint [0]{\href }%
\providecommand \doibase [0]{http://dx.doi.org/}%
\providecommand \selectlanguage [0]{\@gobble}%
\providecommand \bibinfo  [0]{\@secondoftwo}%
\providecommand \bibfield  [0]{\@secondoftwo}%
\providecommand \translation [1]{[#1]}%
\providecommand \BibitemOpen [0]{}%
\providecommand \bibitemStop [0]{}%
\providecommand \bibitemNoStop [0]{.\EOS\space}%
\providecommand \EOS [0]{\spacefactor3000\relax}%
\providecommand \BibitemShut  [1]{\csname bibitem#1\endcsname}%
\let\auto@bib@innerbib\@empty
\bibitem [{\citenamefont {Bereta}\ \emph {et~al.}(2021)\citenamefont {Bereta}, \citenamefont {Caracanhas},\ and\ \citenamefont {Fetter}}]{SM-Bereta2021}%
  \BibitemOpen
  \bibfield  {author} {\bibinfo {author} {\bibfnamefont {S.~J.}\ \bibnamefont {Bereta}}, \bibinfo {author} {\bibfnamefont {M.~A.}\ \bibnamefont {Caracanhas}}, \ and\ \bibinfo {author} {\bibfnamefont {A.~L.}\ \bibnamefont {Fetter}},\ }\bibfield  {title} {\bibinfo {title} {\emph {Superfluid vortex dynamics on a spherical film}},\ }\href {\doibase 10.1103/PhysRevA.103.053306} {\bibfield  {journal} {\bibinfo  {journal} {Phys. Rev. A}\ }\textbf {\bibinfo {volume} {103}},\ \bibinfo {pages} {053306} (\bibinfo {year} {2021})}\BibitemShut {NoStop}%
\bibitem [{\citenamefont {Schulze}\ \emph {et~al.}(2018)\citenamefont {Schulze}, \citenamefont {Hartmann}, \citenamefont {Voges}, \citenamefont {Gempel}, \citenamefont {Tiemann}, \citenamefont {Zenesini},\ and\ \citenamefont {Ospelkaus}}]{SM-Schulze2018}%
  \BibitemOpen
  \bibfield  {author} {\bibinfo {author} {\bibfnamefont {T.~A.}\ \bibnamefont {Schulze}}, \bibinfo {author} {\bibfnamefont {T.}~\bibnamefont {Hartmann}}, \bibinfo {author} {\bibfnamefont {K.~K.}\ \bibnamefont {Voges}}, \bibinfo {author} {\bibfnamefont {M.~W.}\ \bibnamefont {Gempel}}, \bibinfo {author} {\bibfnamefont {E.}~\bibnamefont {Tiemann}}, \bibinfo {author} {\bibfnamefont {A.}~\bibnamefont {Zenesini}}, \ and\ \bibinfo {author} {\bibfnamefont {S.}~\bibnamefont {Ospelkaus}},\ }\bibfield  {title} {\bibinfo {title} {\emph {{Feshbach spectroscopy and dual-species {Bose-Einstein} condensation of $^{23}\mathrm{Na}\text{\ensuremath{-}}^{39}\mathrm{K}$ mixtures}}},\ }\href {\doibase 10.1103/PhysRevA.97.023623} {\bibfield  {journal} {\bibinfo  {journal} {Phys. Rev. A}\ }\textbf {\bibinfo {volume} {97}},\ \bibinfo {pages} {023623} (\bibinfo {year} {2018})}\BibitemShut {NoStop}%
\bibitem [{\citenamefont {Dubessy}\ and\ \citenamefont {Perrin}(2025)}]{SM-Dubessy2025}%
  \BibitemOpen
  \bibfield  {author} {\bibinfo {author} {\bibfnamefont {R.}~\bibnamefont {Dubessy}}\ and\ \bibinfo {author} {\bibfnamefont {H.}~\bibnamefont {Perrin}},\ }\bibfield  {title} {\bibinfo {title} {\emph {Quantum gases in bubble traps}},\ }\href {\doibase 10.1116/5.0242948} {\bibfield  {journal} {\bibinfo  {journal} {AVS Quantum Sci.}\ }\textbf {\bibinfo {volume} {7}},\ \bibinfo {pages} {010501} (\bibinfo {year} {2025})}\BibitemShut {NoStop}%
\bibitem [{\citenamefont {McEwen}\ and\ \citenamefont {collaborators}(2025)}]{SM-SSHT}%
  \BibitemOpen
  \bibfield  {author} {\bibinfo {author} {\bibfnamefont {J.~D.}\ \bibnamefont {McEwen}}\ and\ \bibinfo {author} {\bibnamefont {collaborators}},\ }\href {https://astro-informatics.github.io/ssht/} {\bibinfo {title} {\emph {SSHT: Spin Spherical Harmonic Transforms}}} (\bibinfo {year} {2025}),\ \bibinfo {note} {software package}\BibitemShut {NoStop}%
\bibitem [{\citenamefont {McEwen}\ and\ \citenamefont {Wiaux}(2011)}]{SM-McEwen2011}%
  \BibitemOpen
  \bibfield  {author} {\bibinfo {author} {\bibfnamefont {J.~D.}\ \bibnamefont {McEwen}}\ and\ \bibinfo {author} {\bibfnamefont {Y.}~\bibnamefont {Wiaux}},\ }\bibfield  {title} {\bibinfo {title} {\emph {A novel sampling theorem on the sphere}},\ }\href {\doibase 10.1109/TSP.2011.2166394} {\bibfield  {journal} {\bibinfo  {journal} {IEEE Trans. Signal Process.}\ }\textbf {\bibinfo {volume} {59}},\ \bibinfo {pages} {5876} (\bibinfo {year} {2011})}\BibitemShut {NoStop}%
\bibitem [{\citenamefont {Wolf}\ \emph {et~al.}(2022)\citenamefont {Wolf}, \citenamefont {Boegel}, \citenamefont {Meister}, \citenamefont {Bala\ifmmode~\check{z}\else \v{z}\fi{}}, \citenamefont {Gaaloul},\ and\ \citenamefont {Efremov}}]{SM-Wolf2022}%
  \BibitemOpen
  \bibfield  {author} {\bibinfo {author} {\bibfnamefont {A.}~\bibnamefont {Wolf}}, \bibinfo {author} {\bibfnamefont {P.}~\bibnamefont {Boegel}}, \bibinfo {author} {\bibfnamefont {M.}~\bibnamefont {Meister}}, \bibinfo {author} {\bibfnamefont {A.}~\bibnamefont {Bala\ifmmode~\check{z}\else \v{z}\fi{}}}, \bibinfo {author} {\bibfnamefont {N.}~\bibnamefont {Gaaloul}}, \ and\ \bibinfo {author} {\bibfnamefont {M.~A.}\ \bibnamefont {Efremov}},\ }\bibfield  {title} {\bibinfo {title} {\emph {Shell-shaped {Bose-Einstein} condensates based on dual-species mixtures}},\ }\href {\doibase 10.1103/PhysRevA.106.013309} {\bibfield  {journal} {\bibinfo  {journal} {Phys. Rev. A}\ }\textbf {\bibinfo {volume} {106}},\ \bibinfo {pages} {013309} (\bibinfo {year} {2022})}\BibitemShut {NoStop}%
\bibitem [{\citenamefont {Tononi}\ \emph {et~al.}(2020)\citenamefont {Tononi}, \citenamefont {Cinti},\ and\ \citenamefont {Salasnich}}]{SM-Tononi2020}%
  \BibitemOpen
  \bibfield  {author} {\bibinfo {author} {\bibfnamefont {A.}~\bibnamefont {Tononi}}, \bibinfo {author} {\bibfnamefont {F.}~\bibnamefont {Cinti}}, \ and\ \bibinfo {author} {\bibfnamefont {L.}~\bibnamefont {Salasnich}},\ }\bibfield  {title} {\bibinfo {title} {\emph {Quantum Bubbles in Microgravity}},\ }\href {\doibase 10.1103/PhysRevLett.125.010402} {\bibfield  {journal} {\bibinfo  {journal} {Phys. Rev. Lett.}\ }\textbf {\bibinfo {volume} {125}},\ \bibinfo {pages} {010402} (\bibinfo {year} {2020})}\BibitemShut {NoStop}%
\bibitem [{\citenamefont {Aref}(1995)}]{SM-Aref1995}%
  \BibitemOpen
  \bibfield  {author} {\bibinfo {author} {\bibfnamefont {H.}~\bibnamefont {Aref}},\ }\bibfield  {title} {\bibinfo {title} {\emph {On the equilibrium and stability of a row of point vortices}},\ }\href {\doibase 10.1017/s002211209500245x} {\bibfield  {journal} {\bibinfo  {journal} {J. Fluid Mech.}\ }\textbf {\bibinfo {volume} {290}},\ \bibinfo {pages} {167–181} (\bibinfo {year} {1995})}\BibitemShut {NoStop}%
\bibitem [{\citenamefont {Hernandez-Rajkov}\ \emph {et~al.}(2024)\citenamefont {Hernandez-Rajkov}, \citenamefont {Grani}, \citenamefont {Scazza}, \citenamefont {Pace}, \citenamefont {Kwon}, \citenamefont {Inguscio}, \citenamefont {Xhani}, \citenamefont {Fort}, \citenamefont {Modugno}, \citenamefont {Marino},\ and\ \citenamefont {Roati}}]{SM-Hernandez2024}%
  \BibitemOpen
  \bibfield  {author} {\bibinfo {author} {\bibfnamefont {D.}~\bibnamefont {Hernandez-Rajkov}}, \bibinfo {author} {\bibfnamefont {N.}~\bibnamefont {Grani}}, \bibinfo {author} {\bibfnamefont {F.}~\bibnamefont {Scazza}}, \bibinfo {author} {\bibfnamefont {G.~D.}\ \bibnamefont {Pace}}, \bibinfo {author} {\bibfnamefont {W.~J.}\ \bibnamefont {Kwon}}, \bibinfo {author} {\bibfnamefont {M.}~\bibnamefont {Inguscio}}, \bibinfo {author} {\bibfnamefont {K.}~\bibnamefont {Xhani}}, \bibinfo {author} {\bibfnamefont {C.}~\bibnamefont {Fort}}, \bibinfo {author} {\bibfnamefont {M.}~\bibnamefont {Modugno}}, \bibinfo {author} {\bibfnamefont {F.}~\bibnamefont {Marino}}, \ and\ \bibinfo {author} {\bibfnamefont {G.}~\bibnamefont {Roati}},\ }\bibfield  {title} {\bibinfo {title} {\emph {{Connecting shear flow and vortex array instabilities in annular atomic superfluids}}},\ }\href {\doibase 10.1038/s41567-024-02466-4} {\bibfield  {journal} {\bibinfo  {journal} {Nat. Phys.}\ }\textbf {\bibinfo {volume} {20}},\ \bibinfo {pages} {939}
  (\bibinfo {year} {2024})}\BibitemShut {NoStop}%
\bibitem [{\citenamefont {Caldara}\ \emph {et~al.}(2024)\citenamefont {Caldara}, \citenamefont {Richaud}, \citenamefont {Capone},\ and\ \citenamefont {Massignan}}]{SM-Caldara2024}%
  \BibitemOpen
  \bibfield  {author} {\bibinfo {author} {\bibfnamefont {M.}~\bibnamefont {Caldara}}, \bibinfo {author} {\bibfnamefont {A.}~\bibnamefont {Richaud}}, \bibinfo {author} {\bibfnamefont {M.}~\bibnamefont {Capone}}, \ and\ \bibinfo {author} {\bibfnamefont {P.}~\bibnamefont {Massignan}},\ }\bibfield  {title} {\bibinfo {title} {\emph {{Suppression of the superfluid Kelvin-Helmholtz instability due to massive vortex cores, friction and confinement}}},\ }\href {\doibase 10.21468/SciPostPhys.17.3.076} {\bibfield  {journal} {\bibinfo  {journal} {SciPost Phys.}\ }\textbf {\bibinfo {volume} {17}},\ \bibinfo {pages} {076} (\bibinfo {year} {2024})}\BibitemShut {NoStop}%
\bibitem [{\citenamefont {Richaud}\ \emph {et~al.}(2021)\citenamefont {Richaud}, \citenamefont {Penna},\ and\ \citenamefont {Fetter}}]{SM-Richaud2021PRA}%
  \BibitemOpen
  \bibfield  {author} {\bibinfo {author} {\bibfnamefont {A.}~\bibnamefont {Richaud}}, \bibinfo {author} {\bibfnamefont {V.}~\bibnamefont {Penna}}, \ and\ \bibinfo {author} {\bibfnamefont {A.~L.}\ \bibnamefont {Fetter}},\ }\bibfield  {title} {\bibinfo {title} {\emph {{Dynamics of massive point vortices in a binary mixture of {Bose-Einstein} condensates}}},\ }\href {\doibase 10.1103/PhysRevA.103.023311} {\bibfield  {journal} {\bibinfo  {journal} {Phys. Rev. A}\ }\textbf {\bibinfo {volume} {103}},\ \bibinfo {pages} {023311} (\bibinfo {year} {2021})}\BibitemShut {NoStop}%
\end{thebibliography}
\end{document}